\documentclass[openacc]{rstransa}

\usepackage{multirow}
\begin{document}
\newcommand{\PP}{{\textrm{\tiny P}}}
\newcommand{\AP}{{\textrm{\tiny A$\!$P}}}
\newcommand{\beq} {\begin{equation}}
\newcommand{\eeq} {\end{equation}}
\newcommand{\ber} {\begin{eqnarray}}
\newcommand{\eer} {\end{eqnarray}}
\newcommand{\nn}{\nonumber }
\newcommand{\mat}{\left(\begin{array}{cc}}
\newcommand{\matend}{\end{array}\right)}
\newcommand{\matl}{\left(\begin{array}{c}}
\newcommand{\matlend}{\end{array}\right)}
\newcommand{\eps}{\varepsilon}
\newcommand{\ce}{{\rm e}}

\newcommand{\mbf}[1]{\mbox{\boldmath $#1$}}
\newcommand{\mbfsc}[1]{\mbox{\scriptsize\boldmath $#1$}}
\newcommand{\mvec}[1]{{\mbf{#1}}}
\renewcommand{\vec}[1]{{\mathbf{#1}}}
\newcommand{\vecsc}[1]{{\mbfsc{#1}}}
\newcommand*{\vnabla}{\mvec{\nabla}}
\newcommand{\bv}{{\mbf p}_F}
\newcommand{\pf}{{\mbf p}_F}
\newcommand{\vf}{{\mbf v}_F}
\newcommand{\qpartial}{\vf \! \cdot \vnabla }
\newcommand{{\ra}}{{\scriptscriptstyle R,A}}
\newcommand{{\rak}}{{\scriptscriptstyle R,A,K}}
\newcommand{{\ret}}{{\scriptscriptstyle R}}
\newcommand{{\adv}}{{\scriptscriptstyle A}}
\newcommand{\kel}{{\scriptscriptstyle K}}
\newcommand{\mats}{{\scriptscriptstyle M}}
\newcommand{\ano}{{\scriptstyle a}}
\newcommand{\adj}[1]{#1^\star}
\newcommand{\hc}[1]{#1^\dagger}
\newcommand{\tr}[1]{\underline{#1}^\dagger}
\newcommand{\Nr}{\op{N}^{\ret}}
\newcommand{\Na}{\op{N}^{\adv}}
\newcommand{\Nra}{\op{N}^{\ra}}
\newcommand{\Da}{\Delta}
\newcommand{\Dm}{\op{\Da }}
\newcommand{\Db}{\tilde{\Da}}
\newcommand{\Dar}{\Da^{\ret}}
\newcommand{\Dbr}{\Db^{\ret}}
\newcommand{\Daa}{\Da^{\adv}}
\newcommand{\Dba}{\Db^{\adv}}
\newcommand{\Dara}{\Da^{\ra}}
\newcommand{\Dbra}{\Db^{\ra}}
\newcommand{\Dam}{\Da^{\mats}}
\newcommand{\Dbm}{\Db^{\mats}}
\newcommand{\Dak}{\Da^{\kel}}
\newcommand{\Dbk}{\Db^{\kel}}
\newcommand{\dDa}{\de \! \Da}
\newcommand{\dDb}{\de \! \Db}
\newcommand{\dDar}{\de \! \Dar}
\newcommand{\dDbr}{\de \! \Dbr}
\newcommand{\dDaa}{\de \! \Daa}
\newcommand{\dDba}{\de \! \Dba}
\newcommand{\dDara}{\de \! \Dara}
\newcommand{\dDbra}{\de \! \Dbra}
\newcommand{\dDak}{\de \! \Dak}
\newcommand{\dDbk}{\de \! \Dbk}
\newcommand{\dDaan}{\de \! \Da^{\ano}}
\newcommand{\dDban}{\de \! \Db^{\ano}}
\newcommand{\dDm}{\de \! \op{\Da }}
\newcommand{\Ea}{E}
\newcommand{\Eb}{\tilde{\Ea}}
\newcommand{\Ear}{\Ea^{\ret}}
\newcommand{\Ebr}{\Eb^{\ret}}
\newcommand{\Eaa}{\Ea^{\adv}}
\newcommand{\Eba}{\Eb^{\adv}}
\newcommand{\Eara}{\Ea^{\ra}}
\newcommand{\Ebra}{\Eb^{\ra}}
\newcommand{\Eak}{\Ea^{\kel}}
\newcommand{\Ebk}{\Eb^{\kel}}
\newcommand{\Eam}{\Ea^{\mats}}
\newcommand{\Ebm}{\Eb^{\mats}}
\newcommand{\Earag}{E^{\ra}}
\newcommand{\Earg}{E^{\ret}}
\newcommand{\Eaag}{E^{\adv}}
\newcommand{\Eamg}{E^{\mats}}
\newcommand{\dfk}{\de \! \op{f}^{\kel}}
\newcommand{\dfr}{\de \! \op{f}^{\ret}}
\newcommand{\dfa}{\de \! \op{f}^{\adv}}
\newcommand{\dfra}{\de \! \op{f}^{\ra}}
\newcommand{\fk}{\op{f}^{\kel}}
\newcommand{\fr}{\op{f}^{\ret}}
\newcommand{\fa}{\op{f}^{\adv}}
\newcommand{\fra}{\op{f}^{\ra}}
\newcommand{\hk}{\op{h}^{\kel}}
\newcommand{\hr}{\op{h}^{\ret}}
\newcommand{\ha}{\op{h}^{\adv}}
\newcommand{\hra}{\op{h}^{\ra}}
\newcommand{\gmk}{\op{g}^{\kel}}
\newcommand{\gmr}{\op{g}^{\ret}}
\newcommand{\gma}{\op{g}^{\adv}}
\newcommand{\gmra}{\op{g}^{\ra}}
\newcommand{\ep}{\varepsilon }
\newcommand{\qt}{{\, \scriptstyle \circ}\, }
\newcommand{\op}[1]{\hat{#1}}
\newcommand{\ta}{\op{\tau }_1 }
\newcommand{\tb}{\op{\tau }_2 }
\newcommand{\tc}{\op{\tau }_3 }
\newcommand{\te}{\op{1} }
\newcommand{\tz}{\op{0} }
\newcommand{\plus}{\; \; \, }
\newcommand{\va}{\Sigma}
\newcommand{\vb}{\tilde{\va}}
\newcommand{\vak}{\va^{\kel}}
\newcommand{\vbk}{\vb^{\kel}}
\newcommand{\vara}{\va^{\ra}}
\newcommand{\vbra}{\vb^{\ra}}
\newcommand{\var}{\va^{\ret}}
\newcommand{\vbr}{\vb^{\ret}}
\newcommand{\vaa}{\va^{\adv}}
\newcommand{\vba}{\vb^{\adv}}
\newcommand{\dva}{\de \! \va}
\newcommand{\dvb}{\de \! \vb}
\newcommand{\dvak}{\de \! \vak}
\newcommand{\dvbk}{\de \! \vbk}
\newcommand{\dvara}{\de \! \vara}
\newcommand{\dvbra}{\de \! \vbra}
\newcommand{\dvar}{\de \! \var}
\newcommand{\dvbr}{\de \! \vbr}
\newcommand{\dvaa}{\de \! \vaa}
\newcommand{\dvba}{\de \! \vba}
\newcommand{\dvaan}{\de \! \va^{\ano}}
\newcommand{\dvban}{\de \! \vb^{\ano}}
\newcommand{\gq}{\op{g}}
\newcommand{\gqr}{\gq^{\ret}}
\newcommand{\gqa}{\gq^{\adv}}
\newcommand{\gqra}{\gq^{\ra}}
\newcommand{\gqk}{\gq^{\kel}}
\newcommand{\gqm}{\gq^{\mats}}
\newcommand{\dgq}{\de \! \gq}
\newcommand{\dgqr}{\de \! \gqr}
\newcommand{\dgqa}{\de \! \gqa}
\newcommand{\dgqra}{\de \! \gqra}
\newcommand{\dgqk}{\de \! \gqk}
\newcommand{\dgqan}{\de \! \gq^{\ano}}
\newcommand{\gaq}{U}
\newcommand{\gbq}{\tilde{U}}
\newcommand{\gaqr}{\gaq^{\ret}}
\newcommand{\gbqr}{\gbq^{\ret}}
\newcommand{\gaqa}{\gaq^{\adv}}
\newcommand{\gbqa}{\gbq^{\adv}}
\newcommand{\gaqra}{\gaq^{\ra}}
\newcommand{\gbqra}{\gbq^{\ra}}
\newcommand{\gaqk}{\gaq^{\kel}}
\newcommand{\gbqk}{\gbq^{\kel}}
\newcommand{\gaqm}{\gaq^{\mats}}
\newcommand{\gbqm}{\gbq^{\mats}}
\newcommand{\faq}{W}
\newcommand{\fbq}{\tilde{W}}
\newcommand{\faqr}{\faq^{\ret}}
\newcommand{\fbqr}{\fbq^{\ret}}
\newcommand{\faqa}{\faq^{\adv}}
\newcommand{\fbqa}{\fbq^{\adv}}
\newcommand{\faqra}{\faq^{\ra}}
\newcommand{\fbqra}{\fbq^{\ra}}
\newcommand{\faqk}{\faq^{\kel}}
\newcommand{\fbqk}{\fbq^{\kel}}
\newcommand{\faqm}{\faq^{\mats}}
\newcommand{\fbqm}{\fbq^{\mats}}
\newcommand{\dfaqk}{\de \! \faq^{\kel}}
\newcommand{\dfbqk}{\de \! \fbq^{\kel}}
\newcommand{\haq}{V}
\newcommand{\hbq}{\tilde{V}}
\newcommand{\haqr}{\haq^{\ret}}
\newcommand{\hbqr}{\hbq^{\ret}}
\newcommand{\haqa}{\haq^{\adv}}
\newcommand{\hbqa}{\hbq^{\adv}}
\newcommand{\haqra}{\haq^{\ra}}
\newcommand{\hbqra}{\hbq^{\ra}}
\newcommand{\haqk}{\haq^{\kel}}
\newcommand{\hbqk}{\hbq^{\kel}}
\newcommand{\haqm}{\haq^{\mats}}
\newcommand{\hbqm}{\hbq^{\mats}}
\newcommand{\dhaqk}{\de \! \haq^{\kel}}
\newcommand{\dhbqk}{\de \! \hbq^{\kel}}

\newcommand{\PPa}{\delta \Pa}
\newcommand{\PPb}{\delta \Pb}

\newcommand{\vga}{\gvec{\gamma}}
\newcommand{\vgb}{\tilde{\vga}}
\newcommand{\ppa}{\rho}
\newcommand{\ppb}{\tilde\ppa}
\newcommand{\ppar}{\ppa^{\ret}}
\newcommand{\ppaa}{\ppa^{\adv}}
\newcommand{\ppbr}{\ppb^{\ret}}
\newcommand{\ppba}{\ppb^{\adv}}
\newcommand{\ppara}{\ppa^{\ra}}
\newcommand{\ppbra}{\ppb^{\ra}}
\newcommand{\ga}{\gamma}
\newcommand{\gb}{\tilde{\ga}}
\newcommand{\gar}{\ga^{\ret}}
\newcommand{\gbr}{\gb^{\ret}}
\newcommand{\gaa}{\ga^{\adv}}
\newcommand{\gba}{\gb^{\adv}}
\newcommand{\gara}{\ga^{\ra}}
\newcommand{\gbra}{\gb^{\ra}}
\newcommand{\Ga}{\Gamma}
\newcommand{\Gb}{\tilde{\Ga}}
\newcommand{\Gar}{\Ga^{\ret}}
\newcommand{\Gbr}{\Gb^{\ret}}
\newcommand{\Gaa}{\Ga^{\adv}}
\newcommand{\Gba}{\Gb^{\adv}}
\newcommand{\Gara}{\Ga^{\ra}}
\newcommand{\Gbra}{\Gb^{\ra}}
\newcommand{\gam}{\ga^{\mats}}
\newcommand{\gbm}{\gb^{\mats}}
\newcommand{\dga}{\de \! \ga}
\newcommand{\dgb}{\de \! \gb}
\newcommand{\dgar}{\de \! \gar}
\newcommand{\dgbr}{\de \! \gbr}
\newcommand{\dgaa}{\de \! \gaa}
\newcommand{\dgba}{\de \! \gba}
\newcommand{\dgara}{\de \! \gara}
\newcommand{\dgbra}{\de \! \gbra}
\newcommand{\Gam}{\Ga^{\mats}}
\newcommand{\Gbm}{\Gb^{\mats}}
\newcommand{\dGa}{\de \! \Ga}
\newcommand{\dGb}{\de \! \Gb}
\newcommand{\dGar}{\de \! \Gar}
\newcommand{\dGbr}{\de \! \Gbr}
\newcommand{\dGaa}{\de \! \Gaa}
\newcommand{\dGba}{\de \! \Gba}
\newcommand{\dGara}{\de \! \Gara}
\newcommand{\dGbra}{\de \! \Gbra}

\newcommand{\xa}{x}
\newcommand{\xb}{\tilde{x}}
\newcommand{\dxak}{\de \! x^{\kel}}
\newcommand{\dxaan}{\de \! x^{\ano}}
\newcommand{\dxban}{\de \! \tilde{x}^{\ano}}
\newcommand{\dxbk}{\de \! \tilde{x}^{\kel}}
\newcommand{\dXak}{\de \! X^{\kel}}
\newcommand{\dXbk}{\de \! \tilde{X}^{\kel}}
\newcommand{\Xa}{X^{\kel}}
\newcommand{\Xb}{\tilde{X}^{\kel}}
\newcommand{\Fe}{\Fa_{eq}}
\newcommand{\Fa}{F}
\newcommand{\Fb}{\tilde{\Fa}}
\newcommand{\dxa}{\de \! x^{\ano}}
\newcommand{\dxb}{\de \! \tilde{x}^{\ano}}
\newcommand{\dXa}{\de \! X^{\ano}}
\newcommand{\dXb}{\de \! \tilde{X}^{\ano}}
\newcommand{\dFa}{\de \! \Fa}
\newcommand{\dFb}{\de \! \Fb}
\newcommand{\XX}{\op{X}^{\kel}}
\newcommand{\dXXra}{\de \! \op{X}^{\ra}}
\newcommand{\dYYra}{\de \! \op{Y}^{\ra}}
\newcommand{\dXXan}{\de \! \op{X}^{\ano}}
\newcommand{\dYYan}{\de \! \op{Y}^{\ano}}
\newcommand{\YY}{\op{Y}^{\kel}}
\newcommand{\dxx}{\de \! \op{x}^{\ano}}
\newcommand{\dyy}{\de \! \op{y}^{\ano}}
\newcommand{\dxxra}{\de \! \op{x}^{\ra}}
\newcommand{\dyyra}{\de \! \op{y}^{\ra}}

\newcommand{\state}{coherence }
\newcommand{\State}{COHERENCE }

\title{Theory of Andreev Bound States in S-F-S Junctions and S-F Proximity Devices}

\author{
M. Eschrig$^{1}$}

\address{$^{1}$Department of Physics, Royal Holloway, University of London, Egham, Surrey TW20 0EX, United Kingdom
}

\subject{Solid state physics, superconductivity, spintronics}


\corres{Matthias Eschrig\\
\email{matthias.eschrig@rhul.ac.uk}}

\begin{abstract}
Andreev bound states are an expression of quantum coherence between particles and holes in hybrid structures composed of superconducting and non-superconducting metallic parts. Their spectrum carries important information on the nature of the pairing, and determines the current in Josephson devices. Here I focus on Andreev bound states in systems involving superconductors and ferromagnets with strong spin-polarization. I provide a general framework for non-local Andreev phenomena in such structures in terms of
coherence functions, and show how the latter link
wave-function and Green-function based theories.
\end{abstract}


\begin{fmtext}
\section{Andreev reflection phenomena}

In an isotropic non-magnetic superconductor the normal-state single-particle excitation spectrum $\varepsilon_\vec{k}$ is modified in the superconducting state to $E_\vec{k}=[(\varepsilon_\vec{k}-\mu)^2+\Delta^2]^{1/2}$, acquiring a gap $\Delta$ around the electrochemical potential $\mu $, and the density of states is characterized by a coherence peak just above the gap, accounting for the missing sub-gap states. This spectral signature, predicted by Bardeen-Cooper-Schrieffer (BCS) theory \cite{Cooper56,Bardeen57} has been first observed by infrared absorption spectroscopy \cite{Glover56,Tinkham56} and by tunneling spectroscopy \cite{Giaever60}. 

The spectral features above the gap may show information about electron-phonon interaction (or about interaction with some low-energy bosonic modes, e.g. spin-fluctutations \cite{Eschrig06}), or may exhibit geometric interference patterns. 
Features due to electron-phonon interaction, predicted by 
Migdal-Eliashberg theory \cite{Migdal58,Eliashberg60} and studied in detail by Scalapino {\it et al.} \cite{Scalapino66}, were measured early in tunneling experiments by Giaever {\it et al.} \cite{Giaever62}. They are a consequence of electronic particle-hole coherence in a superconductor
and build the basis for the McMillan-Rowell inversion procedure for determining the Eliashberg effective interaction spectrum $\alpha^2F(\omega )$ \cite{McMillan65}.
Geometric interference effects include oscillations
\end{fmtext}
\maketitle
\noindent
in the density of states, such as
Tomasch oscillations in N-I-S and N-I-S-N' tunnel structures with a superconductor of thickness $d_{\rm S}$ (and with transverse Fermi velocity $v_{\rm F,S}$), giving rise to  voltage peaks at ${e}V_n=[(2\Delta)^2+(n\pi\hbar v_{\rm F,S}/d_{\rm S})^2]^{1/2}$ ($n$ integer)
\cite{Tomasch65,McMillan66,Wolfram68}; and Rowell-McMillan oscillations in N-I-N'-S tunnel structures with a normal metal N' of thickness $d_{\rm N'}$ (transverse Fermi velocity $v_{\rm F,N'}$), giving rise to voltage peaks at ${e}V_n=n\pi \hbar v_{\rm F,N'}/2d_{\rm N'}$ \cite{Rowell66}. 
Possible offsets due to spatial variation of the gap $\Delta $ may occur \cite{Nedellec71}.

Inhomogeneous superconducting states exhibit also features at energies inside the gap. Surface bound states 
in a normal metal overlayer on a superconductor were predicted first by de Gennes and Saint-James \cite{deGennes63,James64} and measured by Rowell \cite{Rowell73} and Bellanger {\it et al.} \cite{Bellanger73}. 
These de Gennes-Saint-James bound states have a natural explanation in terms of the so-called Andreev reflection process, an extremely fruitful physical picture suggested in 1964 by Andreev \cite{Andreev64}.
For example, geometric resonances above the gap appear due to Andreev reflection at N-S interfaces, which describe (for normal impact) scattering of a particle at wavevector $k_{{\rm F},x}+(E^2-\Delta^2)^{1/2}/\hbar v_{{\rm F},x}$ into a hole at wavevector
$k_{{\rm F},x}-(E^2-\Delta^2)^{1/2}/\hbar v_{{\rm F},x}$ or vica versa.
Below the gap Andreev reflections lead
to subharmonic gap structure due to multiple Andreev reflections at voltages $V_n=2\Delta/en$ ($n$ integer) in SIS junctions \cite{Klapwijk82} (for a general treatment in diffusive systems see \cite{Cuevas06}), and
control electrical and thermal resistance of a superconductor/normal-metal interface and the Josephson current in a superconductor/normal-metal/superconductor junction.
The Andreev mechanism also gives rise to bound states in various other systems with inhomogeneous superconducting order parameter, which are named in the general case Andreev bound states.

Transport trough an N-S contact is strongly influenced by Andreev scattering, and is described in the single-channel case by the theory of Blonder, Tinkham, and Klapwijk \cite{Blonder82}, generalized to the multi-channel case by Beenakker \cite{Beenakker92}. Andreev scattering at N-S interfaces is the cause of the superconducting proximity effect \cite{deGennes63a,Werthamer63}.
Interference effects in transport appear also as the result of impurity disorder.
In contrast to unconventional superconductors, where normal impurities are pair breaking,
isotropic $s$-wave superconductors are insensitive to scattering from normal impurities for not too high impurity concentration, which is the content of a theorem by Abrikosov and Gor'kov \cite{Abrikosov58,Abrikosov59}, and by Anderson \cite{Anderson59a}.
In strongly disordered superconductors (weak localization regime) the superconducting transition temperature $T_{\rm c}$ is reduced \cite{Finkelstein87}, accompagnied by
localized tail states (similar to Lifshitz tail states in semiconductors \cite{Lifshitz64}) just below 
the gap edge \cite{Larkin71,Balatsky97}.
An interference effect is the
so-called reflectionless tunneling \cite{Kastalsky91,Wees92,Beenakker92,Zaitsev90,Volkov92}, which leads to a zero-bias conductance peak in a diffusive N-I-S structure. It results from multiple scattering of Andreev-reflected coherent particle-hole pairs at impurities, and from the resulting backscattering to the interface barrier, making the barrier effectively transparent near the electrochemical potential for a pair current even in the tunneling regime.

Abrikosov and Gor'kov developed in 1960 a theory for pair-breaking by paramagnetic impurities, showing that at a critical value for the impurity concentration superconductivity is destroyed, and that gapless superconductivity can exist in a narrow region below this critical value \cite{Abrikosov60}. 
Yu \cite{Luh65}, Shiba \cite{Shiba68},
and Rusinov \cite{Rusinov69} (who happened to work isolated from each other in China, Japan, and Russia) independently discovered within the framework of a full $t$-matrix treatment of the problem that local Andreev bound states (now called the Yu-Shiba-Rusinov states) are present within the BCS energy gap due to multiple scattering between conduction electrons and paramagnetic impurities.
Andreev bound states also exist in the cores of vortices in type II superconductors. These are called Caroli-de Gennes-Matricon bound states \cite{Caroli64}, and carry current in the core region of a vortex \cite{Rainer96}. Their dynamics plays a crucial role in the absorption of electromagnetic waves \cite{Eschrig99,Eschrig02,Sauls09}.

In an S-N-I or S-N-S junction, Andreev bound states appear in the normal metal region at energies below the gaps of the superconductors.
The number and distribution of these bound states depend on details such as interface transmission, mean free path, and length of the normal metal $d_{\rm N}$. In general, there is a characteristic energy, the Thouless energy \cite{Thouless72}
(related to the dwell time between Andreev reflections), given by $\hbar v_{\rm F,N}/d_{\rm N}$ for the clean limit, and by $\hbar D_{\rm N}/d_{\rm N}^2$ for the diffusive limit, with Fermi velocity $v_{\rm F,N}$ and diffusion constant $D_{\rm N}$ of the normal metal. In the diffusve limit, the Andreev states build a quasi-continuum below the superconducting gap, whereas in the case of ballistic junctions bands of Andreev bound states arise.
For the case that no superconducting phase gradient is present in the system, however,
a low-energy gap always arises in the spectrum of Andreev states in the normal metal.
This so-called minigap scales for sufficiently thick normal metal layers approximately with its Thouless energy 
and with the transmission probability (possibly further reduced by inelastic scattering processes). It was found first by McMillan \cite{McMillan68} and can be probed by scanning tunneling microscopy \cite{Gueron96}. 
In chaotic Andreev billiards \cite{Kosztin95}, where disorder is restricted to boundaries,
a second time scale, the Ehrenfest time, competes with the dwell time to set the minigap \cite{Lodder98}.

The importance of Andreev bound states in S-N-S Josephson junctions for current transport was first discussed by Kulik in 1969 \cite{Kulik69}. Andreev bound states form in a sufficiently long normal region, which are doubly degenerate (carrying current in opposite direction) for zero
phase difference between the superconducting banks. For a finite phase difference, this degeneracy is lifted.
The gap or minigap in a Josephson structure is reduced and eventually closes when a supercurrent flows across the junction. This is a result of a ``dispersion'' of the energy $E_{\rm b.s.}$ of the Andreev bound states as function of phase difference $\Delta \chi$ between the superconducting banks \cite{Kulik69,Zhou98,leSeur08}.
The contribution of the bound states to the supercurrent is given by $(2e/\hbar) \partial E_{\rm b.s.}/\partial \Delta \chi $, with $e<0$ the electron charge. 
Apart from the current carried by the Andreev bound states, there is also a contribution from continuum states above the gap \cite{Kulik69}.
For a single-channel weak link between two superconductors with normal-state transmission probability $\tau^2 $ there is one pair of Andreev bound states with dispersion $E_\pm=\pm\Delta [1-\tau^2 \sin^2(\Delta \chi/2)]^{1/2}$.

The large size of a Cooper pair in conventional superconductors leads to a pronounced non-locality of Andreev reflection processes. This allows for interference effects due to
crossed Andreev reflection, in which the particle and hole involved in the process enter different normal-state (typically spin-polarized) terminals, which are both simultaneously accessable to one Cooper pair \cite{Byers95,Deutscher00}. This effect has been first experimentally observed by Beckmann {\it et al.} \cite{Beckmann04}.

Finally, an important role is played by Andreev zero modes as topological surface states. Examples are zero-bias states at the surface of a $d$-wave superconductor \cite{Hu94,Tanaka95}, and Majorana zero modes in topological superconductors \cite{Volovik99,Kitaev01} and superfluids \cite{Volovik03}.

\renewcommand{\i}{{\rm i}}
\section{Andreev bound states at magnetically active interfaces}
\label{ABS}
\subsection{Spin-dependent interface scattering phase shifts}
\begin{figure}[t]
\centering{(a) \hspace{1.6in} (b) \hspace{1.6in} (c) \hspace{1.5in}$\;$}\\
\centering{
\includegraphics[width=1.7in]{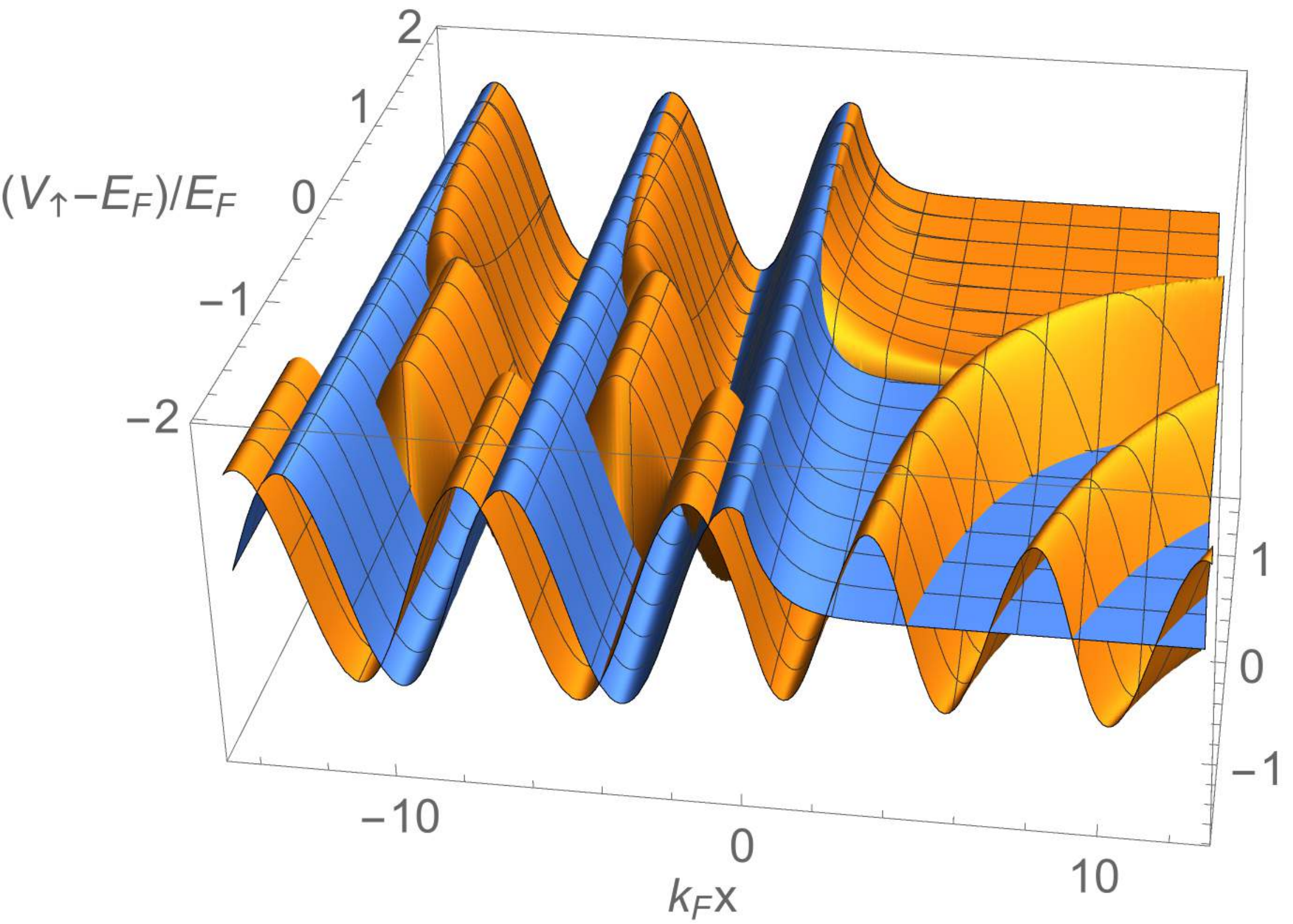}
\includegraphics[width=1.7in]{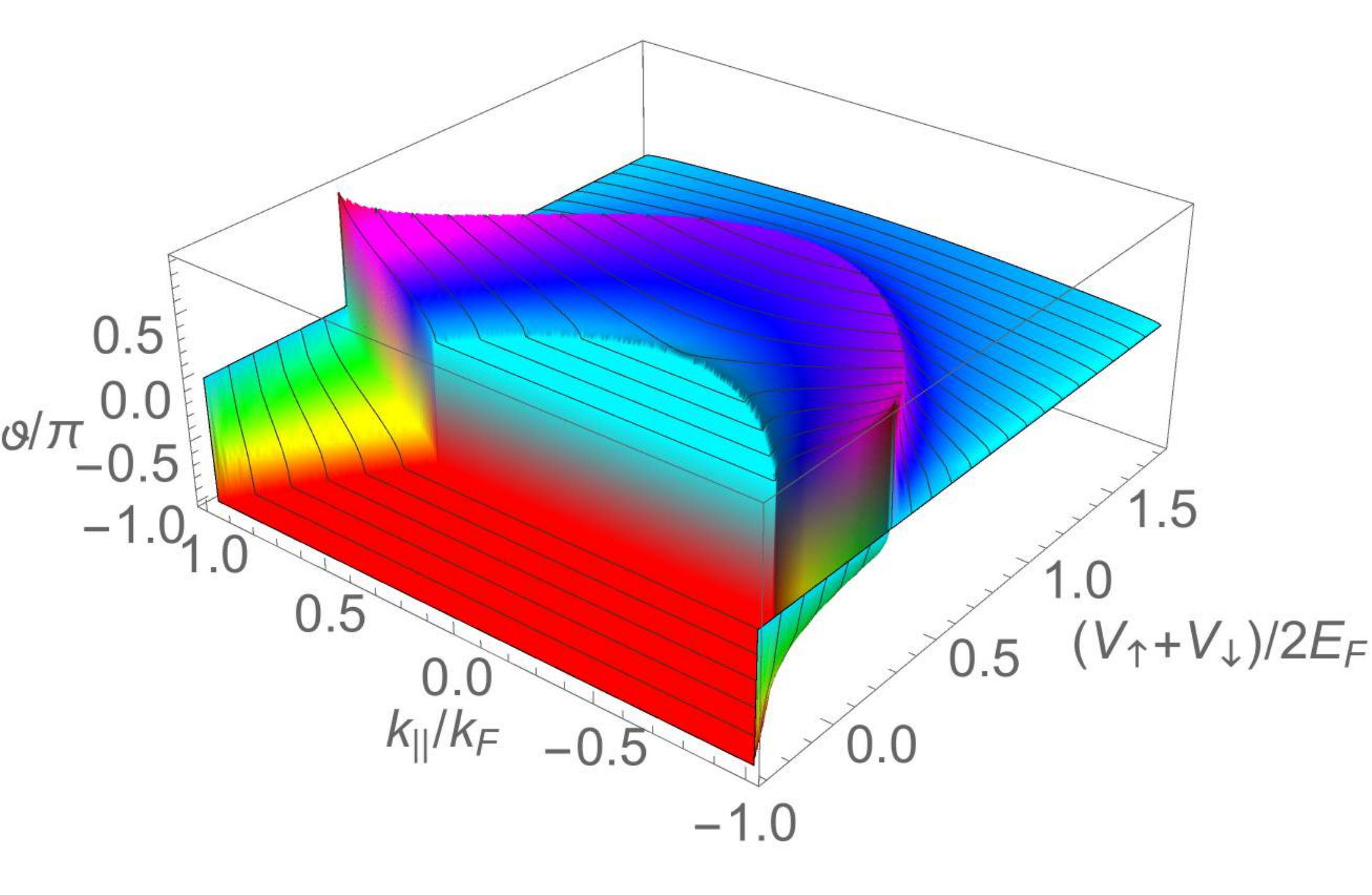}
\includegraphics[width=1.7in]{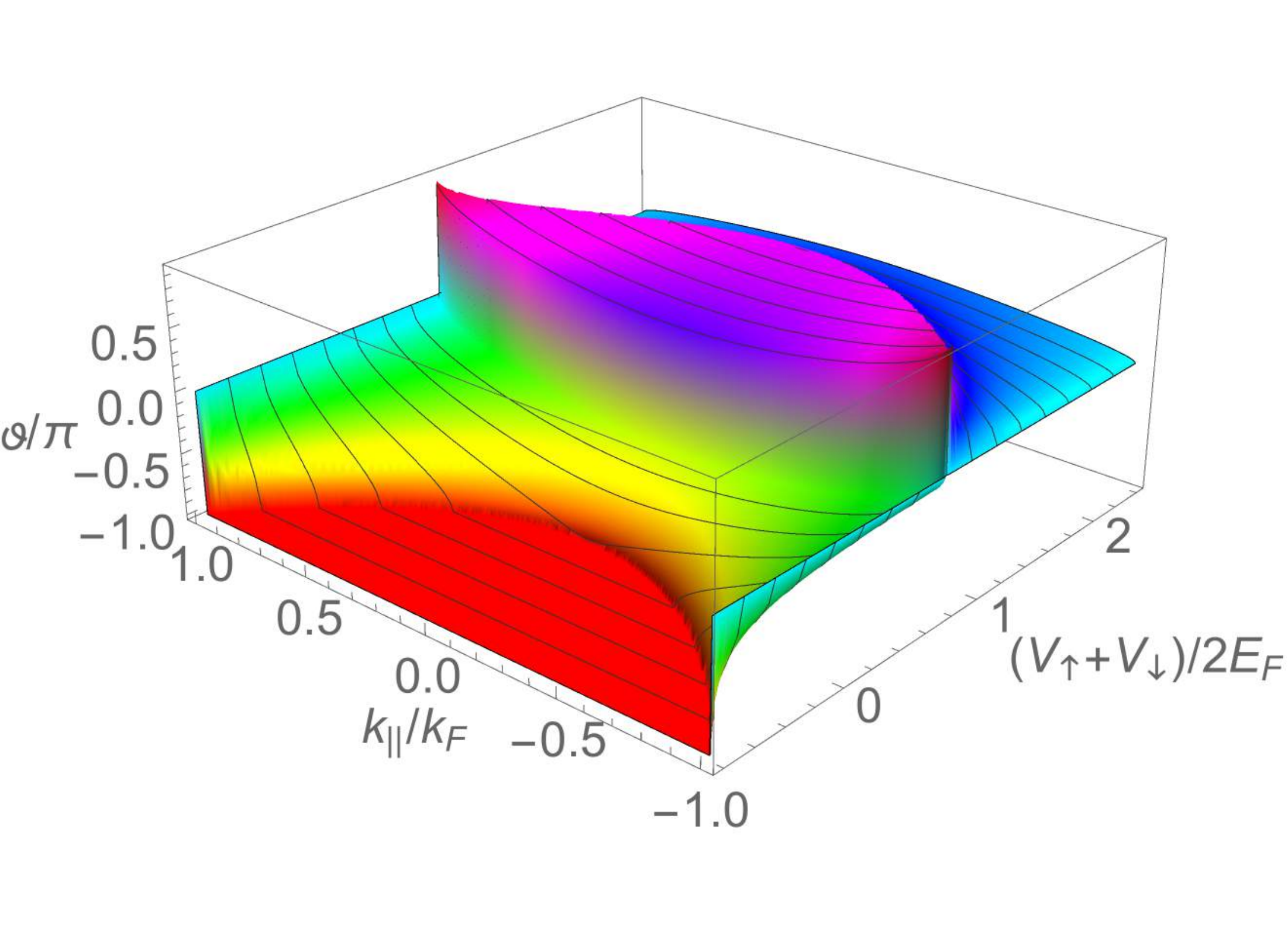}
}
\caption{
(a) Spin dependent scattering phase shifts for Bloch waves with energy $E_{\rm F}$ reflected from an N-FI or N-HM interface (at $x=0$).
Here $V_\downarrow =3E_{\rm F}$ is fixed and $V_\uparrow $ varied from $-E_{\rm F} $ to $3E_{\rm F}$, $k_\parallel =0$. 
For $V_\uparrow>E_{\rm F}$ this describes an FI, for $V_\uparrow<E_{\rm F}$ a HM. For $V_\uparrow<0$ the Fermi surface of the spin-up band in the HM becomes larger than in N.
Shown are for $x<0$ the (normalized) reflected wave, quantified by $-{\rm Im} (r_\sigma\;e^{-\i kx})/|r_\sigma|$, and for $x>0$ the transmitted wave, quantified by $-{\rm Im} (t_\sigma e^{-\kappa_\sigma x})$ (for real $\kappa_\sigma$) or $-{\rm Im}(t_\uparrow e^{\i k_\uparrow x})$ (for real $k_\uparrow$, i.e. $V_\uparrow<E_{\rm F}$). The spin-up wave is shown in orange, the spin-down wave in blue.
(b) Spin mixing angle $\vartheta $ as function of parallel momentum $k_\parallel /k_{\rm F}$ and $\nu=(V_\uparrow+V_\downarrow)/2$ for $J=0.3E_{\rm F}$ (FI for $\nu>1.3$, HM for $\nu>0.7$); (c) the same for $J=0.8E_{\rm F}$ (FI for $\nu>1.8$, HM for $\nu>0.2$).
}
\label{Phasedelay}
\end{figure}

The importance of spin-dependent interface scattering phase shifts for superconducting phenomena has been pioneered in the work of Tokuyasu, Sauls, and Rainer in 1988 \cite{Tokuyasu88}.
Consider an interface between a normal metal (N) at $x<0$ and a ferromagnetic insulator (FI) or a half-metallic ferromagnet (HM) at $x> 0$. 
For simplicity, let us model the FI (or HM) by a single electronic band with energy gap $V_\downarrow$ for spin-down particles and an energy gap $V_\uparrow=V_\downarrow-2J$ for spin-up particles, where $J>0$ denotes an effective exchange field.
The exchange field can be related to an effective magnetic field via $ \mu\vec{B}_{\rm eff}=
\vec{J}$ (for free electrons the magnetic moment is $\mu=\mu_{\rm e}<0$).
Let us assume an incoming Bloch electron with energy $0<E<V_\uparrow $ and spin $\sigma\in\left\{\uparrow,\downarrow\right\}$, reflected back from the interface with amplitude $r_\sigma $.
It is described by a wave function
$\Psi_\sigma (x,\vec{r}_{\parallel })=e^{\i\vec{k}_{\parallel }\vec{r}_{\parallel }} (e^{\i kx}+r_\sigma\;e^{-\i kx})$ at $x<0$ and
$\Psi_\sigma (x, \vec{r}_{\parallel })= t_\sigma e^{\i\vec{k}_{\parallel }\vec{r}_{\parallel}} e^{-\kappa_\sigma x}$ at $x>0$.
For the normal metal 
$\hbar k(E)=[2mE-(\hbar k_{\parallel })^2]^{1/2}$. 
For the FI 
$\hbar \kappa_\sigma(E)=[2m(V_\sigma-E)+(\hbar k_{\parallel })^2]^{1/2}$. The reflection scattering matrix is
\begin{eqnarray}
{\bf S}= \left(\begin{array}{cc} e^{\i\vartheta_\uparrow} &0 \\ 0& e^{\i\vartheta_\downarrow} \end{array}\right), \qquad
e^{\i \vartheta_\uparrow }= r_\uparrow=\frac{k-\i\kappa_\uparrow }{k+\i\kappa_\uparrow}, \quad
e^{\i \vartheta_\downarrow} = r_\downarrow=\frac{k-\i\kappa_\downarrow }{k+\i\kappa_\downarrow}
\label{Smatrix}
\end{eqnarray}
In the range $V_\uparrow<E<V_\downarrow $ the spin-up electron can be transmitted for sufficiently small $k_{\parallel}$ with amplitude $t_\uparrow=2\sqrt{kk_\uparrow }/(k+k_\uparrow)$, where $\hbar k_\uparrow(E)=[2m(E-V_\uparrow)-(\hbar k_{\parallel })^2]^{1/2}$. In this case, the reflection amplitude is also real, and equal to $r_\sigma=(k-k_\uparrow)/(k+k_\uparrow)$. The reflection phase is $-\pi$ for $k<k_\uparrow$ and zero for $k>k_\uparrow$. In figure \ref{Phasedelay} (a)
$V_\downarrow =3E_{\rm F}$ is fixed and $V_\uparrow $ varied from $-E_{\rm F} $ to $3E_{\rm F}$, $k_\parallel =0$. 
A phase shift for reflected waves between the two spin projections appears.
It results from the well-known effect that
reflection from an insulating region results in 
a {\it phase-delay} of the reflected wave with respect to the case of an infinite interface potential. This phase delay appears due to the quantum mechanical penetration of the wave function into the classically forbidden region. The range $V_\uparrow-E_{\rm F}>0$ in figure \ref{Phasedelay} (a) corresponds to an N-FI interface, where both spin-projections are evanescent in FI. Here, the reflected spin-up wave trails that of the spin-down wave, and the effect increases when $V_\uparrow-E_{\rm F}$ approaches zero. The phase $\vartheta =\vartheta_\uparrow-\vartheta_\downarrow $ of the parameter $r_\uparrow r_\downarrow^\ast = |r_\uparrow r_\downarrow| e^{\i\vartheta}$ is called {\it spin-mixing angle} \cite{Tokuyasu88}, or {\it spin-dependent scattering phase shift} \cite{Cottet05}. It is an important parameter for superconducting spintronics. For the N-FI model interface it is given by
\begin{equation}
\tan \frac{\vartheta}{2} =
\tan \frac{\vartheta_\uparrow-\vartheta_\downarrow}{2}= \frac{k(\kappa_\downarrow-\kappa_\uparrow )}{k^2+\kappa_\uparrow\kappa_\downarrow },
\end{equation}
which is positive due to $\kappa_\downarrow>\kappa_\uparrow$.
The range $V_\uparrow-E_{\rm F}<0$ 
corresponds to a N-HM interface, with the spin-up band itinerant in HM. Here, as long as the spin-up Fermi surface in the HM is smaller than that in N (for $-1<(V_\uparrow-E_{\rm F})/E_{\rm F}<0$), the reflection phase in $r_\uparrow $ is zero, and 
\begin{equation}
\tan \frac{\vartheta}{2} =
\tan \frac{-\vartheta_\downarrow}{2}= \frac{\kappa_\downarrow}{k},
\end{equation}
which can aquire large values. Finally, in the range $(V_\uparrow-E_{\rm F})/E_{\rm F}<-1$ the Fermi surface in the HM is larger than in N, which leads to a reflection phase of $\pi$ for spin-up particles, and
\begin{equation}
\tan \frac{\vartheta}{2} =
\tan \frac{\pi-\vartheta_\downarrow}{2}= -\frac{k}{\kappa_\downarrow},
\end{equation}
which now is negative.
In ballistic structures, the spin-mixing angle depends on the momentum $\hbar \vec{k}_\parallel$ parallel to the interface, as illustrated in figure \ref{Phasedelay} (b) and (c) for varying Fermi surface geometry in the ferromagnet, here parameterized by varying $(V_\uparrow+V_\downarrow)/2$ keeping $J$ fixed.
If both spin-bands are itinerant in the ferromagnet (F), then the spin-mixing angle is zero 
(if $k>k_\uparrow, k_\downarrow $ or $k< k_\uparrow, k_\downarrow $)
or $-\pi$ (if $k_\uparrow>k> k_\downarrow $, red areas in figure \ref{Phasedelay} b and c),
unless an interface potential exists, rendering the reflection amplitudes complex valued. In general, the spin-mixing angle should be considered as material parameter, which in addition depends on the impact angle of the incoming electron or on transport channel indices.

Note that the parameter $r_\uparrow r_\downarrow^\ast $ has become well-known in the spintronics community, as it governs the {\it spin mixing conductance} \cite{Brataas00} in spintronics devices.

It is also instructive to study an incoming Bloch-electron polarized in a direction different from the magnetization direction in the ferromagnet. Let us consider the case of a FI.
For a Bloch electron polarized in a direction $\vec{n}(\alpha, \phi)$, parameterized by polar and azimuthal angles, $\alpha $ and $\phi $,
\begin{equation}
\uparrow_{\alpha,\phi} e^{\i \vec{k}_\parallel \vec{r}_\parallel } e^{\i kx} = \left[\cos \frac{\alpha}{2} e^{-\i\frac{\phi}{2}} \uparrow_z + \sin \frac{\alpha}{2} e^{\i\frac{\phi}{2}} \downarrow_z 
\right] e^{\i \vec{k}_\parallel \vec{r}_\parallel } e^{\i kx}
\end{equation}
the reflected wave will have the form
\begin{equation}
\left[\cos \frac{\alpha}{2} e^{-\i\frac{\phi-\vartheta}{2}} \uparrow_z + \sin \frac{\alpha}{2} e^{\i\frac{\phi-\vartheta}{2}} \downarrow_z \right] 
e^{\i \frac{\vartheta_\uparrow+\vartheta_\downarrow}{2}}
e^{\i \vec{k}_\parallel \vec{r}_\parallel } e^{-\i kx}
\equiv \uparrow_{\alpha,\phi-\vartheta} 
e^{\i \bar\vartheta }
e^{\i \vec{k}_\parallel \vec{r}_\parallel } e^{-\i kx}
\label{SpinRot}
\end{equation}
with $\bar\vartheta=(\vartheta_\uparrow+\vartheta_\downarrow)/2$. Similarly, $\downarrow_{\alpha,\phi}$ scatters into $\downarrow_{\alpha,\phi-\vartheta} e^{\i \bar\vartheta}$.
This
means that scattering leads, apart from an unimportant spin-independent phase factor $e^{\i \bar\vartheta}$, to a precession of the spin around the magnetization axis \cite{Tokuyasu88}. The direction of precession depends on the Fermi surface geometries, and is determined by the sign of the spin-mixing angle $\vartheta$.

\subsection{Andreev reflection in an S-N-FI structure}
\label{ABS1}

An important consequence of spin-mixing phases is the appearence of Andreev bound states at magnetically active interfaces, predicted theoretically \cite{Fogelstrom00,Eschrig03,Krawiec04,Zhao04,Annett06,Lofwander10,Metalidis10}, and verified experimentally \cite{Huebler12}.

Consider a superconductor near an interface with a ferromagnetic insulator. Let us assume that the superconducting order parameter is suppressed to zero in a layer of thickness $d$ next to the FI 
interface, such that the structure can be described as a S-N-FI junction with identical normal state parameters in S and N. For simplicity I consider here a spatially constant order parameter in S (extending to the half space $x<0$). The FI ($x>d$) will be parameterized by reflection phases $\vartheta_\uparrow$ and $\vartheta_\downarrow $, with spin-mixing angle $\vartheta =\vartheta_\uparrow-\vartheta_\downarrow $. Solving the corresponding Bogoliubov-de Gennes equations in the superconductor ($\sigma_i$ are spin Pauli matrices, $\sigma_0$ a 2$\times$2 unit spin matrix)
\begin{align}\label{BdG}
\left(\begin{array}{cc}
(-\frac{\hbar^2 \nabla^2}{2m}-\mu) \sigma_0 & \Delta \i\sigma_2\\ -\Delta^\ast \i \sigma_2 & (\frac{\hbar^2 \nabla^2}{2m}+\mu) \sigma_0
\end{array}\right)
\left(\begin{array}{c}
u \\v
\end{array}\right)
=\varepsilon
\left(\begin{array}{c}
u \\v
\end{array}\right)
\end{align}
with spinors $u$ and $v$,
the (still unnormalized) eigenvectors for given energy $\varepsilon $ and $\vec{k}_\parallel=0$ are
\begin{align}\label{sols}
\left(\begin{array}{r}
1\\0\\0\\\tilde \gamma
\end{array}\right) e^{\pm \i k_+x},
\left(\begin{array}{r}
0\\1\\-\tilde \gamma\\0
\end{array}\right) e^{\pm \i k_+x},
\left(\begin{array}{r}
0\\ \gamma\\1\\0
\end{array}\right) e^{\pm \i k_-x},
\left(\begin{array}{r}
-\gamma\\0\\0\\1
\end{array}\right) e^{\pm \i k_-x},
\end{align}
where to first order in $|\Delta|/E_{\rm F}$ the wavevectors are
$k_\pm (\varepsilon) = k_{\rm F} \pm \i  \sqrt{|\Delta|^2 - \varepsilon^2}/(\hbar v_{\rm F})$
and
\begin{align}\label{sols1}
\gamma (\varepsilon) = -\frac{\Delta }{ \varepsilon+\i \sqrt{|\Delta|^2 - \varepsilon^2}}, \quad
\tilde\gamma (\varepsilon) = +\frac{\Delta^\ast }{ \varepsilon+\i \sqrt{|\Delta|^2 - \varepsilon^2}}. 
\end{align}
For $|\epsilon | > |\Delta |$ the expression $\i \sqrt{|\Delta|^2 - \varepsilon^2}$
is replaced by $\varepsilon\sqrt{1-(|\Delta|/\varepsilon)^2}$ (which corresponds to $\varepsilon \to \varepsilon+\i 0^+$ with infinitesimally small positive $0^+$).
In the N layer the solutions are obtained by setting $\Delta=0$. In the FI only evanescent solutions of the form $e^{-\kappa_\uparrow x}$ and $e^{-\kappa_\downarrow x}$ are allowed. The reflection coefficients connect incoming ($e^{\i k_+x}$, $e^{-\i k_-x}$) solutions with outgoing ($e^{-\i k_+x}$, $e^{\i k_-x}$) solutions.
For scattering from electron-like to electron-like Bogoliubov quasiparticles and for electron-like to hole-like Bogoliubov quasiparticles in leading order in $|\Delta|/E_{\rm F}$ they are 
\begin{align}\label{refl}
r_{e\uparrow \to e\uparrow} = \frac{e^{2\i d k_{\rm F}}e^{2\i d\frac{ \varepsilon}{\hbar v_{\rm F}}}e^{\i \vartheta_\uparrow}(1+\gamma \tilde \gamma )}{1+\gamma \tilde \gamma e^{4\i d\frac{ \varepsilon}{\hbar v_{\rm F}}}e^{\i \vartheta }}, &\quad
r_{e\uparrow \to h\downarrow} = \frac{-\tilde \gamma (1-e^{4\i d\frac{ \varepsilon}{\hbar v_{\rm F}}}e^{\i \vartheta })}{1+\gamma \tilde \gamma e^{4\i d\frac{ \varepsilon}{\hbar v_{\rm F}}}e^{\i \vartheta }},\\
r_{h\uparrow \to h\uparrow} = \frac{e^{-2\i d k_{\rm F}}e^{2\i d\frac{ \varepsilon}{\hbar v_{\rm F}}}e^{-\i \vartheta_\uparrow}(1+\gamma \tilde \gamma )}{1+\gamma \tilde \gamma e^{4\i d\frac{ \varepsilon}{\hbar v_{\rm F}}}e^{-\i \vartheta }}, &\quad
r_{h\uparrow \to e\downarrow} = \frac{-\gamma (1-e^{4\i d\frac{ \varepsilon}{\hbar v_{\rm F}}}e^{-\i \vartheta })}{1+\gamma \tilde \gamma e^{4\i d\frac{ \varepsilon}{\hbar v_{\rm F}}}e^{-\i \vartheta }},
\end{align}
and similar relations hold for $\uparrow \leftrightarrow \downarrow $ and simultaneously $\vartheta \to -\vartheta $, $\gamma \to -\gamma $, $\tilde \gamma \to -\tilde \gamma $.
These relations have a simple interpretation. 
The {\it coherence functions} $\gamma $ and $\tilde \gamma $ represent probability amplitudes for hole-to-particle conversion ($-\gamma $) or particle-to-hole conversion ($-\tilde \gamma $), whereas the factors $e^{\i d\frac{ \varepsilon}{\hbar v_{\rm F}}}$ represent the electron-hole dephasing when crossing the N layer. Thus, the factors $\gamma \tilde \gamma e^{4\i d\frac{ \varepsilon}{\hbar v_{\rm F}}}e^{\i \vartheta }$ represent a Rowell-McMillan process of four times crossing N with two reflections from FI (once as particle and once as hole, contributing $e^{\i \vartheta_\uparrow }$ and $e^{-\i \vartheta_\downarrow }$), and two Andreev conversions. 
When this factor equals $-1$, which happens for energies below the gap,
a bound state appears in N due to constructive interference between particles and holes.
Note that for $|\varepsilon |\le |\Delta|$ the coherence functions have unit modulus: $|\gamma|=|\tilde \gamma| = 1$, such that with $\Delta =|\Delta|e^{\i\chi}$ one can write
$\gamma = \i e^{\i \Psi(\varepsilon) }e^{\i \chi} $ and
$\tilde\gamma = -\i e^{\i \Psi(\varepsilon) } e^{-\i \chi}$, with $\sin \Psi = \varepsilon/|\Delta|$, $\cos \Psi >0$.
For $|\varepsilon |< |\Delta|$ only outgoing wavevectors $-k_+$ and $k_-$ lead to normalizable solutions in the superconductor, which are restricted to the bound state energies, given by the solution of $\varepsilon = |\Delta| \cos(\frac{ 2d\varepsilon}{\hbar v_{\rm F}} \pm \frac{\vartheta}{2})\mbox{sign}[\sin(\frac{ 2d\varepsilon}{\hbar v_{\rm F}} \pm \frac{\vartheta}{2})] $.

\begin{figure}[t]
\centering{(a) \hspace{1.5in} (b) \hspace{1.5in} (c) \hspace{1.5in}$\;$}\\
\centering{\includegraphics[width=1.7in]{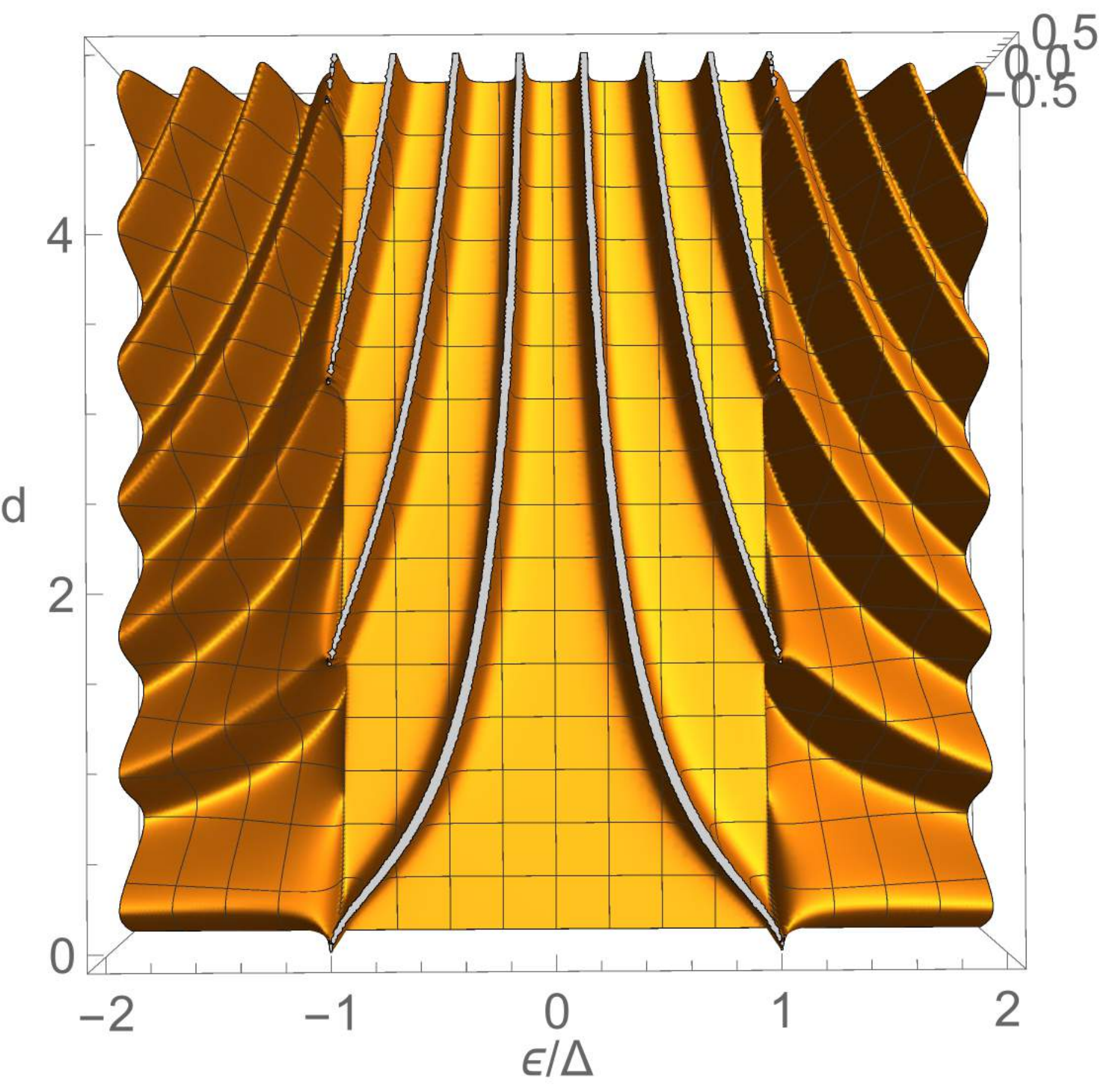}
\includegraphics[width=1.7in]{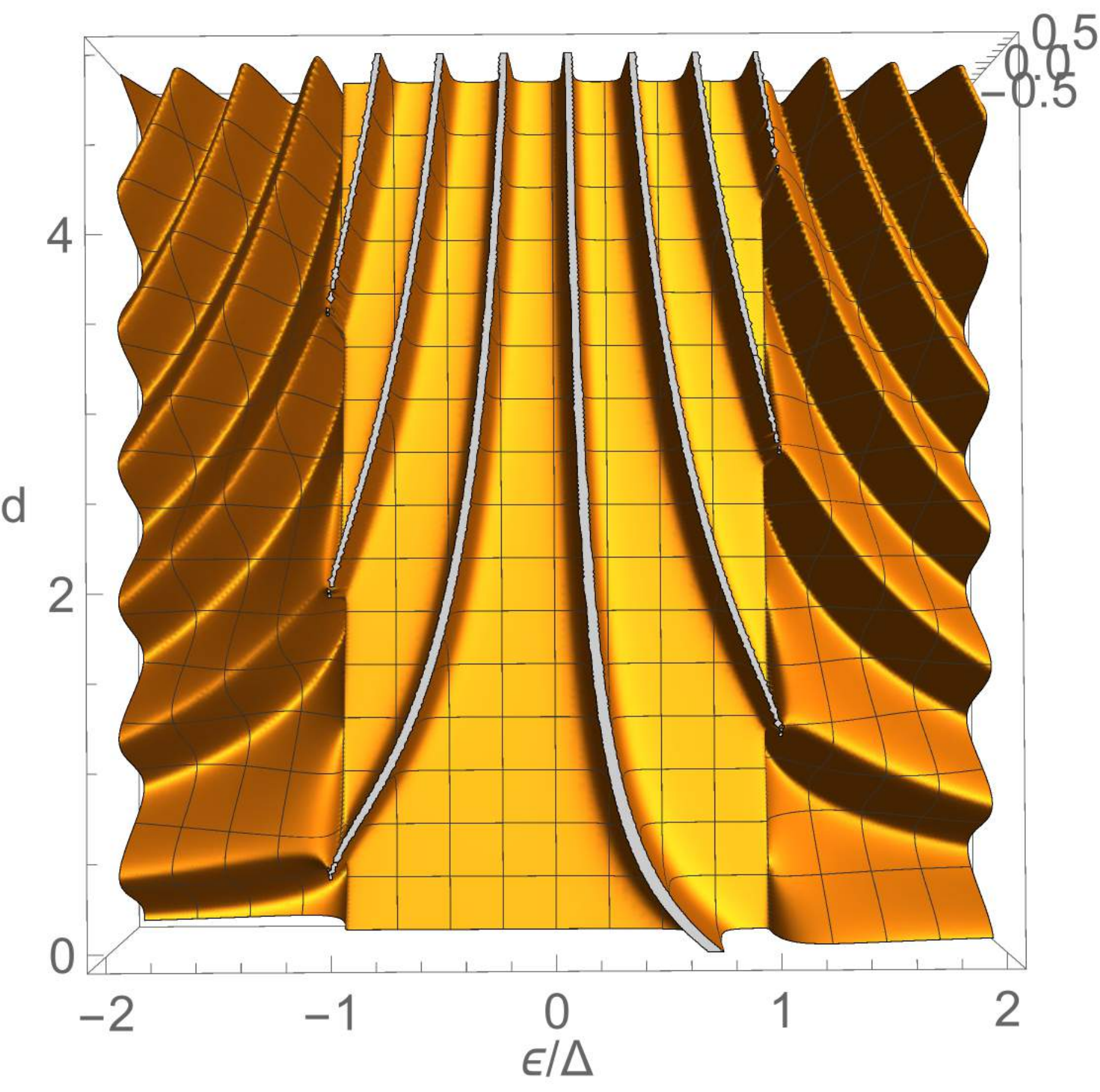}
\includegraphics[width=1.7in]{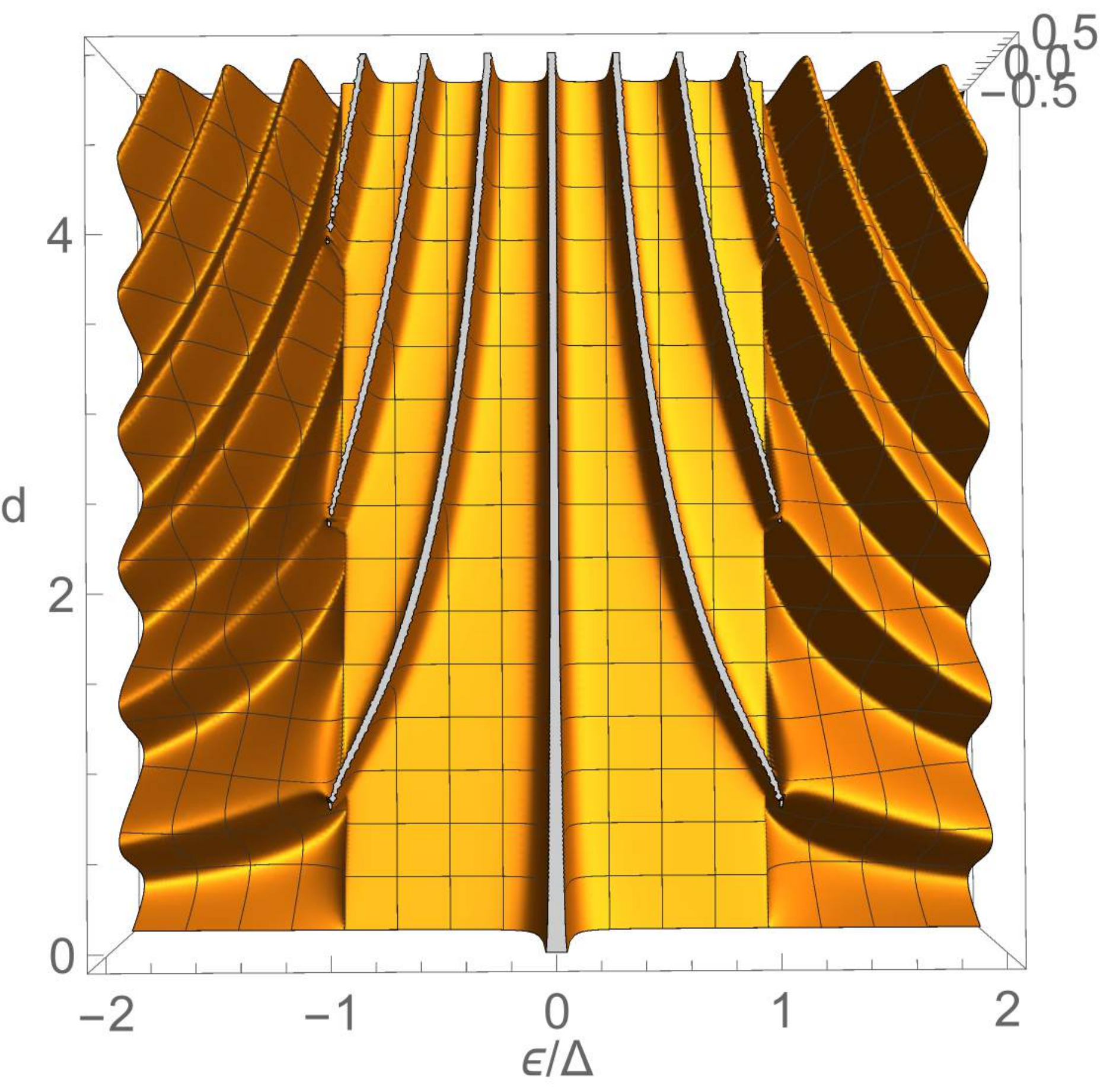}
}
\centering{(d) \hspace{1.5in} (e) \hspace{1.5in} (f) \hspace{1.5in}$\;$}\\
\centering{\includegraphics[width=1.75in]{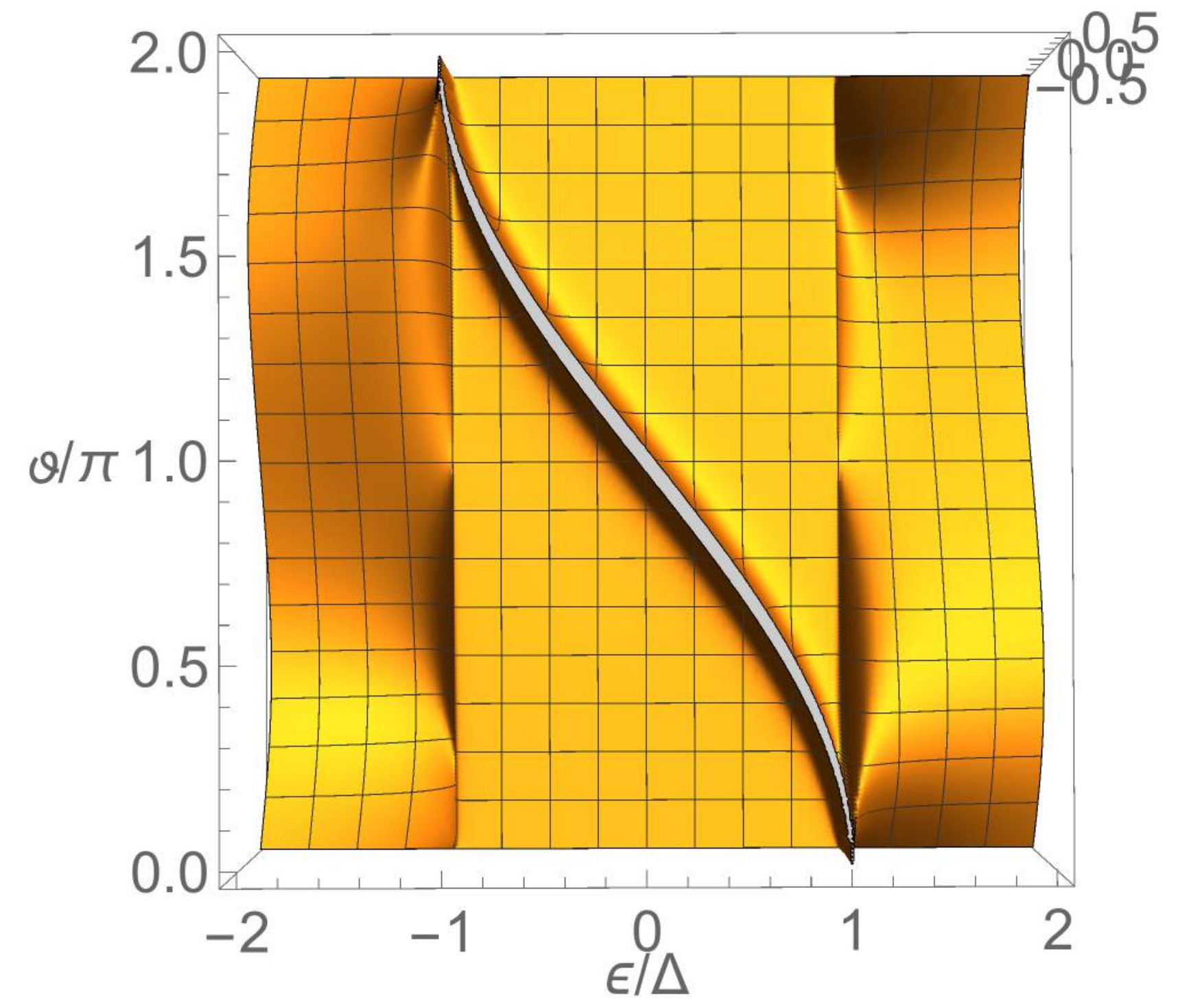}
\includegraphics[width=1.75in]{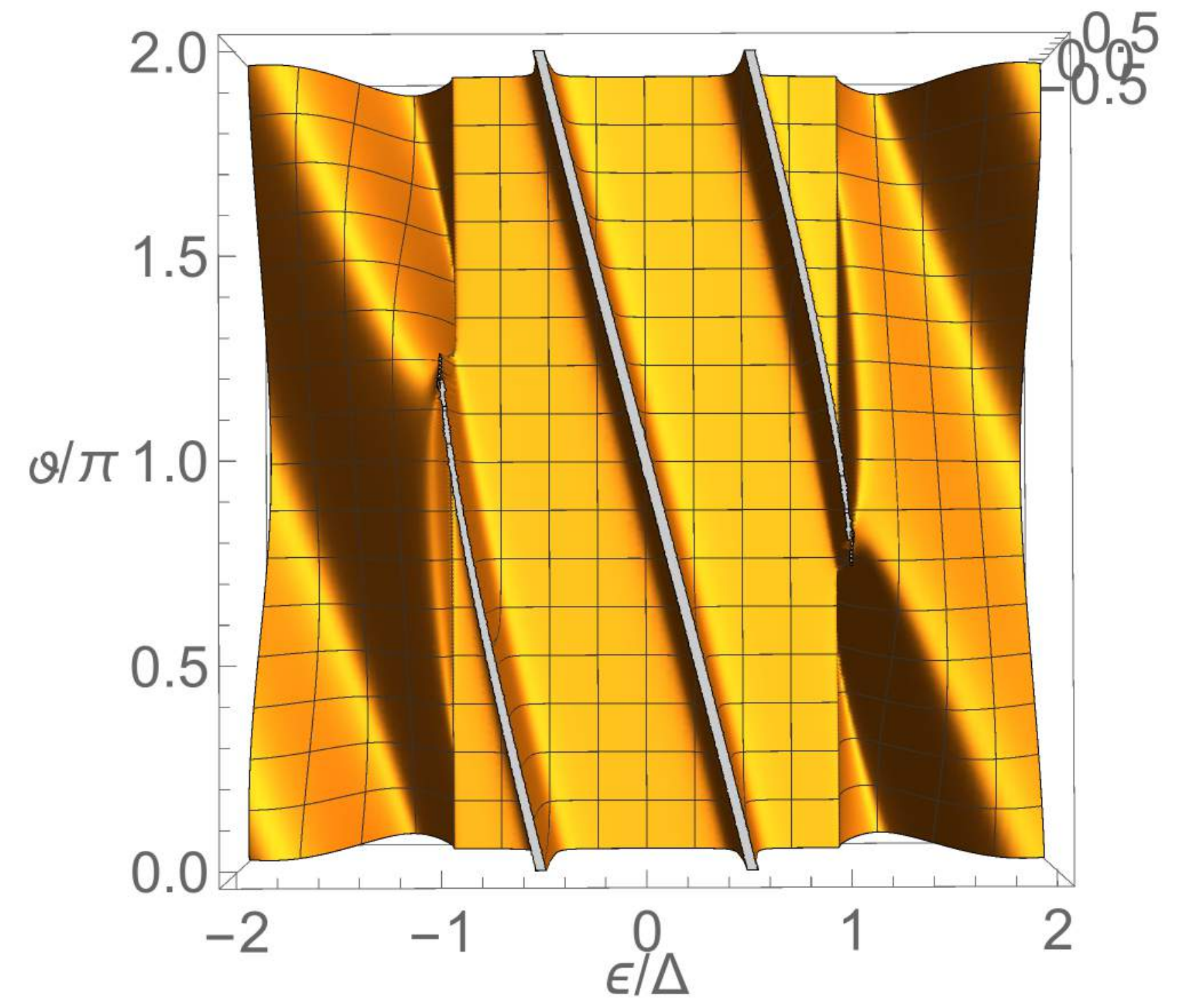}
\includegraphics[width=1.75in]{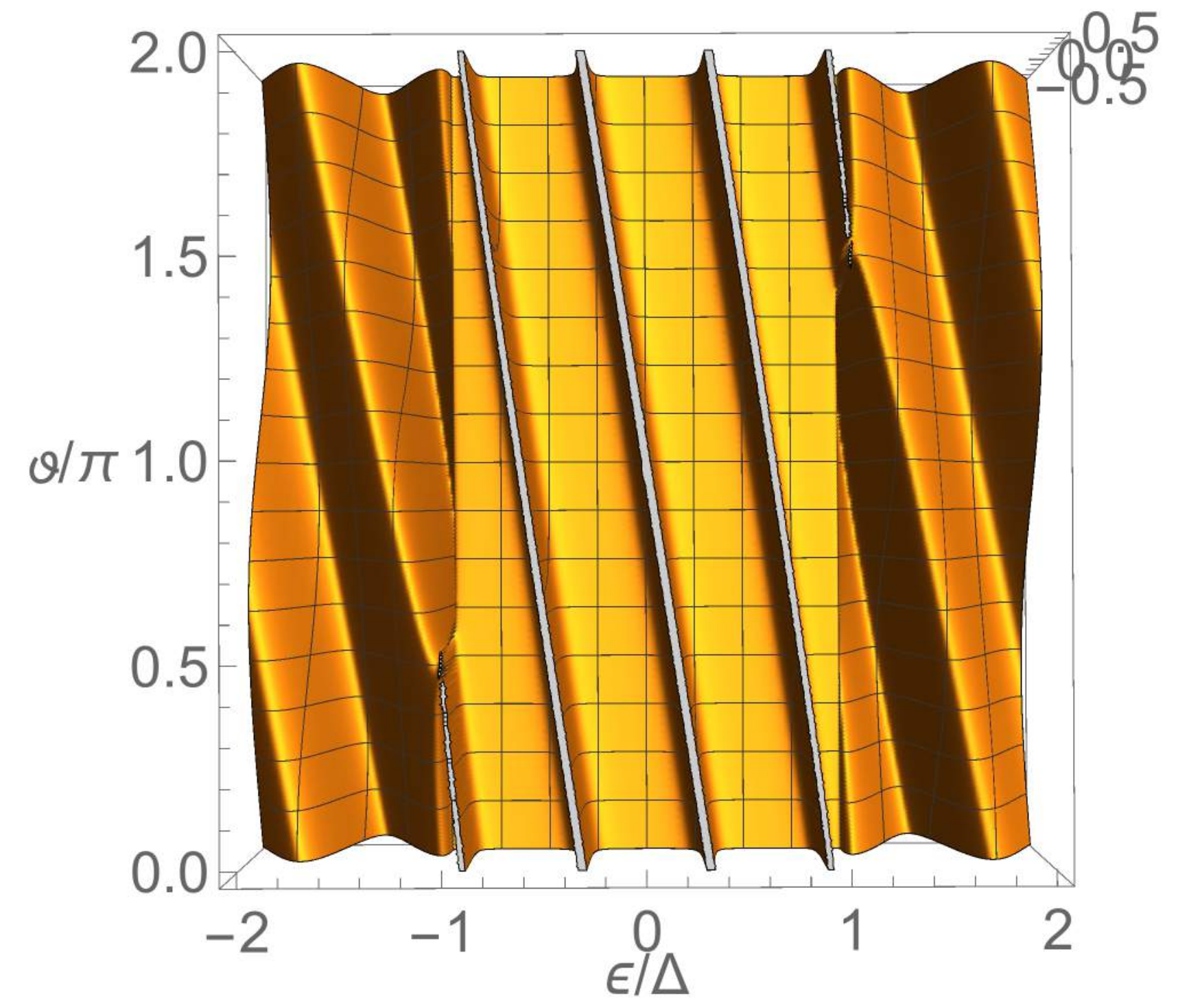}
}
\caption{Imaginary part of Andreev reflection amplitudes for 
spin-up Bogoliubov quasiparticle to spin-down Bogoliubov quasihole, 
Im$(r_{e\uparrow \to h\downarrow })$, at normal impact, for an S-N-FI structure with a normal region of thickness $d$, as function of energy $\varepsilon$, and of $d$ (in units of 
$\xi_0=\hbar v_{F,z}/\Delta $ with $v_{F,z}$ the projection of the Fermi velocity on the surface normal).
The reflection amplitude at the N-FI interface is for spin-up $e^{\i\vartheta_\uparrow}$ and for spin-down $e^{\i\vartheta_\downarrow}$. The spin-mixing angle is defined as $\vartheta=\vartheta_\uparrow-\vartheta_\downarrow$. It has the values (a) $\vartheta=0$, (b) $\vartheta=\pi/2$, (c) $\vartheta=\pi$. 
In (d)-(f) the thickness of the normal layer is fixed to (d) $d=0$, (e) $d=\xi_0$, (f) $d=2\xi_0$, and vary the spin-mixing angle $\vartheta $.
The negative reflection amplitude for spin-down quasiparticle to spin-up quasihole, $-$Im$(r_{e\downarrow \to h\uparrow })$, is obtained by inverting the energy axis, $\varepsilon \leftrightarrow -\varepsilon$. 
}
\label{Reh_th}
\end{figure}
In figure \ref{Reh_th}
results for the quantity Im$(r_{e\uparrow \to h\downarrow})$ for normal impact are shown. 
For energies above the gap the typical Rowell-McMillan oscillations are visible. Below the gap sharp bound states exist, the energy of which depends on both $\vartheta $ and $d$. In (a)-(c) the influence of varying $d$ is illustrated. 
The special case $\vartheta=0$, shown in (a), corresponds to the classical Rowell-McMillan S-N-I structure. For $\vartheta=\pi$, shown in (c), a midgap bound state exists for all $d$.
For $0<\vartheta <\pi$ particle-hole symmetry is broken, as seen in figure \ref{Reh_th} (b) and (d-f). 
A corresponding bound state at negative energy exists for $r_{e\downarrow \to h\uparrow}$. The two corresponding bound states in the density of states have opposite spin polarization. With increasing $d$ more and more bound states enter the sub-gap region, emerging from the continuum Rowell-McMillan resonances.
For $d$=0 a bound state exists for any nonzero $\vartheta $, as shown in (d), and as discussed in Ref. \cite{Fogelstrom00}. 
The influence of $\vartheta $ is shown for various $d$ in (d)-(f). 
These results can be interpreted as
one spin-polarised chiral branch crossing the gap region with increasing $\vartheta $. For $d=0$ this happenes when varying $\vartheta $ from zero to $2\pi$. For $d\ne 0$ one needs a variation exceeding $2\pi$ (up to multiple times) until the branch crosses the entire gap.
The figure shows results for normal impact, $k_\parallel=0$. In general, an integration over $k_\parallel $ will lead to Andreev bands instead of sharp bound states, similar as in the case of de Gennes-Saint-James bound states in S-N-I structures.

Finally, note that with the reflection matrix \eqref{Smatrix} the resulting coherence function developes a spin-triplet component from a singlet component $\gamma^{\rm in}=\gamma_0 \i\sigma_2$: 
\begin{align}
\left(\begin{array}{cc} 0 &\gamma^{\rm out}_{\uparrow \downarrow}\\ \gamma^{\rm out}_{\downarrow \uparrow} & 0 \end{array}\right) &=
\left(\begin{array}{cc} e^{\i\vartheta_\uparrow} &0 \\ 0& e^{\i\vartheta_\downarrow} \end{array}\right)
\left(\begin{array}{cc} 0 &\gamma_0\\ -\gamma_0 & 0 \end{array}\right) 
\left(\begin{array}{cc} e^{-\i\vartheta_\uparrow} &0 \\ 0& e^{-\i\vartheta_\downarrow} \end{array}\right) \nonumber \\
&=
\cos(\vartheta) 
\left(\begin{array}{cc} 0 &\gamma_0\\ -\gamma_0 & 0 \end{array}\right) 
+ \i \sin(\vartheta )
\left(\begin{array}{cc} 0 &\gamma_0\\ \gamma_0 & 0 \end{array}\right) 
\label{mixing0}
\end{align}
which implies that a singlet pair is scattered into a superposition of a singlet and a triplet pair:
\begin{align}
(\uparrow\downarrow-\downarrow\uparrow) \to
(\uparrow\downarrow e^{\i\vartheta }-\downarrow\uparrow e^{-\i\vartheta }) =
\cos(\vartheta) (\uparrow\downarrow-\downarrow\uparrow) +\i \sin(\vartheta) (\uparrow\downarrow+\downarrow\uparrow) .
\label{mixing}
\end{align}

\subsection{Andreev bound states in an S-FI-N structure}
As next example I summarize some results from Refs. \cite{Linder09,Linder10} and section IV of Ref. \cite{Eschrig09} for an S-FI-N junction, consisting of a bulk superconductor coupled via a thin ferromagnetic insulator (such as EuO) of thickness $d_{\rm I}$ to a normal layer of thickness $d_{\rm N}$. I assume here the ballistic case, and refer for the diffusive case to Refs. \cite{Linder09,Linder10,Cottet11}.
The interface is characterized by potentials $V_\uparrow $ and $V_\downarrow=V_\uparrow+2J$, such that the energy dispersions in the superconductor is $\hbar^2\vec{k}^2/2m$, in the normal metal $V_{\rm N}+\hbar^2\vec{k}^2/2m$, and in the barrier $V_\sigma+\hbar^2\vec{k}^2/2m$, $\sigma\in\left\{\uparrow,\downarrow\right\}$. The parameter $V_{\rm N}$ is used to vary the Fermi surface mismatch. 
The Fermi wave vectors and Fermi velocities in S and N are $\vec{k}_{\rm F,S}$, $\vec{v}_{\rm F,S}$ and $\vec{k}_{\rm F,N}$, $\vec{v}_{\rm F,N}$, respectively. The Fermi energy is $E_{\rm F}=\hbar^2\vec{k}_{\rm F,S}^2/2m$.
For shorter notation we
introduce a directional vector for electrons moving in positive $x$-direction, $\vec{v}_{\rm F,N}=|\vec{v}_{\rm F,N}|\hat{\vec{n}}$ (i.e. $\hat n_x\ge 0$), 
and the corresponding $x$-component of the Fermi velocity, $v_{{\rm F,N},x}\equiv v_{x}\ge 0$,  in the normal metal (situated at $0\le x\le d_{\rm N}$).
It is convenient to define coherence functions $\gamma $ and $\tilde \gamma $ as 2$\times$2 spin matrices. 
The coherence functions in the superconductor are given by $\gamma=\gamma_0 \i \sigma_2$ and $\tilde\gamma=\tilde\gamma_0 \i \sigma_2$, where $\gamma_0$ and $\tilde \gamma_0$ are given by the expressions in \eqref{sols1}. 
The solutions in the normal metal are (for simplicity of notation I also suppress the arguments $\vec{k}_\parallel $ and $\varepsilon $ in $\gamma $ and $\tilde \gamma $)
\begin{align}
\gamma(\hat n_x,x)&= \gamma(\hat n_x,0) e^{2\i \varepsilon x/\hbar v_x}, \quad
\gamma(-\hat n_x,x)= \gamma(-\hat n_x,d_{\rm N}) e^{-2\i \varepsilon (x-d_{\rm N})/\hbar v_x} \\
\tilde \gamma(-\hat n_x,x)&= \tilde \gamma(-\hat n_x,0) e^{2\i \varepsilon x/\hbar v_x}, \quad 
\tilde \gamma(\hat n_x,x)= \tilde \gamma(\hat n_x,d_{\rm N}) e^{-2\i \varepsilon (x-d_{\rm N})/\hbar v_x} 
\end{align}
At $x=d_{\rm N}$ one obtains $\gamma(\hat n_x,d_{\rm N})=\gamma(-\hat n_x,d_{\rm N})\equiv \gamma_{\rm B}$ and
$\tilde \gamma(\hat n_x,d_{\rm N})=\tilde \gamma(-\hat n_x,d_{\rm N})\equiv \tilde \gamma_{\rm B}$, 
with
\begin{align}
\gamma_{\rm B}= \left(\begin{array}{cc} 0& \gamma_+\\ -\gamma_-&0\end{array}\right),\quad
\tilde\gamma_{\rm B}= \left(\begin{array}{cc} 0& \tilde \gamma_+\\ -\tilde \gamma_-&0\end{array}\right).
\end{align}
The scattering parameters are the modulus of the transmission amplitudes, $t_\uparrow$ and $t_\downarrow$, the modulus of the reflection amplitudes $r_\uparrow=(1-t_\uparrow^2)^{1/2}$ and $r_\downarrow=(1-t_\downarrow^2)^{1/2}$ (equal on both sides of the FI), as well as the phase factors of the scattering parameters (all these parameters depend on $\vec{k}_\parallel$).
The relevant energy scale in the normal metal for given direction $\hat{\vec{n}}(\vec{k}_\parallel)$ is
\begin{align}
\delta(\vec{k}_\parallel)=t_{\uparrow}(\vec{k}_\parallel)\; t_{\downarrow}(\vec{k}_\parallel ) \; \hat n_x(\vec{k}_\parallel ) \;\varepsilon_{\rm Th}, \qquad
\varepsilon_{\rm Th}=\hbar v_{\rm F,N}/2d_{\rm N},
\end{align}
with the Thouless energy $\varepsilon_{\rm Th}$.
Matching the wavefunctions at $x=0$ to the thin FI layer and the superconductor, leads to $\tilde \gamma_-=-\gamma_+$ and $\tilde\gamma_+=-\gamma_-$, as well as \cite{Linder09,Linder10}
\begin{align}\label{eq:gamma}
\gamma_\sigma = - \frac{\delta }{\nu_\sigma + \i \sqrt{\delta^2-(\nu_\sigma+\i 0^+)^2}}
\end{align}
where $\sigma\in \{+,-\}$, and
the function $\nu_\sigma (\varepsilon )$ is defined as
\begin{align}\label{eq:u}
\nu _\sigma (\varepsilon) &= \hat n_x \varepsilon_{\rm Th} \left[\sin\left(\frac{\varepsilon }{\hat n_x\varepsilon_{\rm Th}} + 
\sigma\vartheta_+ 
+ \Psi \right)
+r_\uparrow r_\downarrow \sin\left(\frac{\varepsilon }{\hat n_x\varepsilon_{\rm Th}} + 
\sigma\vartheta_- 
- \Psi\right)\right]
\end{align}
with $\vartheta_\pm = \frac{1}{2}(\vartheta_{\rm N}\pm \vartheta_{\rm S})$, where $\vartheta_{\rm N}$ and $\vartheta_{\rm S}$ are the spin-mixing angles for reflection at the FI-N interface and the S-FI interface, respectively, and the
variable $\sigma $ is to be understood as a factor $\pm 1$ for $\sigma=\pm$. Note that \eqref{eq:gamma} has the same form as \eqref{sols1} with the role of $\Delta $ and $\varepsilon $ taken over by 
$\delta $ and $\nu_\sigma $, respectively. 
Note also that $|\gamma_\sigma |=1$ for $\nu_\sigma <\delta $, even in the tunneling limit. This is an example of reflectionless tunneling at low energies and resuls from multiple reflections within the normal layer. Quasiparticles in the normal layer stay fully coherent in this energy range. 

\begin{figure}[t]
\centering{
\includegraphics[width=1.7in]{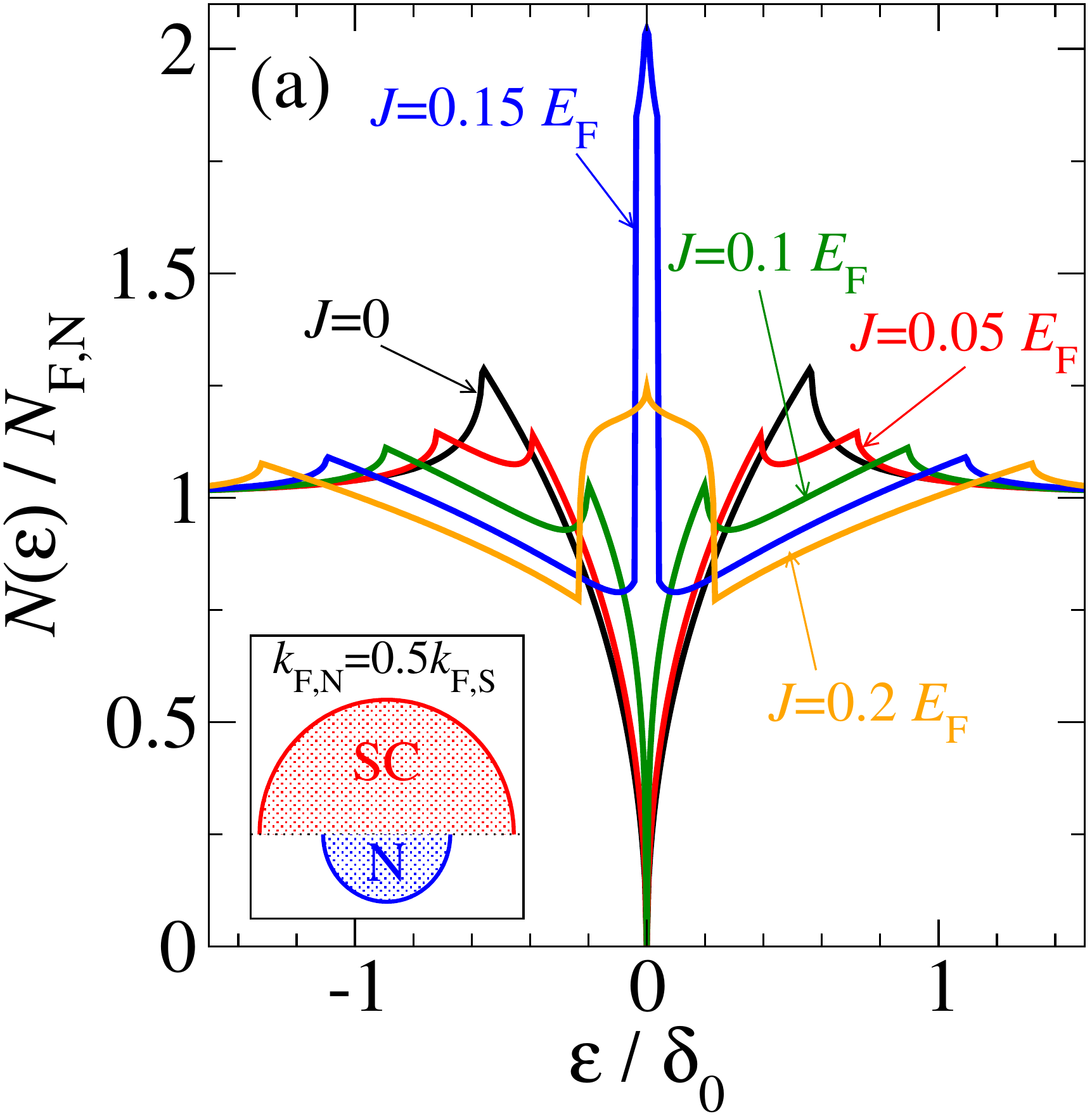}
\includegraphics[width=1.7in]{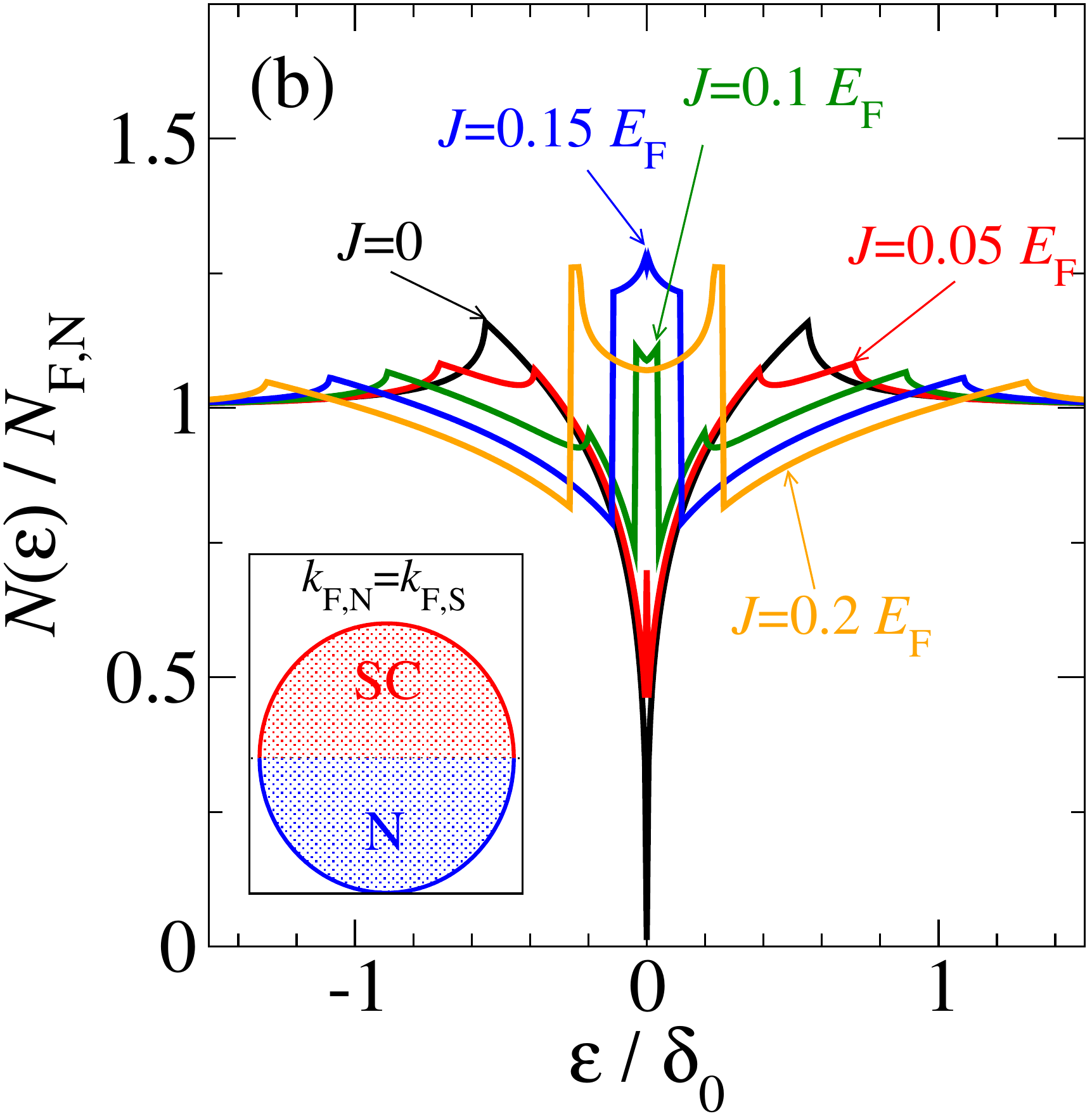}
\includegraphics[width=1.7in]{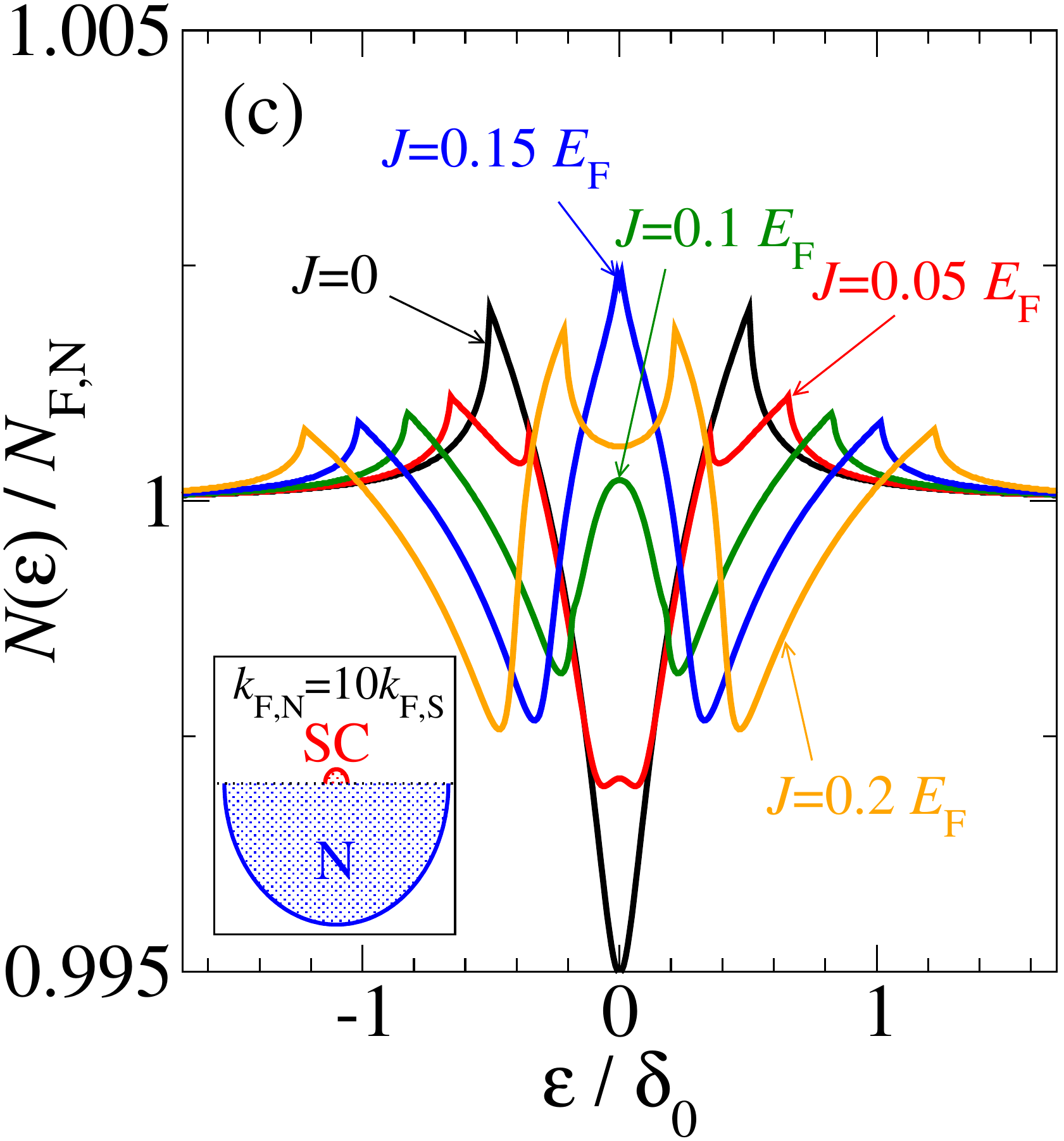}
}
\caption{
Energy-resolved DOS in the normal metal for different values of the interface exchange field $J$.
The energy scale is $\delta_0=(t_{\uparrow}t_{\downarrow} \varepsilon_{\rm Th})_{\vec{k}_\parallel=0}$,
with the Thouless energy $\varepsilon_{\rm Th}=\hbar v_{\rm F,N}/2d_{\rm N}$.
The interlayer thickness is $d_\text{I}=2/k_{\rm F,S}$
and the interface potentials are $V_\uparrow=1.2 E_{\rm F}$, $V_\downarrow=V_\uparrow+2J$.
The width of the normal layer is $d_{\rm N}=\hbar v_{\rm F,N} /\Delta$.
The inset in the lower left corner of each panel illustrates the Fermi-surface mismatch:
in (a) $k_{\rm F,N}=0.5 k_{\rm F,S}$, in (b) $k_{\rm F,N}=k_{\rm F,S}$, and in (c) $k_{\rm F,N}=10 k_{\rm F,S}$. Adapted from \cite{Linder10}. 
Copyright (2010) by the American Physical Society.
}
\label{SFIN}
\end{figure}

The density of states at the outer surface of the N layer is obtained as
\begin{align}\label{eq:dos}
\frac{N_{\rm B}(\varepsilon)}{N_{\rm F,N}} = \mbox{Re} \sum_\sigma \left\langle \frac{1+\gamma_\sigma^2}{1-\gamma_\sigma^2} \right\rangle= 
\mbox{Re} \sum_\sigma \left\langle \frac{|\nu_\sigma(\varepsilon )|}{\sqrt{\nu_\sigma(\varepsilon)^2-\delta^2}}\right\rangle
\end{align}
where $\langle \ldots \rangle $ denotes Fermi surface averaging, and $N_{\rm F,N}$ is the density of states at the Fermi level of the bulk normal metal. 
Results for this density of states are shown in figure \ref{SFIN}.
The various panels (a)-(c) show examples for various Fermi surface mismatches. In (c) there are non-transmissive channels present in the normal layer ($|\vec{k}_\parallel|>k_{\rm F,S}$), leading to a large constant background density of states.
In each panel, the curve for $J=0$
corresponds to the case of an non-spinpolarized SIN junction. There is a critical value $J_{\rm crit}$ (independent from the Fermi surface mismatch and equal to $\approx 0.15E_{\rm F}$ in the figure) above which the system is in a state where no singlet correlations are present in the normal metal at the chemical potential ($\varepsilon=0$), and pure odd-frequency spin-triplet correlations remain. In this range the density of states is enhanced above its bulk normal state value \cite{Linder10}.
On either side of this critical value the density of states decreases as function of $J$, however stays always above $N_{\rm F,N}$ for $J>J_{\rm crit}$. 
In the diffusive limit a similar scenario arises, with a peak centered at zero energy in the density of states \cite{Linder09, Linder10}.
A zero-energy peak in the density of states has been suggested as a signature of odd-frequency spin-triplet pairing also in hybrid structures with an itinerant ferromagnet or a half-metallic ferromagnet coupled to a superconductor \cite{Yokoyama07,Asano07,Braude07}.

It is interesting to study the tunneling limit, $t_\uparrow\ll 1$,  $t_\downarrow \ll 1$, for small excitation energies $\varepsilon\ll \mbox{min}(\varepsilon_{\rm Th},\Delta )$ and small spin-mixing angles $\vartheta_{\rm N}$, $\vartheta_{\rm S}$. 
Then $\nu_\sigma \approx 2\varepsilon+\sigma \hat n_x \varepsilon_{\rm Th} \vartheta_{\rm N}$, i.e. $\nu_\sigma $ depends only on the spin mixing angle at the FI-N interface, which acts in this case as an (anisotropic) effective exchange field $b=\hat n_x \varepsilon_{\rm Th}\vartheta_{\rm N}/2$ on the quasiparticles. For diffusive structures, a similar connection between an effective exchange field and the spin-mixing angle has been made \cite{Hernando02}.
The parameter $\delta/2=t_\uparrow t_\downarrow \hat n_x \varepsilon_{\rm Th}/2 $ on the other hand acts as effective (anisotropic) gap function. For each direction $\hat{\vec{n}}$, the gap closes at a critical value of effective exchange field, $b=\delta/2$, which happens for $\vartheta_{\rm N}=t_\uparrow t_\downarrow $.

\section{Andreev bound states in Josephson junctions with strongly spin-polarized ferromagnets}

\subsection{Triplet rotation}

Interfaces with strongly spin-polarized ferromagnets polarize the superconductor in proximity with it, as shown in the previous section. However, inorder for superconducting correlations to penetrate the ferromagnet, it is necessary to turn the triplet correlations of the form $\uparrow \downarrow + \downarrow \uparrow $ into equal spin pair correlations of the form $\uparrow \uparrow $ and $\downarrow \downarrow $. The reason is that correlations involving spin-up and spin-down electrons involve a phase factor $k_{\rm F\uparrow}-k_{\rm F\downarrow}$, which in strongly spin-polarized ferromagnets oscillates on a short length scale. This leads to destructive interference and allows to neglect such pair correlations on the superconducting coherence length scale \cite{Eschrig07}.

The way to achieve this is to allow for a non-trivial magnetization profile at the interface between the ferromagnet and the superconductor. This can include for example strong spin-orbit coupling, or a misaligned (with respect to the bulk magnetization) magnetic moment in the interface region. For strongly spin-polarized ferromagnets this has been suggested in Ref. \cite{Eschrig03,Kopu04,Eschrig08}. For weakly spin-polarized ferromagnets a theory was developed in 2001 involving a spiral inhomogeneity on the scale of the superconducting coherence length \cite{Bergeret01,Kadigrobov01}. A multilayer arrangement was subsequently also suggested \cite{Volkov03,Houzet07}.
For various reviews of this field see Refs. \cite{Izyumov02,Fominov03,Eschrig04,Golubov04,Buzdin05,Bergeret05,Eschrig07,Lyuksyutov07,Eschrig11,Blamire14,Linder15,Eschrig15}.

The idea is to rotate the triplet component, once created by spin-mixing phases in the S-F interfaces, into equal-spin triplet amplitudes with respect to the bulk magnetization of the ferromagnet \cite{Eschrig11}.
This is achieved by writing a triplet component with respect to a new axis 
\begin{align}
(\uparrow\downarrow+\downarrow\uparrow)_{\alpha,\phi} &=
-\sin (\alpha) \left[e^{-\i\phi} (\uparrow\uparrow)_z - e^{\i\phi} (\downarrow\downarrow)_z\right]
+ \cos (\alpha) (\uparrow\downarrow+\downarrow\uparrow)_z ,
\label{triplet_z}
\end{align}
where $\alpha $ and $\phi $ are polar and azimuthal angles of the new quantization axis.
Then, if a thin FI layer oriented along the $(\alpha,\phi)$ direction is inserted between the superconductor and the strongly spin-polarized ferromagnet with magnetization in $z$ direction, equal spin-correlations can penetrate with amplitudes $-\sin (\alpha) e^{-\i \phi} $ and $\sin (\alpha) e^{\i \phi}$, respectively. These correlations are long-range and not affected by dephasing on the short length scale associated with $k_{\rm F\uparrow}-k_{\rm F\downarrow}$.

\subsection{Pair amplitudes at an S-FI-F interface}

It is instructive to consider the scattering matrix of an S-FI-F interface between a superconductor and an itinerant ferromagnet, with a thin FI interlayer of width $d$, in the tunneling limit. In this case one can achieve an intuitive understanding of the various spin-mixing phases involved in the reflection and transmission processes. Denoting
wavevector components perpendicular to the interface as
$k$ in the superconductor, $q_\uparrow $ and $q_\downarrow $ in the ferromagnet (I assume $\vec{k}_\parallel $ such that both spin directions are itinerant), and imaginary wavevectors $i\kappa_\uparrow$ and $i\kappa_\downarrow$ in the FI, the FI magnetic moment aligned in direction $(\sin \alpha \cos \varphi , \sin \alpha \sin \varphi, \cos \alpha) $, and a F magnetization aligned with the $z$-direction in spin space ($\alpha=0$), matching of wavefunctions leads to a scattering matrix
\begin{align}
{\bf S}= \left( \begin{array}{cc} 
\bar D_\varphi D_\alpha \Phi_{\rm S}^{\frac{1}{2}} &0\\ 0& \i \bar D_\varphi D_\beta \Phi_{\rm F}^{\frac{1}{2}}
\end{array} \right)
\left( \begin{array}{cc} 
1& 2 \nu_{\rm S} {\cal T} \nu_{\rm F} \\
2 \nu_{\rm F} {\cal T}^\dagger \nu_{\rm S} & -1
\end{array} \right)
\left( \begin{array}{cc} 
\Phi_{\rm S}^{\frac{1}{2}} D_\alpha^\dagger \bar D_\varphi^\dagger &0\\ 0& \i \Phi_{\rm F}^{\frac{1}{2}} D_\beta^\dagger \bar D_\varphi^\dagger
\end{array} \right)
\label{srot}
\end{align}
where $\Phi_{\rm S,F}$ are phase matrices which include
the spin-mixing phase factors, $\bar D_\varphi $, $D_\alpha $, $D_\beta $ are spin-rotation matrices, $\nu_{S,F}$ carry information about S-FI and FI-F wavevector mismatch, and ${\cal T}$ contains the tunneling amplitudes including wavevector mismatch between S and F.
In particular, if one denotes diagonal matrices with diagonal elements $a$, $b$ by diag$(a,b)$, then
$K=\mbox{diag}(\kappa_\uparrow/k,\kappa_\downarrow/k)$, $Q=\mbox{diag}(q_\uparrow/k,q_\downarrow/k)$,
the spin-rotation matrices $\bar D_\varphi$, $D_\alpha $ between the quantization axis in the FI and the $z$ axis, and the phase matrices $\Phi_{\rm S,F}$ are
\begin{align}
\bar D_\varphi = \left( \begin{array}{cc} 
e^{\frac{\i}{2}\varphi}&0\\0& e^{-\frac{\i}{2}\varphi} \end{array} \right), \;
D_\alpha = \left( \begin{array}{cc} \cos \frac{\alpha}{2} & -\sin \frac{\alpha}{2} \\
\sin \frac{\alpha}{2} & \cos \frac{\alpha}{2} \end{array} \right), \;
\Phi_{\rm S,F}=\left( \begin{array}{cc} e^{\i\vartheta_\uparrow^{\rm S,F}} &0\\
0&e^{\i\vartheta_\downarrow^{\rm S,F}} \end{array} \right),
\end{align}
and the spin-rotation matrix $D_\beta $ at the FI-F interface results from
$Q^{-\frac{1}{2}}D_\alpha K D_\alpha^\dagger Q^{-\frac{1}{2}} = D_\beta Z D_\beta^\dagger$ with $Z=\mbox{diag}(\zeta_\uparrow,\zeta_\downarrow)$.  The angle $\beta $ vanishes for $\alpha=0$, and $\zeta_\uparrow $ varies from $\kappa_\uparrow/q_\uparrow $ at $\alpha=0$ to $\kappa_\downarrow/q_\uparrow $ at $\alpha=\pi$, correspondingly $\zeta_\downarrow $ varies from $\kappa_\downarrow/q_\downarrow$ to $\kappa_\uparrow/q_\downarrow$.
Also,
$\Phi_{\rm S} =(1-\i K)/(1+\i K)$,
$\Phi_{\rm F} =(1-\i Z)/(1+\i Z)$, 
$\nu_{\rm S}= \sqrt{2/(1+K^2)}$, $\nu_{\rm F}=\sqrt{2/(1+Z^2)}$, and the tunneling amplitude is 
${\cal T}=VA$ with $V=\mbox{diag}(e^{-\kappa_\uparrow d},e^{-\kappa_\downarrow d})$ and the real-valued mismatch matrix
$A=KD_\alpha^\dagger Q^{-\frac{1}{2}} D_\beta= D_\alpha^\dagger Q^{\frac{1}{2}}D_\beta Z$, of which the off-diagonal elements appear for $\alpha \ne 0, \pi$ only.

One can see from equation \eqref{srot} that the spin-mixing phases, which appear in the reflection amplitudes, also enter the transmission amplitudes; in the tunneling limit they contribute from each side of the interface one half \cite{Eschrig03}. Furthermore, one should notice that the interface is described by two spin-rotation matrices: one given by the misalignment of the FI magnetic moment with the $z$ axis in spin space, and one which is combined from the magnetization in F and the magnetic moment in FI. The latter appears because the wave function at the FI-F interface is delocalized over the FI-F interface region on the scale of the Fermi wavelength and experiences an averaged effective exchange field, which lies in the plane spanned by the $z$ axis and the direction of the FI magnetic moment (same $\bar D_\varphi $ in equation \eqref{srot}).

Pair correlation functions $f$ are related to coherence functions by $f=-2\pi i \gamma (1-\tilde \gamma \gamma)^{-1}$, with $f$, $\gamma$ and $\tilde \gamma$ 2$\times$2 matrices in spin space \cite{Eschrig00}.
When both are small (near $T_{\rm c}$ or induced from a reservoir by tunneling through a barrier), $f$ and $\gamma $ are proportional.
Assuming an incoming singlet coherence function $\gamma_0$ in S,
the coherence functions reflected back into S and the ones transmitted to F can be calculated to linear order in the pair tunneling amplitude according to
\begin{align}
\gamma^{\rm (S)}_{\rm out} = 
{\bf S}_{11} 
\left( \begin{array}{cc} 0&\gamma_0\\-\gamma_0 &0\end{array} \right)
 {\bf S}_{11}^\ast 
, \quad
\gamma^{\rm (F)}_{\rm out} = {\bf S}_{21} 
\left( \begin{array}{cc} 0&\gamma_0\\-\gamma_0 &0 \end{array} \right)
{\bf S}_{12}^\ast .
\end{align}
For the reflected amplitude in S one obtains ($\vartheta_{\rm S}=\vartheta_\uparrow^{\rm S}-\vartheta_\downarrow^{S}$)
\begin{align}
\frac{\gamma^{\rm (S)}_{\rm out}}{\gamma_0}=\left( \begin{array}{cc} -\i\sin \vartheta_{\rm S} \sin \alpha e^{-\i\varphi}
& \cos \vartheta_{\rm S} +\i\cos \alpha \sin \vartheta_{\rm S}\\
-\cos \vartheta_{\rm S} +\i\cos \alpha \sin \vartheta_{\rm S} & \i\sin \vartheta_{\rm S} \sin \alpha e^{\i\varphi}
\end{array} \right),
\end{align}
which is just equation \eqref{mixing0} rotated in spin-space by the spherical angles $\alpha $ and $\varphi $. For the equal-spin coherence functions (or pair amplitudes) in F follows up to leading order in the misalignent angles $\alpha $, $\beta $ (denoting $\vartheta_{\rm F}=\vartheta^{\rm F}_\uparrow-\vartheta^{\rm F}_\downarrow $)
\begin{align}
\frac{\gamma^{\rm (F)}_{\rm out \uparrow\uparrow}}{\gamma_0}\approx
-\i C e^{-\i\varphi } 
\left\{ \frac{\nu_{\rm F\uparrow}}{\nu_{\rm F\downarrow} } 
\sin \left(\frac{\vartheta_{\rm S}}{2} \right)
\left[
\sqrt{\frac{q_\downarrow }{q_\uparrow}}
\sin (\alpha )- \sin (\beta )\right]
+\sin \left(\frac{\vartheta_{\rm S}+\vartheta_{\rm F}}{2} \right)
\sin (\beta )
\right\}
\label{eqspin1}
\\
\frac{\gamma^{\rm (F)}_{\rm out \downarrow\downarrow}}{\gamma_0}\approx
+\i C e^{+\i\varphi } 
\left\{ \frac{\nu_{\rm F\downarrow}}{\nu_{\rm F\uparrow}}
\sin \left(\frac{\vartheta_{\rm S}}{2} \right)
\left[ \sqrt{\frac{q_\uparrow }{q_\downarrow}}
\sin (\alpha ) -\sin (\beta )\right]
+\sin \left(\frac{\vartheta_{\rm S}+\vartheta_{\rm F}}{2}\right)
\sin (\beta )
\right\}
\label{eqspin2}
\end{align}
with $C=4e^{-(\kappa_\uparrow+\kappa_\downarrow)d}(\nu_{\rm S\uparrow} \nu_{\rm S\downarrow}
\nu_{\rm F\uparrow}\nu_{\rm F\downarrow}) \kappa_\uparrow \kappa_\downarrow/[k(q_\uparrow q_\downarrow )^{\frac{1}{2}}]$.
The transmitted $\uparrow\downarrow $ and $\downarrow\uparrow $ coherence functions
$\gamma^{\rm (F)}_{\rm out \uparrow\downarrow}\approx 
C\gamma_0 e^{\frac{\i }{2}(\vartheta_{\rm S}+\vartheta_{\rm F})}$ 
and 
$\gamma^{\rm (F)}_{\rm out \downarrow\uparrow}\approx
-C\gamma_0e^{-\frac{\i }{2}(\vartheta_{\rm S}+\vartheta_{\rm F})}$ spatially oscillate with a wavevector 
$e^{\i\vec{k}_\parallel \vec{r}_\parallel} e^{\pm \i (q_\uparrow-q_\downarrow )x}$ in F, and are suppressed (except in ballistic one-dimensional channels) due to dephasing after a short distance 
$1/|q_\uparrow-q_\downarrow|$ away from the interface.
Importantly, from \eqref{eqspin1} and \eqref{eqspin2} it is visible that the equal-spin amplitudes acquire phases $\pm \varphi $ from the azimuthal angle in spin space, which play an important role in Josephson structures with half-metallic ferromagnets \cite{Eschrig07,Eschrig08} and with strongly spin-polarized ferromagnets when two interfaces with different azimuthal angles $\varphi_1$ and $\varphi_2$ are involved \cite{Grein09,Grein13}.

The misalignment of FI with F also induces a spin-flip term during reflection on the ferromagnetic side of the interface, which creates for an in F incoming amplitude $\gamma_{\uparrow\uparrow}$ a reflected amplitude $\gamma_{\downarrow\downarrow}=\gamma_{\uparrow\uparrow}e^{2\i\varphi}\sin^2\frac{\vartheta_{\rm F}}{2} \sin^2\beta $, and for an incoming amplitude $\gamma_{\downarrow\downarrow} $ a reflected amplitude $\gamma_{\uparrow\uparrow}=\gamma_{\downarrow\downarrow}e^{-2\i\varphi}\sin^2\frac{\vartheta_{\rm F}}{2} \sin^2\beta $. In this case, twice the azimutal angle $\pm \varphi $ enters.

\subsection{Andreev bound states in S-FI-HM-FI'-S junctions}

For an S-FI-HM interface with a half-metallic ferromagnet (HM)
in which one spin-band (e.g spin-down) is insulating and the other itinerant, equation \eqref{eqspin1} is modified to \cite{Eschrig08}
\begin{align}
\frac{\gamma^{\rm (F)}_{\rm out \uparrow\uparrow}}{\gamma_0}\approx
-4\i e^{-\i\varphi } e^{-(\kappa_\uparrow+\kappa_\downarrow)d}
\nu_{\rm S\uparrow} \nu_{\rm S\downarrow}
\nu_{\rm F\uparrow}^2 \frac{\kappa_\uparrow \kappa_\downarrow}{kq_\uparrow}
\sin \left(\frac{\vartheta_{\rm S}}{2}\right) \sin (\alpha ) .
\label{eqspinHM}
\end{align}
The same equation also holds at an S-FI-F interface when the conserved
wavevector $\vec{k}_\parallel $ is such that only one spin-projection on the magnetization axis is itinerant in F. For strongly spin-polarized F this is an appeciable contribution to the transmitted pair correlations.

In order to describe Josephson structures, it is necessary to handle the spatial variation (and possibly phase dynamics) of both coherence amplitudes and superconducting order parameter (which are coupled to each other). For this, a powerful generalization of the 2$\times$2 spin-matrix coherence functions has been introduced \cite{Eschrig00,Eschrig09}, based on previous work for spin-scalar functions in equilibrium \cite{Nagato93,Schopohl95} and nonequilibrium \cite{Eschrig99}.
The coherence functions $\gamma (\vec{n},\vec{R},\varepsilon, t)$ fulfil transport equations 
\begin{align}
\label{cricc1}
\i\hbar \, \qpartial \ga +
2\ep \ga
&= \ga \qt \Db \qt \ga +
\Big( \va \qt \ga - \ga \qt \vb \Big) - \Da, \\
\label{cricc2}
\i\hbar \, \qpartial \gb -
2\ep \gb
&= \gb \qt \Da \qt \gb +
\Big( \vb \qt \gb - \gb \qt \va  \Big) - \Db 
\end{align}
with $\vec{n}$ a unit vector in direction of $\vf$, $\Sigma $ and $\Delta $ include particle-hole diagonal and off-diagonal self-energies and mean fields (e.g for impurity scattering, $\Delta $ includes the superconducting order parameter), and external potentials. The time-dependent case is included by the convolution over the internal energy-time variables in Wigner coordinate representation,
\begin{equation}
(A \qt B)(\ep,t) \equiv
e^{\frac{\i}{2} (\partial_\ep^A\partial_t^B-\partial_t^A\partial_\ep^B)} A(\ep,t) B(\ep,t).
\end{equation}
In the time-independent case it reduces to a simple spin-matrix product. Furthermore, the particle-hole conjugation operation is defined by $\tilde A(\vec{n},\vec{R},\varepsilon, t)=A^\ast(-\vec{n},\vec{R},-\varepsilon,t)$.
Particle-hole diagonal [$g=-\i \pi(2{\cal V}-\sigma_0)$ and $\tilde g=\i \pi(2\tilde{\cal V}-\sigma_0)$] and off-diagonal [$f=-2\pi\i {\cal F}$ and $\tilde f=2\pi \i \tilde{\cal F}$] pair correlation functions (quasiclassical propagators) are obtained in terms of these coherence functions by solving the following algebraic (or in the time-dependent case, differential) equations
\begin{align}
{\cal V}= \sigma_0+\ga\qt\gb\qt {\cal V}, \quad
\tilde{\cal V}= \sigma_0+\gb\qt\ga\qt \tilde{\cal V}, \quad
{\cal F}=\ga+\ga\qt\gb\qt {\cal F}, \quad
\tilde{\cal F}=\gb+\gb\qt\ga\qt \tilde{\cal F} .
\end{align}

\begin{figure}[t]
\centering{
\includegraphics[width=2.4in]{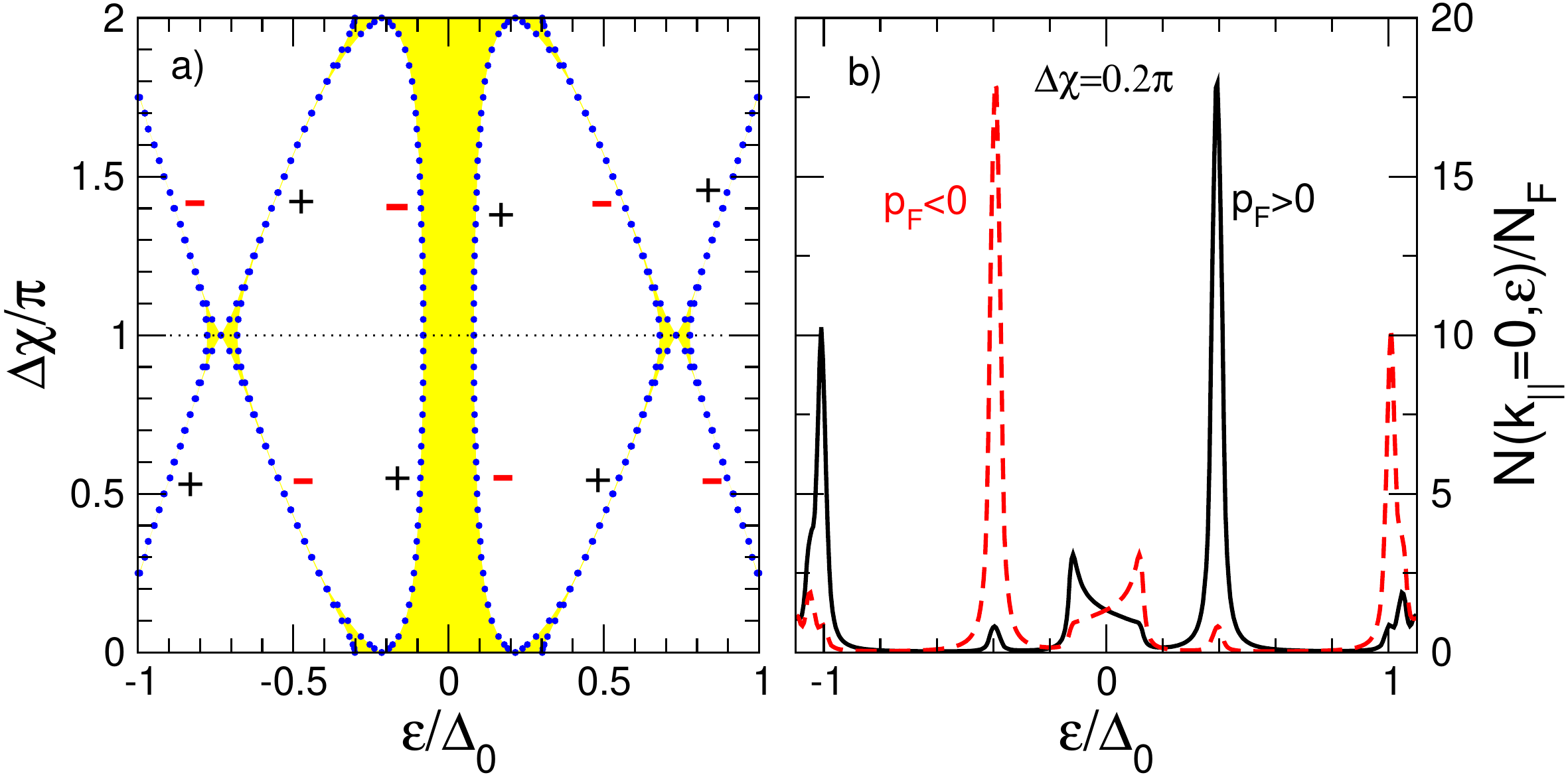}
\hspace{0.2in}
\includegraphics[width=2.4in]{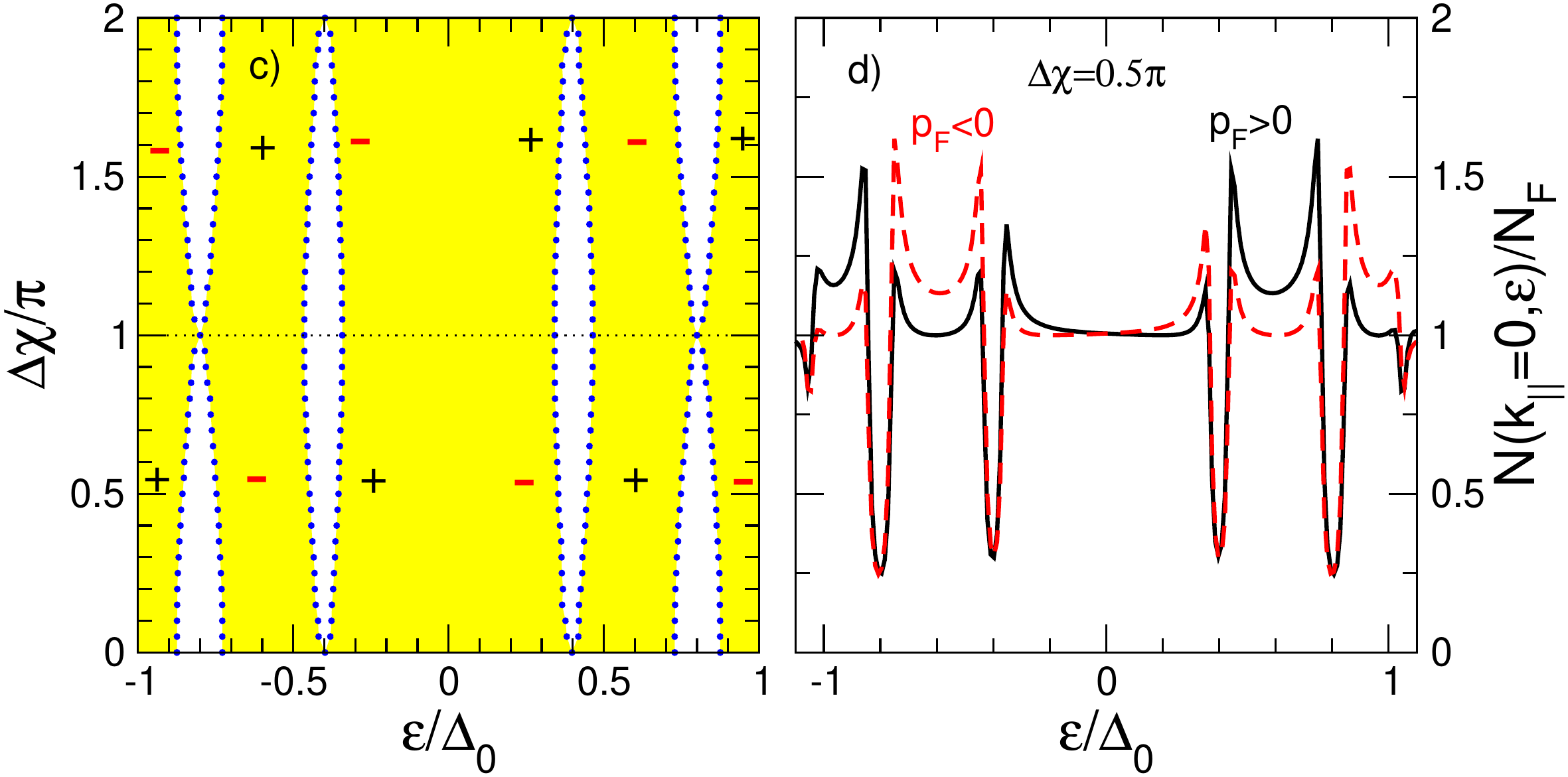}
}
\caption{
Local density of Andreev states (LDOS)
at the HM side of the S-FI-HM interface in a S-FI-HM-FI'-S Josephson structure, as function of phase difference $\Delta\chi $ over the junction,
for quasiparticles with normal impact, at $T = 0.05T_{\rm c}$.
All states are fully spin-polarized.
(a) and (c) Dispersion of the
maxima of the LDOS as function of phase difference. Regions with low LDOS are
white, regions of high LDOS (bands of Andreev bound states) are shaded. The signs indicate the direction of the current
carried by the Andreev states. (b) and (d) show spectra for a fixed
phase difference, both for positive (full lines) and negative (dashed lines) propagation direction. (a)-(b) is for a large misalignment between FI and HM, and (c)-(d) for a small misalignment.
(b) and (d) from \cite{Eschrig03}.
Copyright (2003) by the American Physical Society.
}
\label{SHM}
\end{figure}
In figure \ref{SHM} an example for a fully self-consistent calcuation of the spectrum of subgap Andreev states in a S-FI-HM-FI'-S junction is shown, obtained by solving equations \eqref{cricc1}, \eqref{cricc2} as well as the self-consistency equation for the superconducting order parameter in S.
For details of the calcuation and parameters see \cite{Eschrig03,Eschrig04}.
The ferromagnetic insulating barriers FI and FI' are taken identical in this calculation, and the spectra are shown at the half-metallic side of the S-FI-HM interface.
The most prominent feature in these spectra is an Andreev quasiparticle band centered at zero energy \cite{Eschrig03,Eschrig09,Halterman09}. Further bands at higher excitation energies are separated by gaps. 
The zero-energy band is a characteristic feature of the nature of superconducting correlations in the half metal: they are spin-triplet with a phase shift of $\pm \pi/2$ with respect to the singlet correlations they are created from.
The dispersion of the Andreev states with applied phase difference between the two S banks determine the direction of current carried by these states. This direction is indicated in the figure by $+$ and $-$ signs. In panels (b) and (d) for a selected phase difference the spectra for positive and negative propagation direction are shown. These spectra, multiplied with the equilibrium distribution function (Fermi function) determine the contribution of the Andreev bound states to the Josephson current in the system.
These spectra are shown for normal impact direction ($\vec{k}_\parallel=\vec{0}$). An integration over $\vec{k}_\parallel$ gives the local density of states.

\begin{figure}[b]
\centering{
\includegraphics[width=1.25in]{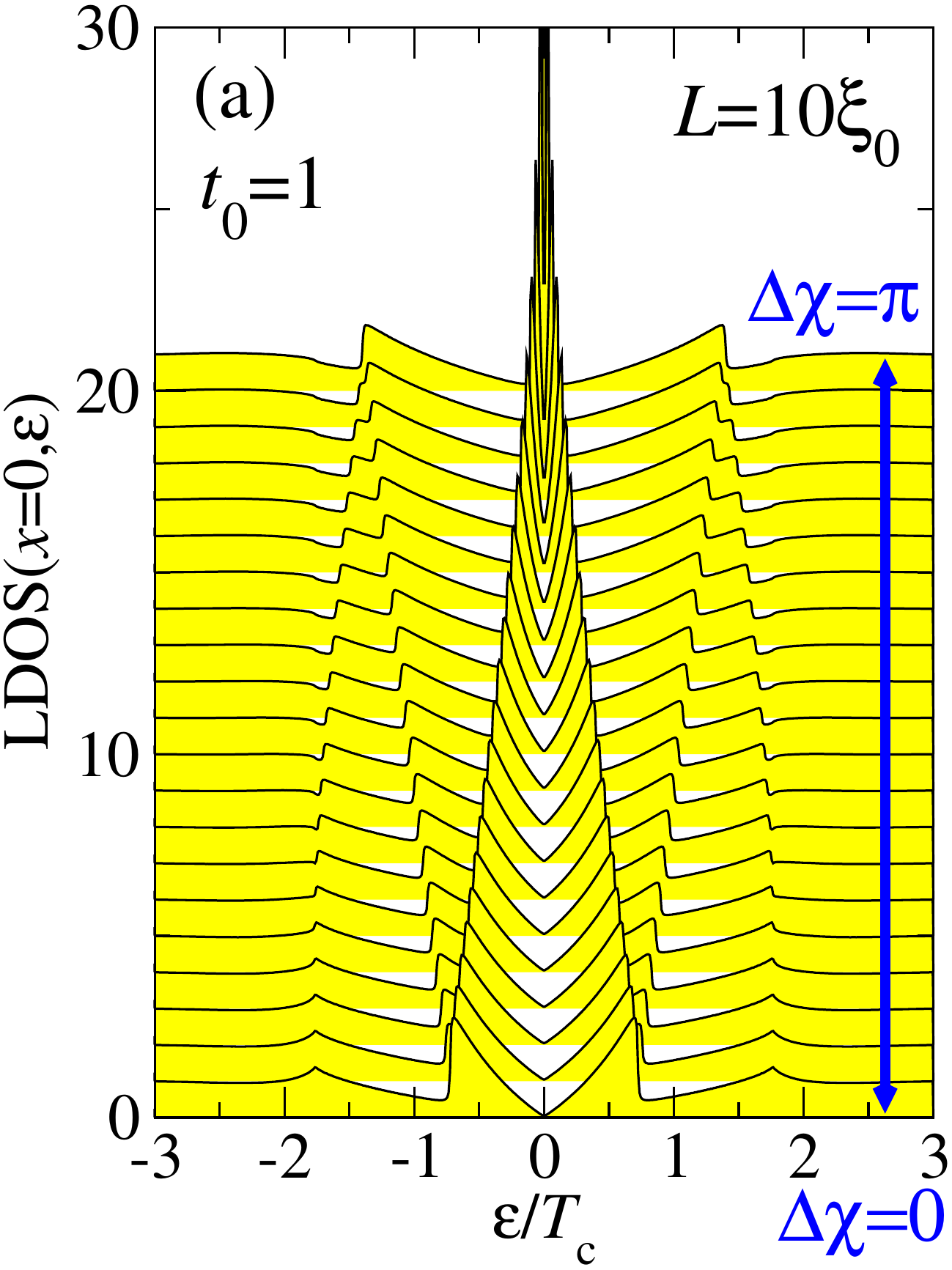}
\includegraphics[width=1.25in]{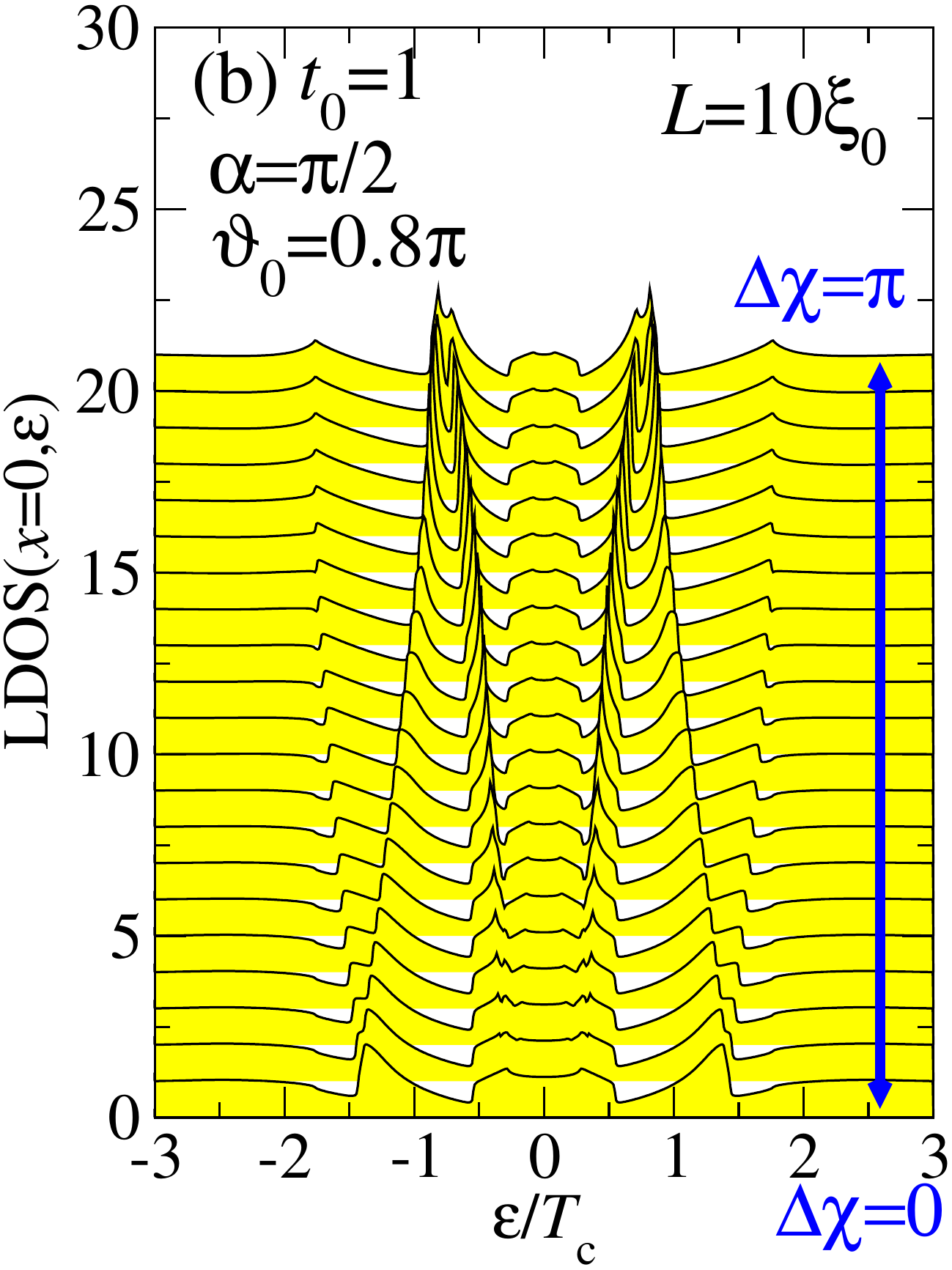}
\includegraphics[width=1.25in]{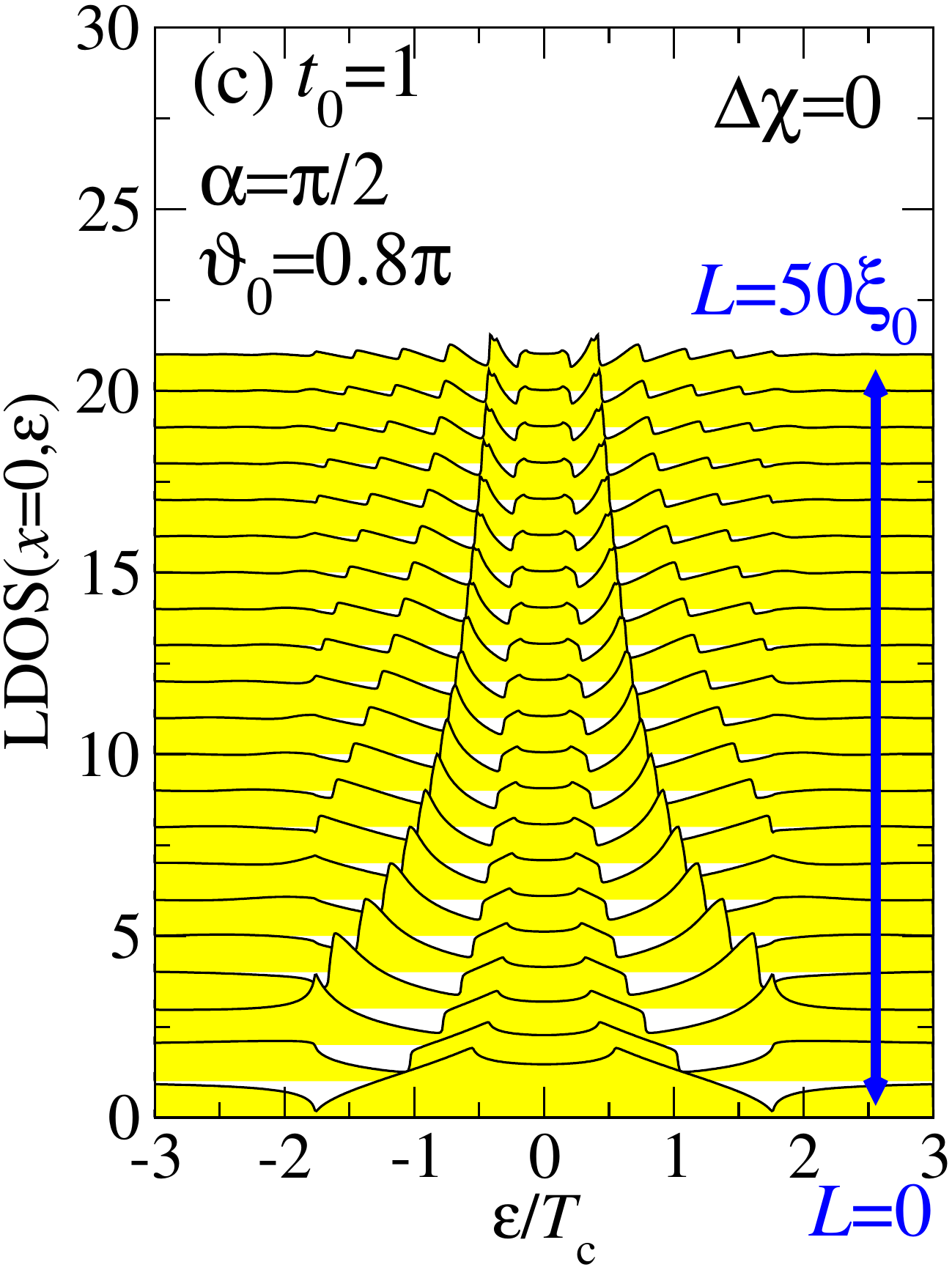}
\includegraphics[width=1.25in]{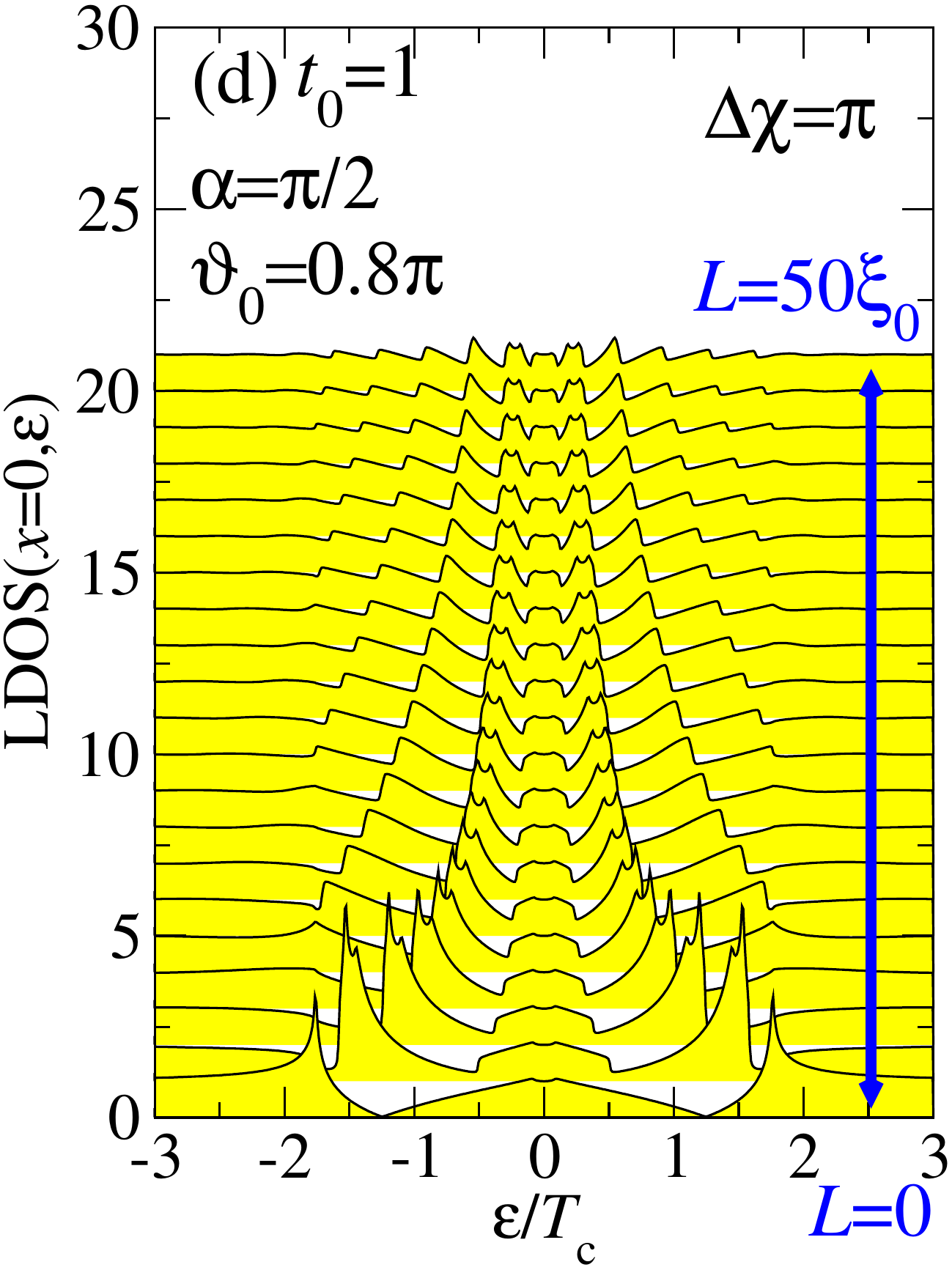}
}
\caption{
Local density of states in the center of a
current biased high-transmissive symmetric Josephson junction for
(a) an S-N-S junction and (b)-(d) an S-FI-HM-FI-S junction.
In (a) and (b) the phase difference $\Delta \chi$ over the junction is varied from 0 to $\pi$.
In (b) and (c) the length of the junction is varied for a zero-junction and a $\pi$-junction.
The temperature is $T=0.1T_{\rm c}$, the coherence length of the half metal $\xi_0=\hbar |\vf |/2\pi T_{\rm c}$.
The FI misalignment angle is $\alpha =\pi/2$.
The transmission parameter $t$ and spin-mixing angle $\vartheta_{\rm S}$ depend on the impact angle $\Psi_{\vec{n}}$ measured from the surface normal; this is modeled here by
$t(\Psi_{\vec{n}})=t_0 \cos \Psi_{\vec{n}}/(1-t_0^2\sin^2 \Psi_{\vec{n}})^{\frac{1}{2}}$ and $\vartheta_{\rm S}=\vartheta_0 \cos \Psi_{\vec{n}}$.
Adapted from \cite{Eschrig09}.
Copyright (2009) by the American Physical Society.
}
\label{SHM1}
\end{figure}
For the case that one can neglect the variation of the order parameter $\Delta $ in S, one can derive quite a number of analytical expressions \cite{Eschrig07,Galaktionov08,Eschrig09}.
Examples for integrated spectra are shown in figure \ref{SHM1}, taken from Ref. \cite{Eschrig09}. In (a) the well-known spectrum of de Gennes-Saint James bound states is seen for an S-N-S junction \cite{Deutscher05}, showing a dispersion with phase bias $\Delta \chi $ between the two superconductors. At $\Delta \chi=\pi$ a zero energy bound state is present, which is a topological feature of the particular Andreev differential equations describing this system, for real-valued order parameters that change sign when going from the left S reservoir to the right S reservoir. The origin is the same as for the midgap state in polyacetylene \cite{Heeger88}, which is governed by similar differencial equations. Such midgap states have been studied in more general context by Jackiw and Rebbi \cite{Jackiw76} and ultimately have their deep mathematical foundation in the Atiyah-Patodi-Singer index theorem \cite{Atiyah75}.
For the S-FI-HM-FI-S junction, shown in (b)-(d), the prominent feature for all values of $\Delta \chi$ is the band of Andreev states centered around zero energy. 
The width $W$ of this low-energy Andreev band depends on the parameter $P=\sin \left(\vartheta_{\rm S}/2\right) \sin (\alpha)$ and
can be calculated for the limit of short junctions ($L\to 0$) for $t=1$ as \cite{Eschrig09}
\begin{align}
W(\Delta \chi=0) = 2|\Delta | \sqrt{1-P^2}, \quad
W(\Delta \chi=\pi ) = |\Delta |(\sqrt{2-P^2}-P).
\end{align}
In the limit $P\to 0$ this gives 
$W(\Delta \chi=0) = 2|\Delta |$ and $W(\Delta \chi=\pi ) = \sqrt{2} |\Delta | $.
Note that compared to the S-N-S junction, the low-energy features disperse in opposite direction when increasing $\Delta \chi$ for the S-FI-HM-FI-S junction. This means that the current flows in opposite direction, and typically a $\pi $-junction is realised for identical interfaces. If the azimuthal interface misalignment angles $\varphi $ differ by $\pi $ in FI and FI', then this phase would add to $\Delta \chi$ according to equation \eqref{eqspinHM} and a zero-junction would be  realized. In the general case, a $\phi$-junction appears, both for ballistic and diffusive structures \cite{Eschrig07,Eschrig08,Eschrig15a}. 

\subsection{Spin torque in S-FI-N-FI'-S' structures}

Andreev states play also an important role in the non-equilibrium spin torque and in the spin-transfer torque in S-F structures \cite{Slonc96,Ralph08,Brataas12,Locatelli14}. Zhao and Sauls found that 
in the ballistic limit the equilibrium torque is related to the spectrum of spin-polarized Andreev bound states, while the ac component, for small bias voltages, is determined by the nearly adiabatic dynamics of the Andreev bound states \cite{Zhao07,Zhao08}.
The equilibrium spin-transfer torque $\tau_{\rm eq} $ in an S-FI-N-FI'-S' structure 
is related to the Josephson current $I_{\rm e}$, the phase difference between S and S', $\Delta \chi$, and the angle $\Delta \alpha $ between FI and FI', by
\cite{Waintal02}
\begin{align}
\hbar \frac{\partial I_{\rm e}}{\partial \Delta \alpha } = 2{\rm e} \frac{\partial \tau_{\rm eq}}{\partial \Delta \chi }.
\end{align}
Similarly, as the dispersion of the Andreev bound states with superconducting phase difference $\Delta \chi $ yields the contribution of the bound state to the Josephson current, the dispersion of the Andreev states with $\Delta \alpha $ yields the contribution of this state to the spin current for spin polarisation in direction of the spin torque.
The dc spin current shows subharmonic gap structure due to multiple Andreev reflections (MAR), similar as for the charge current in voltage biased Josephson junctions \cite{Kummel85,Arnold87}. For high transmission junctions the main contribution to the dc spin current comes from consecutive spin rotations according to equation \eqref{SpinRot} when electrons and holes undergo MAR 
\cite{Zhao08}.

Turning to ac effects, for a voltage $eV\ll\Delta $ 
the time evolution of spin-transfer torque is governed by the nearly adiabatic dynamics of the Andreev bound states. However, the dynamics of the bound state spectrum leads to non-equilibrium population of the Andreev bound states, for which reason the spin-tranfer torque does not assume its instantaneous equilibrium value \cite{Zhao08}.
For the occupation to change, the bound state energy must evolve in time to the continuum gap edges, where it can rapidly equilibrate with the quasiparticles, similar as in the adiabatic limit of ac Josephson junctions \cite{Averin95}.

The effect of rough interfaces and of spin-flip scattering on spin-transfer torque in the presence of Andreev reflections has been discussed by Wang, Tang, and Xia \cite{Wang10}. For diffusive structures see \cite{Shomali11}.
Magnetization dynamics has been also addressed recently \cite{Linder11, Mai11, Linder14}. 

Andreev sidebands in a system with two superconducting leads coupled by a precessing spin proved important to study spin-transfer torques acting on the precessing spin \cite{Holmqvist11}. Spin-polarized Shapiro steps were studied in \cite{Holmqvist14}.

\section{Andreev spectroscopy in F-S and F-S-F' structures}

Andreev point contact spectra in S-F structures are modified with respect to those in S-N structures due to spin-filtering effects and the spin-sensitivity of Andreev scattering \cite{Jong95,Kashiwaya99,Zutic00,Mazin01,Kopu04,Perez04,Eschrig13}.
Spin-dependent phase shifts also crucially affect Andreev point contact spectra \cite{Lofwander10,Grein10,Piano11,Kupferschmidt11,Wilken12,Yates13,Sun15}. 
Point contacts have lateral dimensions much smaller than the superconducting coherence lengths of the materials on either side of the contact. 
Typically, a voltage will be applied over the contact, which makes it necessary to study in addition to the coherence amplitudes $\ga $ and $\gb $ also distribution functions. I follow the definition in Ref. \cite{Eschrig99,Eschrig00}, where 2$\times $2 distribution function spin-matrices $\xa $ and $\xb $ for particles and holes are introduced which obey a transport equation 
\begin{align}
\label{keld1}
&\i\hbar \, (\qpartial + \partial_t )\xa
-(\ga \qt \Db +\va ) \qt \xa +
\xa \qt (\ga \qt \Db +\va )^\dagger
= {\cal I}^{\rm coll}
\\
\label{keld2}
&i\hbar \, (\qpartial - \partial_t )\xb
-( \gb \qt \Da +\vb ) \qt \xb +
\xb \qt ( \gb \qt \Da +\vb )^\dagger
= \tilde{\cal I}^{\rm coll}
\end{align}
The distribution functions matrices are hermitian, $\xa =\xa^\dagger $ and $\xb =\xb^\dagger $.
The right hand sides of equation \eqref{keld1} and \eqref{keld2} contain collision terms (see Ref. \cite{Eschrig00} for details), which vanish in ballistic structures. In general,
these distribution functions can be related to quasiclassical Keldysh propatators 
$g^{\rm K}=-2\pi\i ({\cal V}\qt\xa \qt {\cal V}^\dagger - {\cal F}\qt\xb \qt {\cal F}^\dagger )$ 
and 
$f^{\rm K}=-2\pi\i ({\cal V}\qt\xa \qt {\cal F}^\dagger - {\cal F}\qt\xb\qt {\cal V}^\dagger )$.
The Fermi distribution function for particles, $f_{\rm p}$, and holes, $f_{\rm h}$, is related to $\xa $ in the normal state by $f_{\rm p}=(1-\xa )/2$ and $f_{\rm h}=(1-\xb )/2$.

\subsection{Andreev processes in point contact geometry}

In this section I consider point contacts
of dimensions large compared to the Fermi wavelength and small compared to the superconducting coherence lengths. In this case, the wavevector $\vec{k}_\parallel $ parallel to the contact interface is approximately conserved.
The current on the ferromagnetic side of a point contact,
being directed along the interface normal,
can be decomposed into
\begin{equation}
I=I_{\rm I}-I_{\rm R} + I_{\rm AR}
\end{equation}
where the various terms are the incoming current, $I_{\rm I}$, the
normally reflected part, $I_{\rm R}$,
and the Andreev reflected part, $I_{\rm AR}$.
The sign convention here is such that a positive current denotes
a current into the superconductor. Thus, in the normal state the current $I$ is positive
when the voltage in the ferromagnet is positive.
The various currents can be expressed as 
\begin{equation}
\label{def1}
I_{\rm X}= -
\frac{{\cal A} }{2\pi \hbar }
\int_{\cal A_{\rm cF}}
\frac{{\rm d}^2 S(\vec{k}_\parallel ) }{(2\pi )^2}
\int\limits_{-\infty }^{\infty } \frac{{\rm d}\eps}{2}
\ce \; j_{\rm X}
,
\end{equation}
where $\ce=-|\ce|$ is the charge of the electron, ${\cal A} $ is the contact area, and ${\cal A}_{\rm cF}$ is the projection of the Fermi surfaces in the ferromagnet on the contact plane.
For each value $\vec{k}_\parallel $ there will be a number of (spin-polarized) Fermi surface sheets involved in the interface scattering (in the simplest case spin-up and spin-down, or only spin-up), and the dimension and structure of the scattering matrix will depend on how many Fermi surface sheets are involved. The sum over $\alpha $ and $\beta $ runs over those Fermi surface sheets $1,...,\nu $ for each given value of $\vec{k}_\parallel $. In the superconductor I assume for simplicity that only one Fermi surface sheet is involved for each $\vec{k}_\parallel $.
The reflection and transmission amplitudes for each $\vec{k}_\parallel $ are related to the scattering matrix as
\begin{equation}
{\bf S}(12;34)=\left(
\begin{array}{cc}
R_{12} & \plus T_{14}\\T_{32} & -R_{34}
\end{array}
\right)
\end{equation}
where directions 1 and 2 refer to the superconductor, and 3 and 4 to the ferromagnet.
$R_{12}$ is a 2$\times $2 spin matrix, $R_{34}$ is a $\nu \times \nu$ matrix with elements $R_{\alpha \beta}$, $T_{14} $ is a $2\times \nu $ matrix with elements $T_{1\beta }$, and $T_{32}$ is a $\nu \times 2$ matrix with elements $T_{\alpha 2}$.
The spectral current densities $j_{\rm X}$ are given by
\begin{align}
j_{\rm I} &= \sum_{\beta } \delta x_{\beta},\qquad
j_{\rm R} = \sum_{\alpha \beta } |R_{\alpha \beta} - T_{\alpha 2} \; v_2 \; \gamma_2 \; \tilde R_{2 1} \; \tilde \gamma_1 \; T_{1 \beta}|^2 \delta x_{\beta}
\label{IR}
\\
j_{\rm AR}&=
\sum_{\alpha \underline\alpha} |T_{\alpha 2} \; v_2 \; \gamma_2\;  \tilde T_{2 \underline\alpha } |^2 \delta \tilde x_{\underline\alpha },\qquad
v_2=(1-\gamma_2 \; \tilde R_{2 1} \; \tilde \gamma_1 \; R_{1 2} )^{-1},
\label{AR}
\end{align}
where $\delta x_\beta $ and $\delta \tilde x_\beta $ are the differences between the distribution functions in the ferromagnet and in the superconductor. If there is no spin-accummulation present, they are independent on the index $\beta $ and given by
\begin{align}
\delta x(V,T;\eps) &= \tanh \left( \frac{\eps -\ce V}{2k_{\rm B}T} \right)
-\tanh \left( \frac{\eps }{2k_{\rm B}T_{\rm S}} \right) , \qquad
\delta \tilde x (V, T; \eps )= \delta x(V,T;-\eps ),
\end{align}
with $T_{\rm S}$ the temperature in the superconductor.
Equations \eqref{IR}-\eqref{AR} are valid for general normal-state scattering matrices $S$, and can be applied to non-collinear magnetic structures.
For the case that all reflection and transmission amplitudes are spin-diagonal, and assuming on the superconducting side of the interface a spin-mixing angle $\vartheta $, these expressions are explicitely given by
\begin{align}
j_{\rm R} &= \left[|v_{0+}|^2 \; |r_\uparrow-r_\downarrow e^{\i\vartheta } \gamma_0^2|^2
+ |v_{0-}|^2 \; |r_\downarrow -r_\uparrow e^{-\i \vartheta} \gamma_0^2|^2\right] \delta x\\
j_{\rm I} &= 2 \delta x, \qquad
j_{\rm AR} = (t_\uparrow t_\downarrow)^2 |\gamma_0|^2 \left[|v_{0+}|^2 +|v_{0-}|^2 \right] \delta \tilde x
\end{align}
with
$\gamma_0= -\Delta /(\varepsilon+\i\Omega)$, $\Omega=\sqrt{\Delta^2-\varepsilon^2}$, 
$v_{0\pm}=( 1-\gamma_0^2 r_\uparrow r_\downarrow e^{\pm \i\vartheta} )^{-1}$,
and the energy $\varepsilon $ is assumed to have an infinitesimally small positive imaginary part.
Andreev resonances arise for energies fulfilling $1=\gamma_0(\eps )^2 r_\uparrow r_\downarrow e^{\pm \i \vartheta }$ (in agreement with the discussion in section \ref{ABS}\ref{ABS1} for $d=0$). 

On the other hand, for half-metallic ferromagnets the Andreev reflection contribution is zero in collinear magnetic structures. In non-collinear structures, however,
the process of {\it spin-flip Andreev reflection} takes place, introduced in reference \cite{Grein10}, and illustrated there in figure 11.
Spin-flip Andreev reflection is the only process providing particle-hole coherence in a half-metallic ferromagnet.
Such structures are described by the theory deceloped in appendix C of \cite{Eschrig09}. 
Application of this theory to experiment on CrO$_2$ is provided in \cite{Lofwander10,Yates13}.
A generalization to strongly spin-polarized ferromagnets with two itinerant bands is given in \cite{Grein10} with application to experiment in \cite{Piano11}. 

\begin{figure}[t]
\centering{\hspace{0.8in} {\tiny (c)} \hspace{1.7in} {\tiny (d)} \hspace{1.0in}{\tiny (e)} \hspace{0.7in}$\;$}\\
\centering{
\begin{minipage}[b]{0.8in}
\includegraphics[width=0.8in]{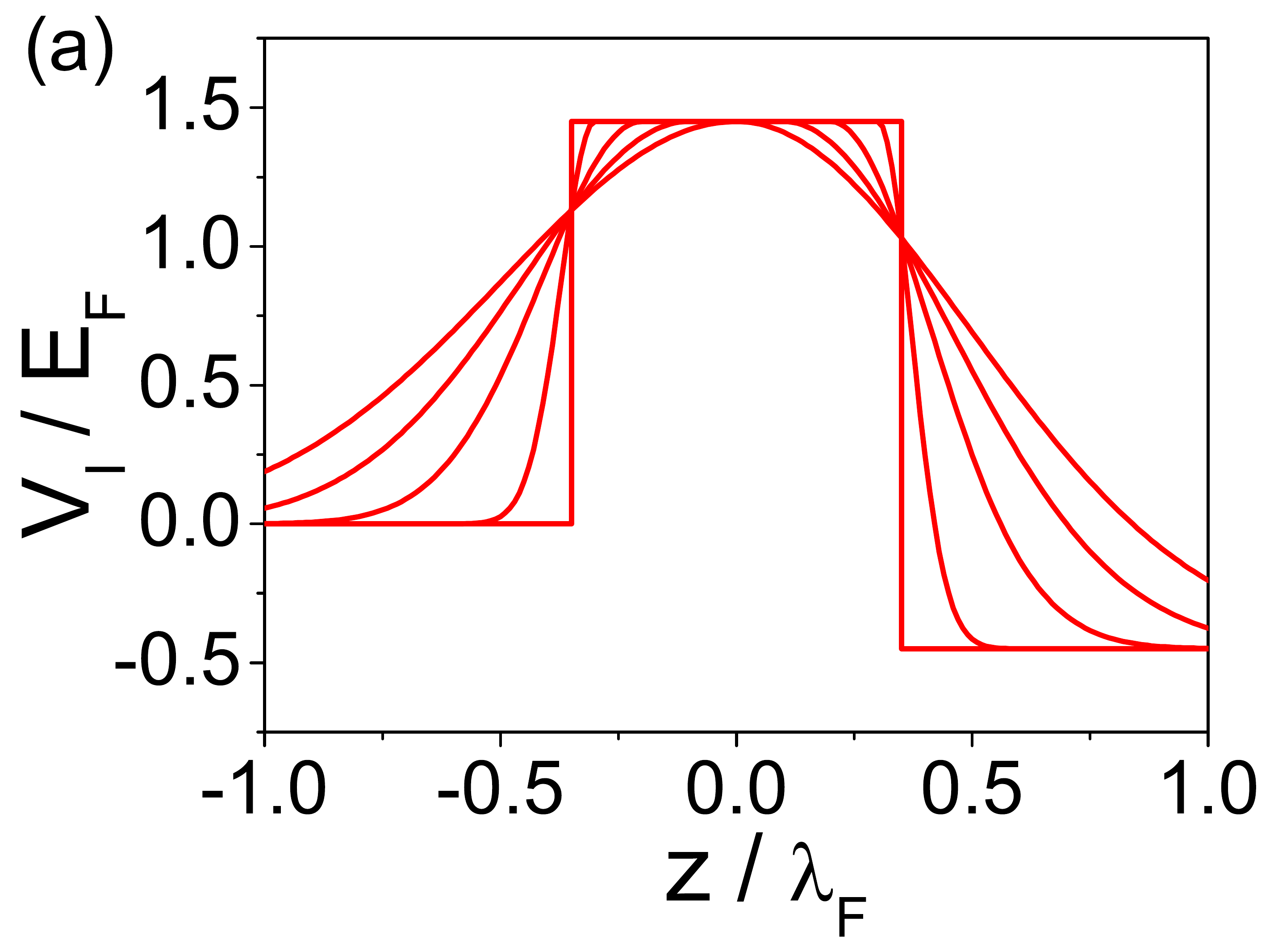}
\includegraphics[width=0.8in]{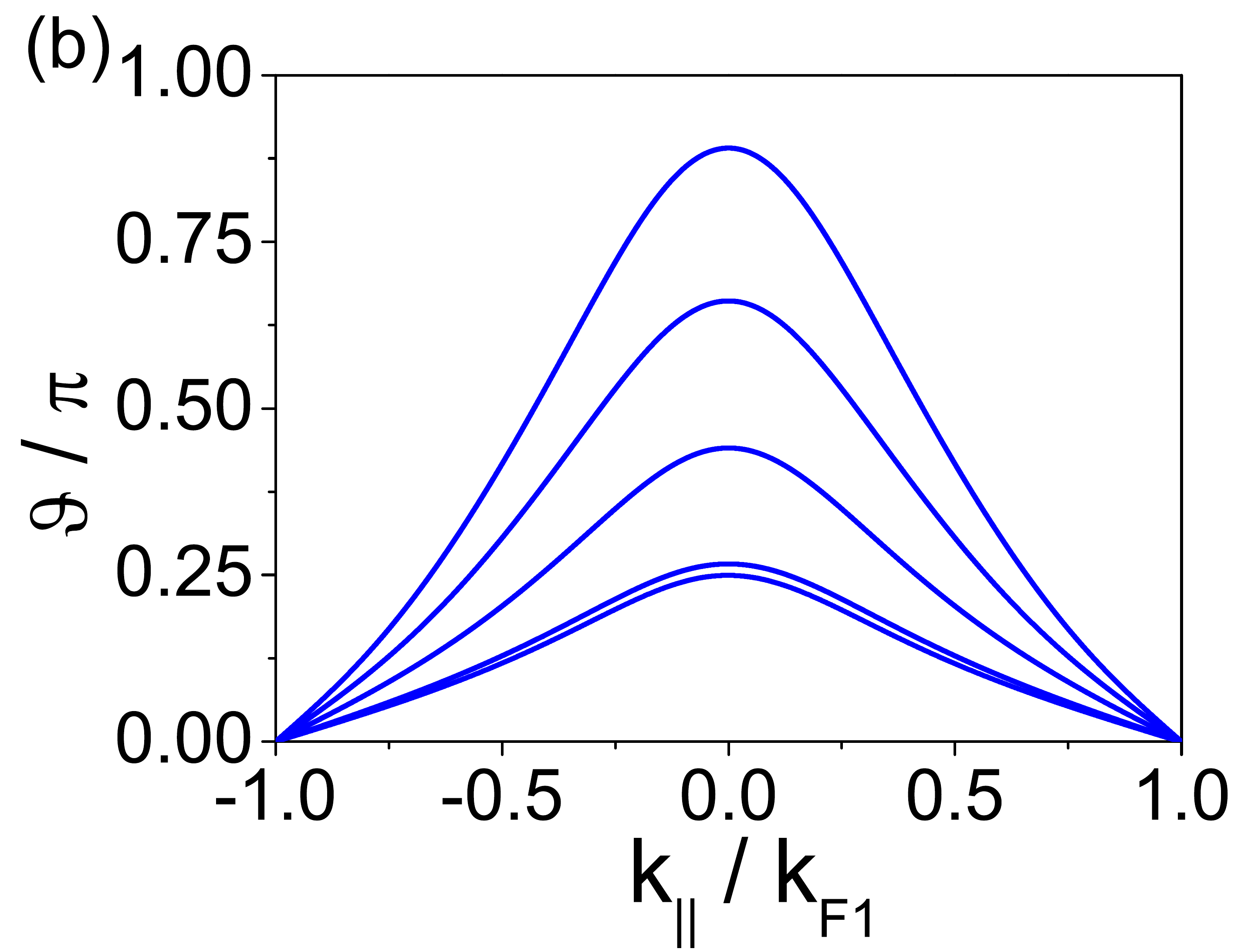}
\end{minipage}
\begin{minipage}[b]{4.2in}
\includegraphics[width=1.7in]{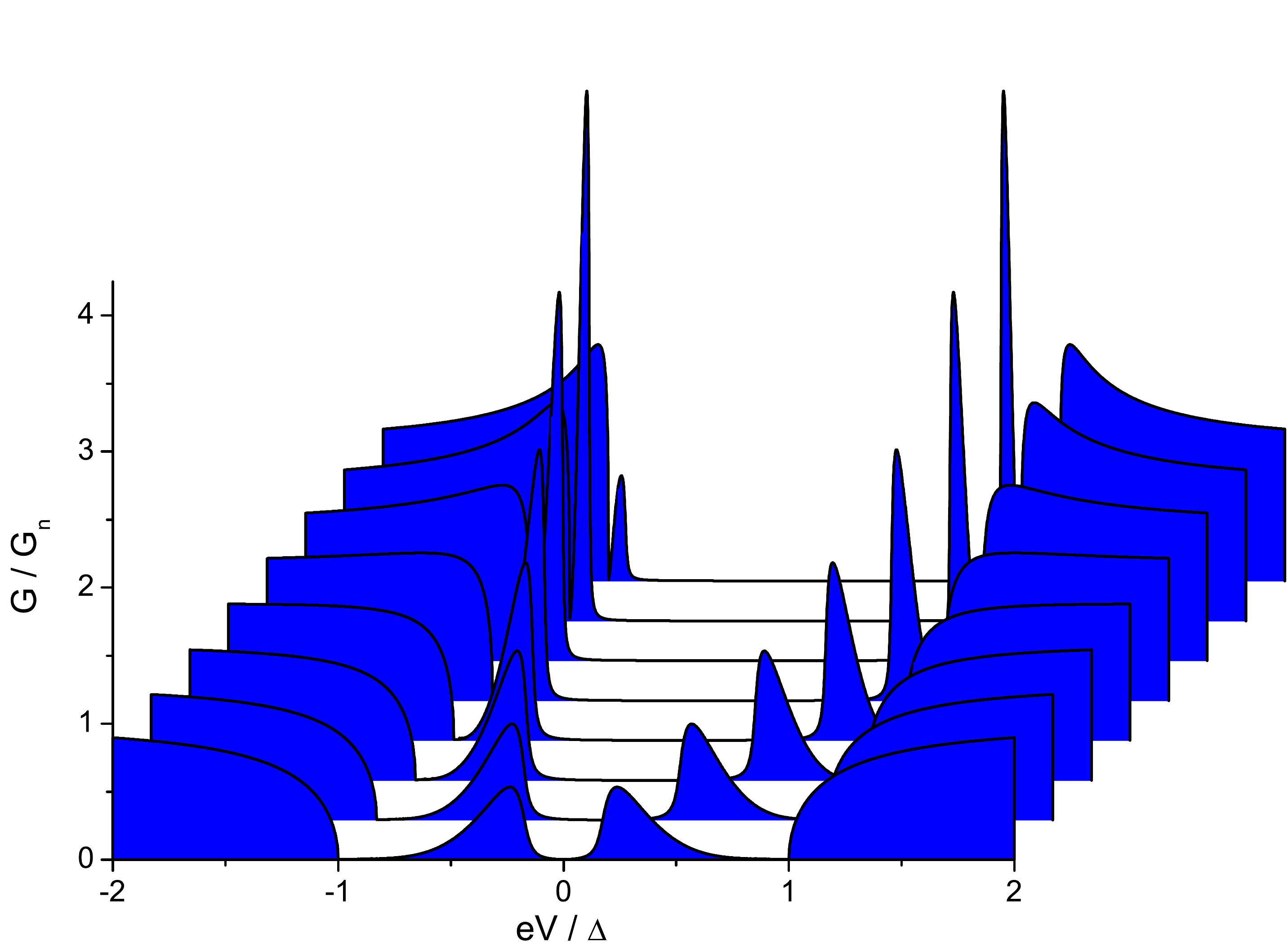}
\includegraphics[width=2.5in]{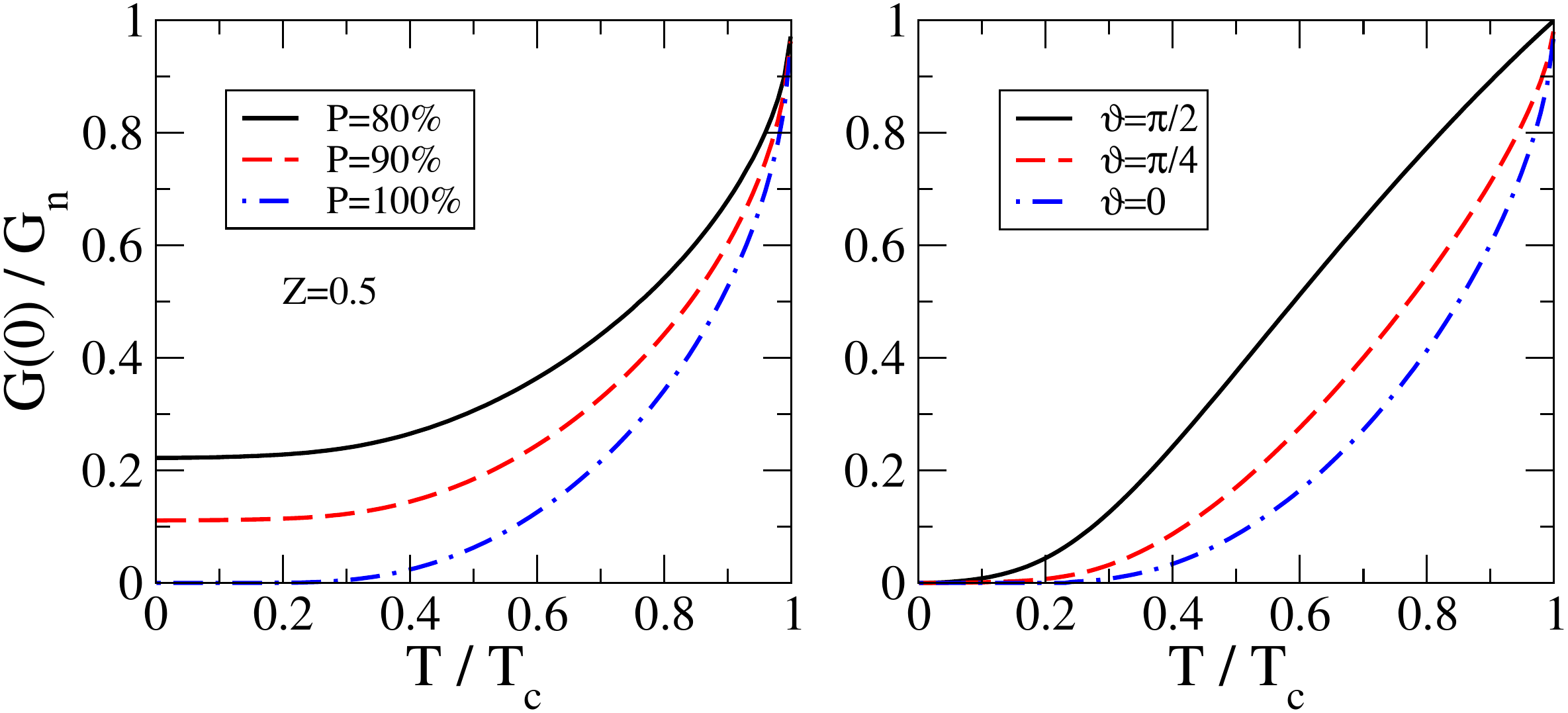}
\end{minipage}
}
\caption{
(a)
Shape function of the (spin-averaged) interface barrier potential.
A shape parameter $\sigma=0\ldots 0.7\ \lambda_{\rm{F}}$ increases for increasing smoothness.
(b)
Spin-mixing angle $\vartheta_{\rm S}$ as a function of $|\vec{k}_\parallel|$
for the shape functions in (a). $\vartheta_{\rm S} $ increases with increasing $\sigma $.
(c)
The differential conductance of an F-FI-S structure with various degrees of smoothness if the FI barrier, at $T=0$;
$\sigma$ increases from back to front in steps of 0.1 $\lambda_{\rm F}$. 
(d)-(e)
Temperature dependence of the zero-voltage
conductance of an HM-FI-S point contact
as predicted by (d) the modified BTK model \cite{Mazin01}
and (e) the spin-active interface model \cite{Eschrig09,Grein10}.
Adapted from \cite{Grein10,Lofwander10}.
Copyright (2010) by the American Physical Society.
}
\label{PC}
\end{figure}
In figure \ref{PC} selected results are shown. In (a) and (b) it is demonstrated that the spin-mixing angle can acquire large values if a smooth spatial interface profile is used instead of an atomically clean interface. Correspondingly, in (c) Andreev resonances are more pronounced for smoother interfaces. In (d) and (e) a comparision of the model by Mazin {\it et al.} \cite{Mazin01} for various spin polarizations $P$ with the spin-mixing model for P=100\% and various spin-mixing angles $\vartheta $ shows that the two can be experimentally differentiated by studying the low-temperature behavior \cite{Lofwander10}. 

In an experiment by Visani {\it et al.}
\cite{Visani12} geometric resonances (Tomasch resonances and Rowell-McMillan resonances)
in the conductance across a La$_{0.7}$Ca$_{0.3}$Mn$_3$O/YBa$_2$Cu$_3$O$_7$ interface were studied, demonstrating long-range propagation of superconducting correlations across the half metal La$_{0.7}$Ca$_{0.3}$Mn$_3$O. 
The effect is interpreted in terms of spin-flip Andreev reflection (or, as named by the authors of \cite{Visani12}, ``equal-spin Andreev reflection'').

Spin-dependent scattering phases qualitatively affect the
zero- and finite-frequency current noice in S-F point contacts \cite{Cottet08a,Cottet08}. 
It was found that for weak transparency noise steps appear at frequencies
or voltages determined directly by the spin dependence of scattering phase shifts.

\subsection{Andreev bound states in non-local geometry}

A particular interesting case is that of two F-S point contacts separated by a distance $L$ of the order of the superconducting coherence length.
This is effectively an F-S-F' system, or if barriers are included, an F-FI-S-FI'-F' system.
In this case, for a ballistic superconductor, one must consider separately trajectories connecting the two contacts \cite{Kalenkov07}. Along these trajectories the distribution function is out of equilibrium, and equations \eqref{keld1}-\eqref{keld2} must be solved. In addition, the coherence functions at these trajectories experience both ferromagnetic contacts, and are consequenctly different from the homogeneous solutions $\gamma_0$, $\tilde \gamma_0$ of all other quasiparticle trajectories.

The current on the ferromagnetic side of one particular interface (positive in direction of the superconductor),
can be decomposed in an exact way,
\begin{equation}
I=I_{\rm I}-I_{\rm R} + I_{\rm AR} - I_{\rm EC} + I_{\rm CAR}
\end{equation}
where the various terms are the incoming current, $I_{\rm I}$, the
normally reflected part, $I_{\rm R}$,
the Andreev reflected part, $I_{\rm AR}$, and the two non-local contributions
due to elastic co-tunneling, $I_{\rm EC}$ and crossed Andreev reflection,
$I_{\rm CAR}$. 

\begin{figure}[t]
\centering{
\begin{minipage}[b]{3.0in}
\includegraphics*[width=2.1in,clip]{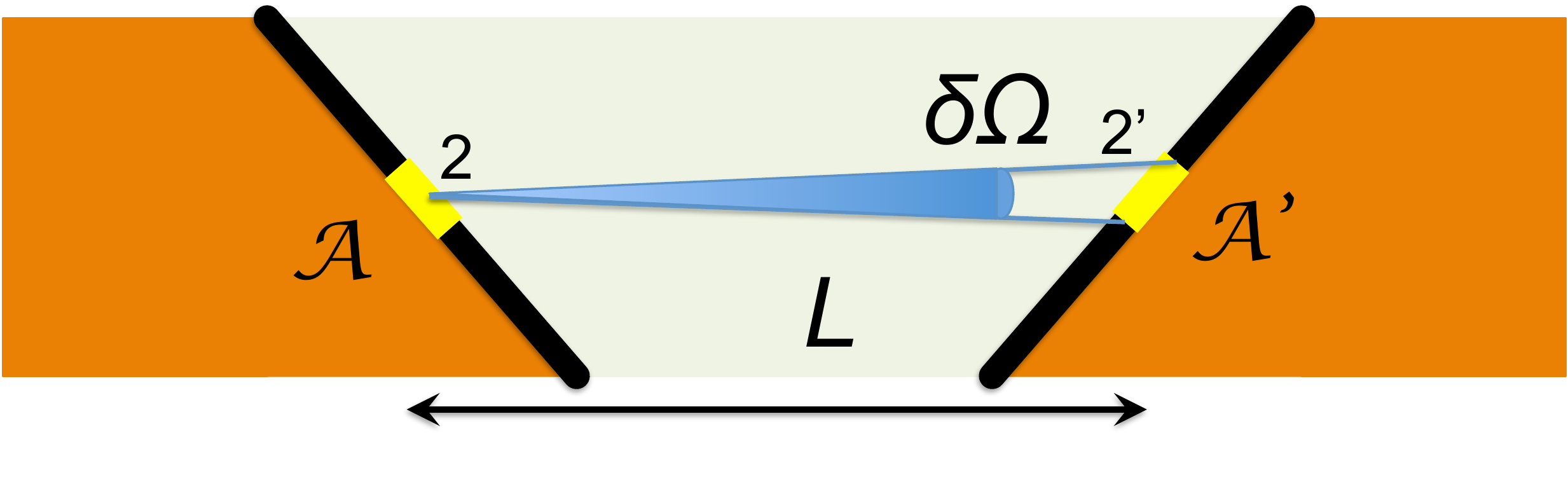}
\includegraphics*[width=0.8in,clip]{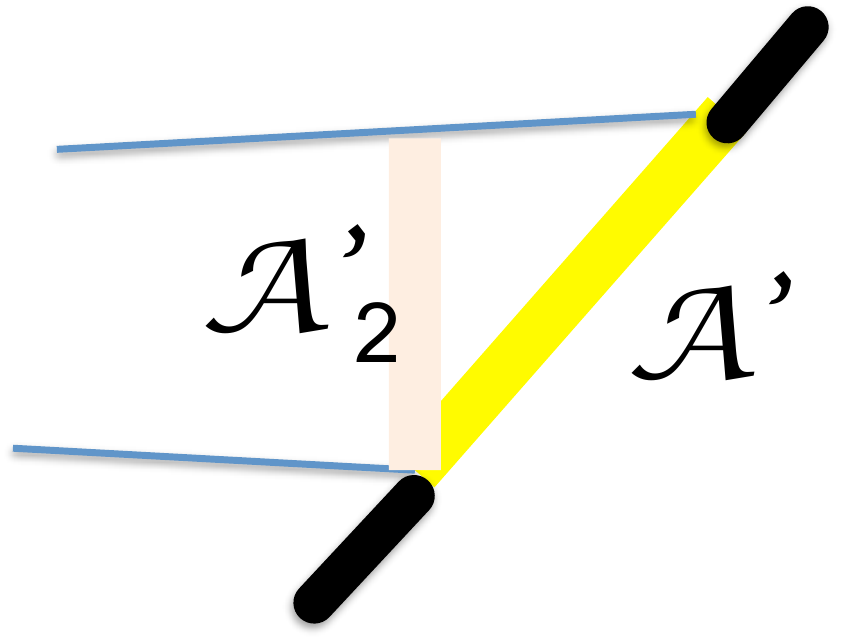}\\
\includegraphics*[width=3.0in,clip]{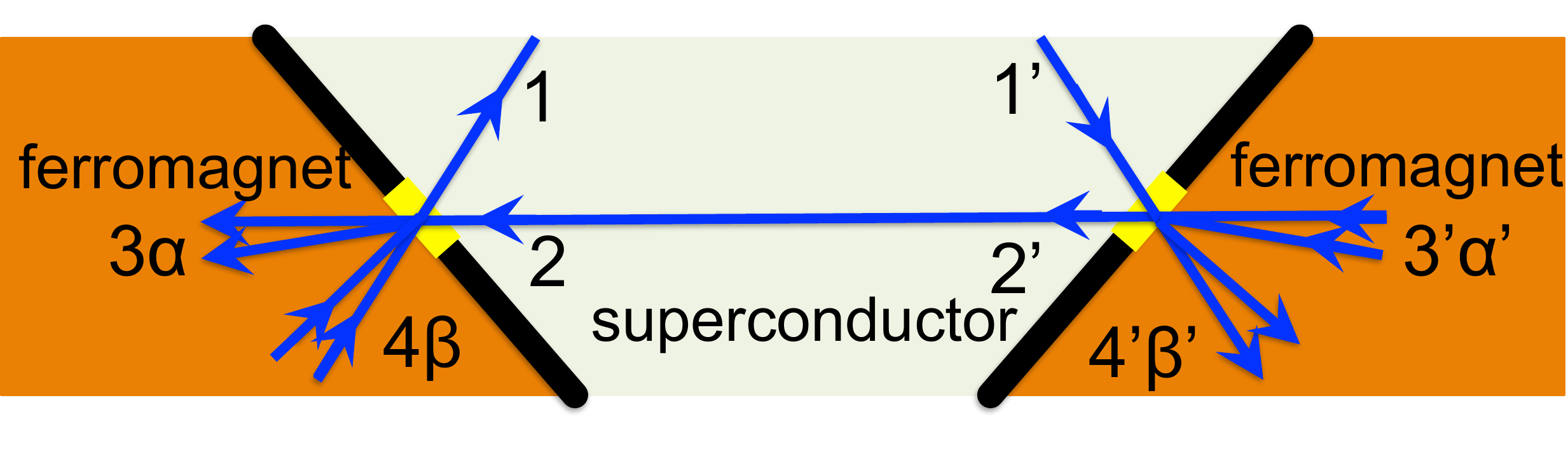}
\end{minipage}
\begin{minipage}[b]{2.0in}
\includegraphics*[width=1.8in,clip]{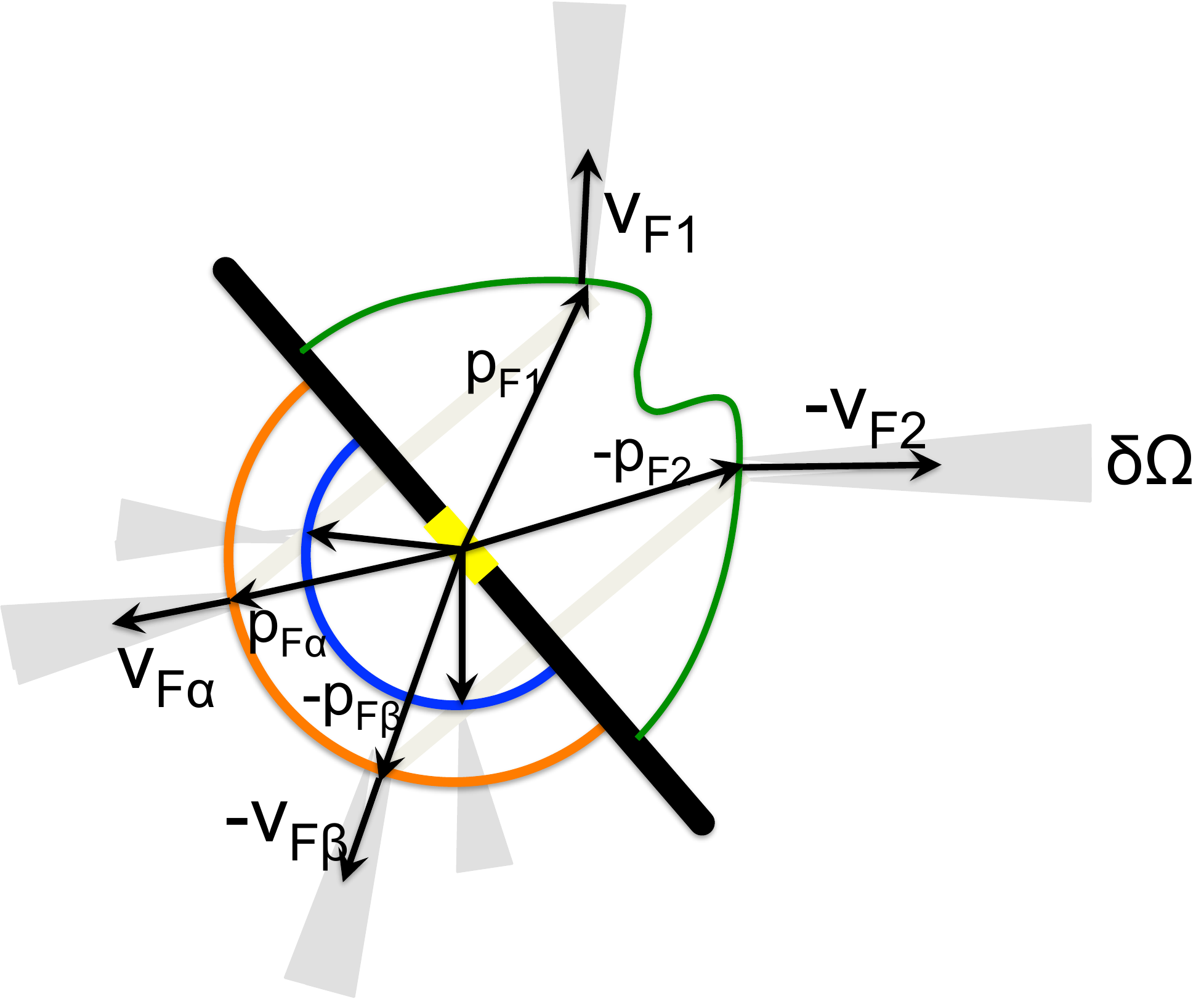}
\end{minipage}
}
\caption{
Illustration of notation used in text. 
For brevity of notation I sometimes omit the label 3 and 4, implying that $\alpha $ then means $3\alpha $ and $\beta $ means $4\beta $. E.g. in the right picture
the ferromagnet has two spin Fermi surfaces (red and blue), labeled by 
$\alpha \in \left\{3\uparrow,3\downarrow\right\}$ and 
$\beta \in \left\{4\uparrow,4\downarrow\right\}$ etc. The superconductor's Fermi surface is drawn in green.
}
\label{setup}
\end{figure}
There will be contributions from trajectories in the superconductor which do not connect the two contacts. These will be described by equations \eqref{IR}-\eqref{AR} above. Here, I will concentrate on the non-local contributions, which arise from the particular trajectories connecting the two contacts. Assuming the area of each contact much smaller than the superconducting coherence lenght (however larger than the Fermi wavelength, such that the momentum component parallel to the contact interfaces are approximately conserved), one can identify all trajectories connecting the two contacts, treating only one and scaling the result with the contact area.
The solid angle from a point at the first contact to the area ${\cal A}'$
of the second contact
is given by $\delta \Omega= {\cal A}'_2/L^2$, where ${\cal A}'_2$ is the projection
of the area of the second contact to the plane perpendicular to the
line $2,2'$ which connects the contacts (see figure \ref{setup} for the notation).

\begin{figure}[t]
\centering{
\hspace{0.5in}{\tiny (a)} \hspace{0.95in} {\tiny (b)} 
\hspace{0.95in}{\tiny (c)} \hspace{0.95in} {\tiny (d)} \hspace{1.0in}$\;$}\\
\centering{
\includegraphics*[width=0.2\linewidth,clip]{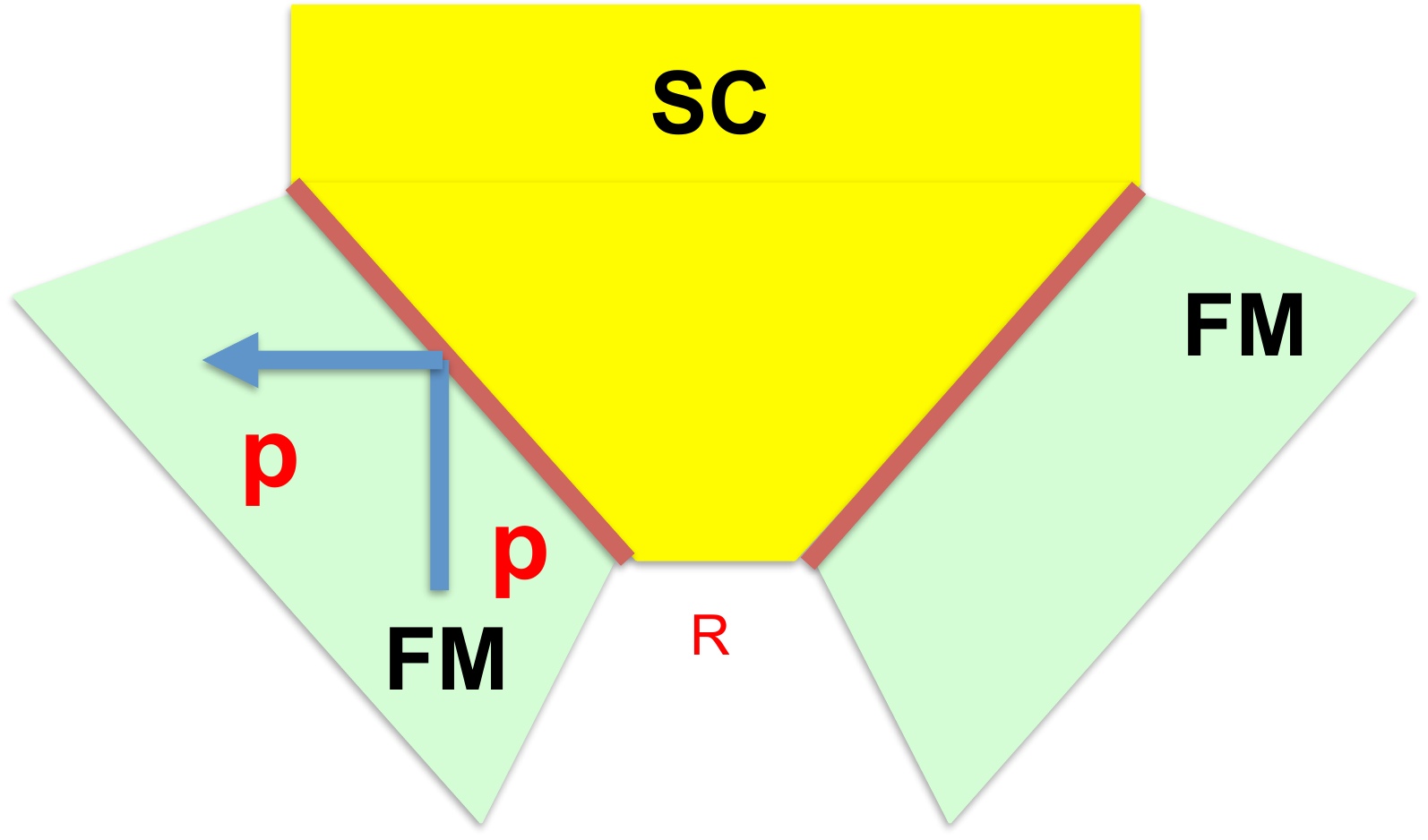}
\includegraphics*[width=0.2\linewidth,clip]{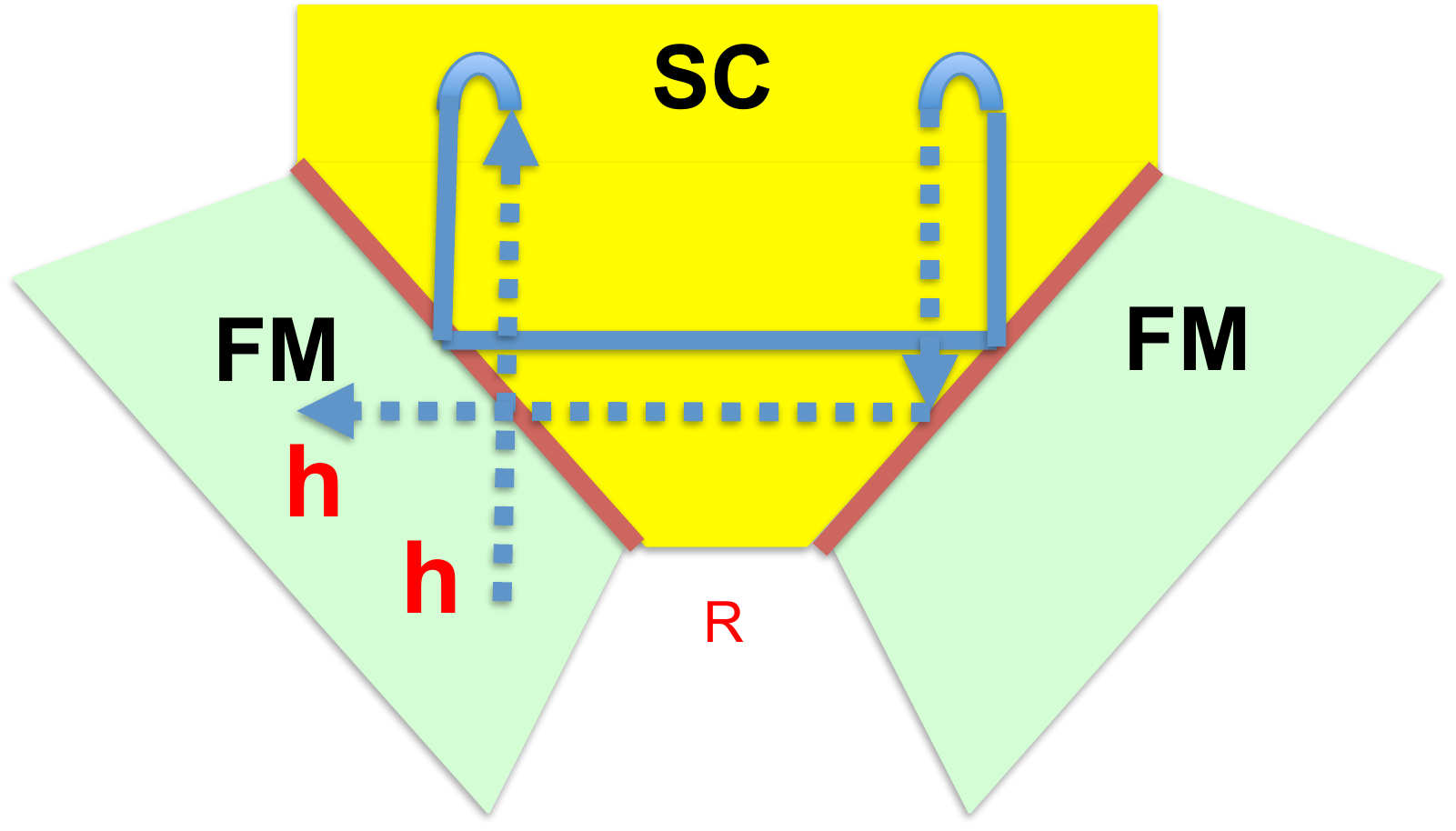}
\includegraphics*[width=0.2\linewidth,clip]{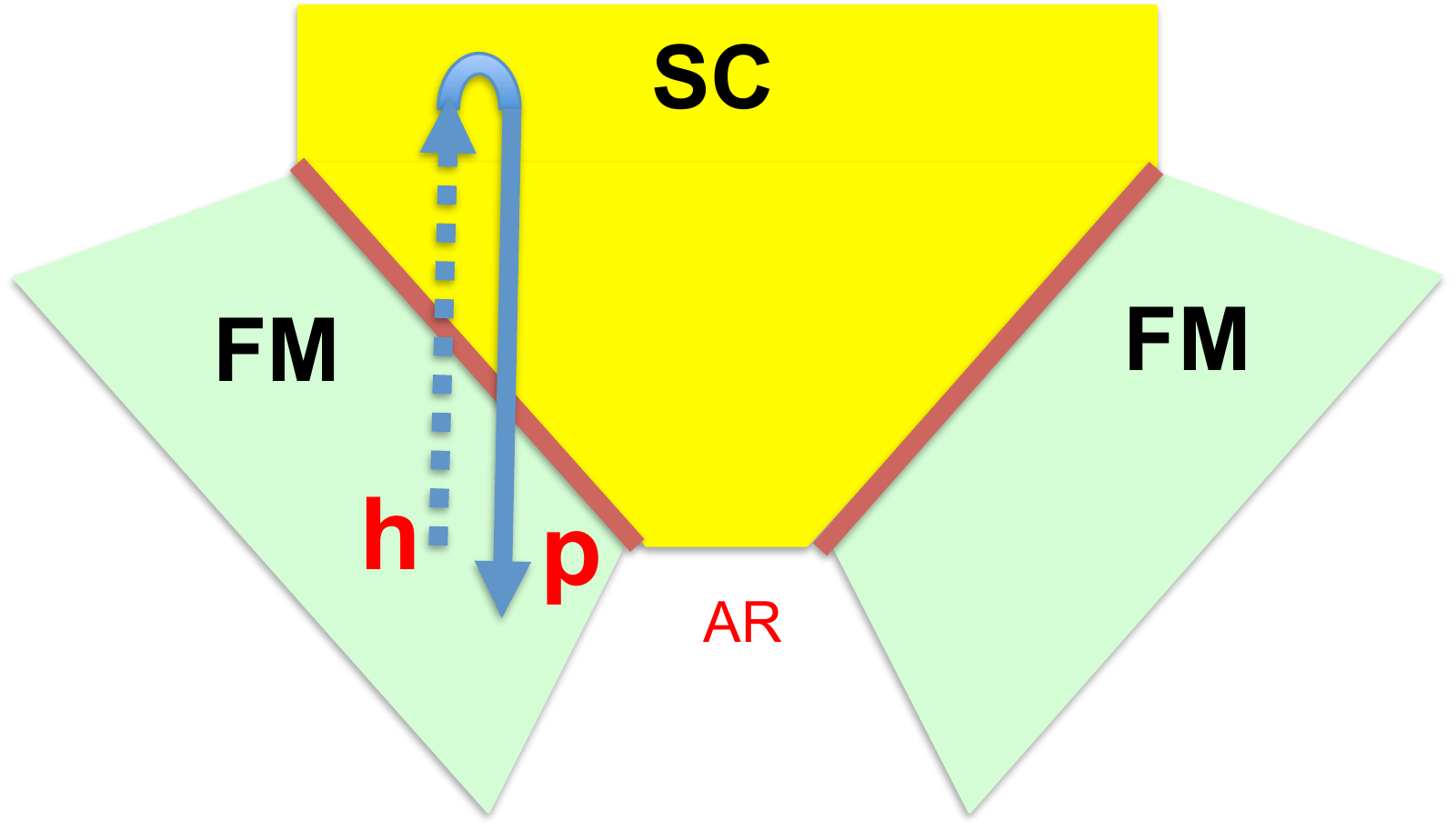}
\includegraphics*[width=0.2\linewidth,clip]{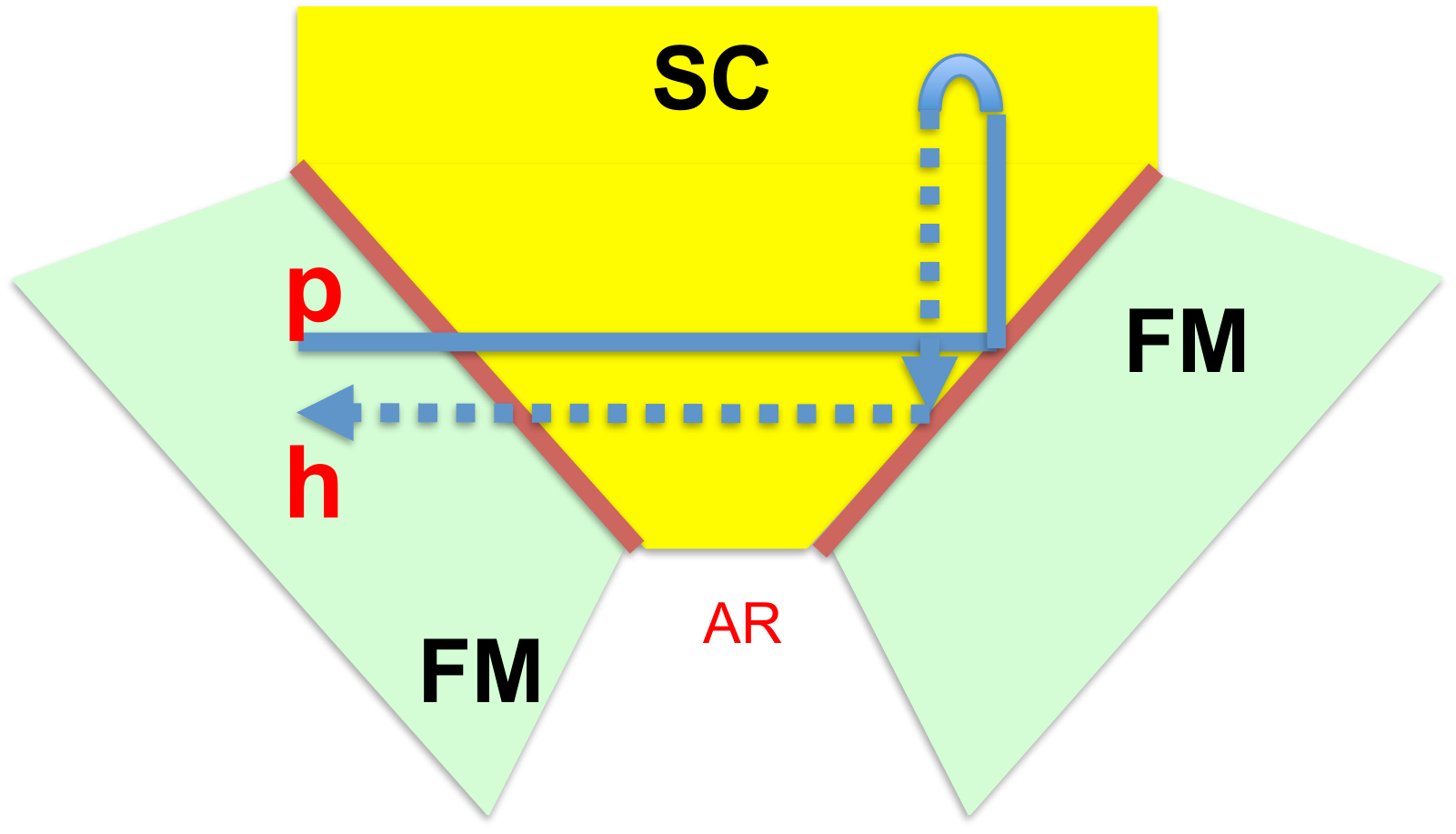}
}\\
\centering{
\hspace{0.5in}{\tiny (e)} \hspace{0.95in} {\tiny (f)} 
\hspace{0.95in}{\tiny (g)} \hspace{0.95in} {\tiny (h)} \hspace{1.0in}$\;$}\\
\centering{
\includegraphics*[width=0.2\linewidth,clip]{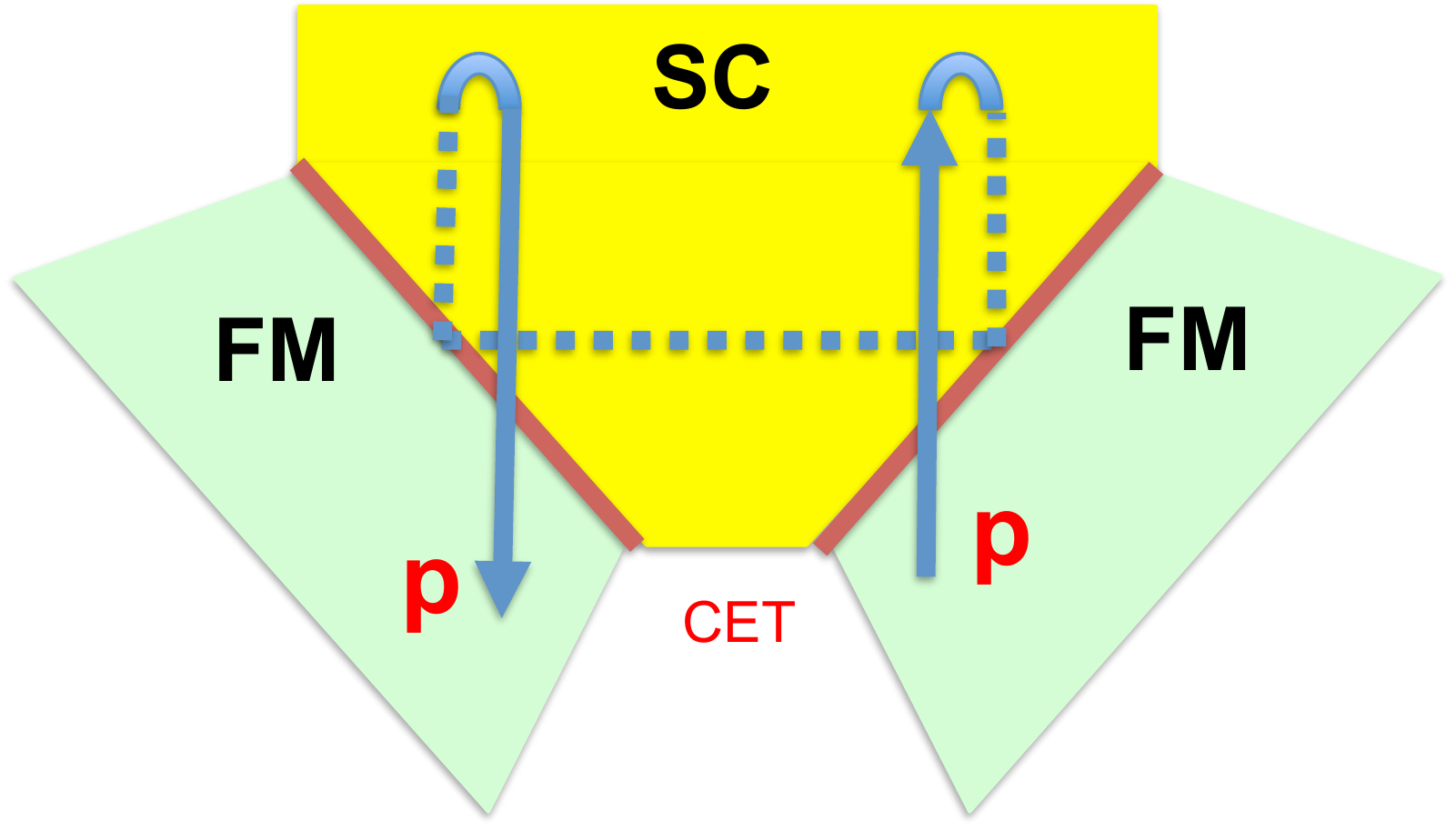}
\includegraphics*[width=0.2\linewidth,clip]{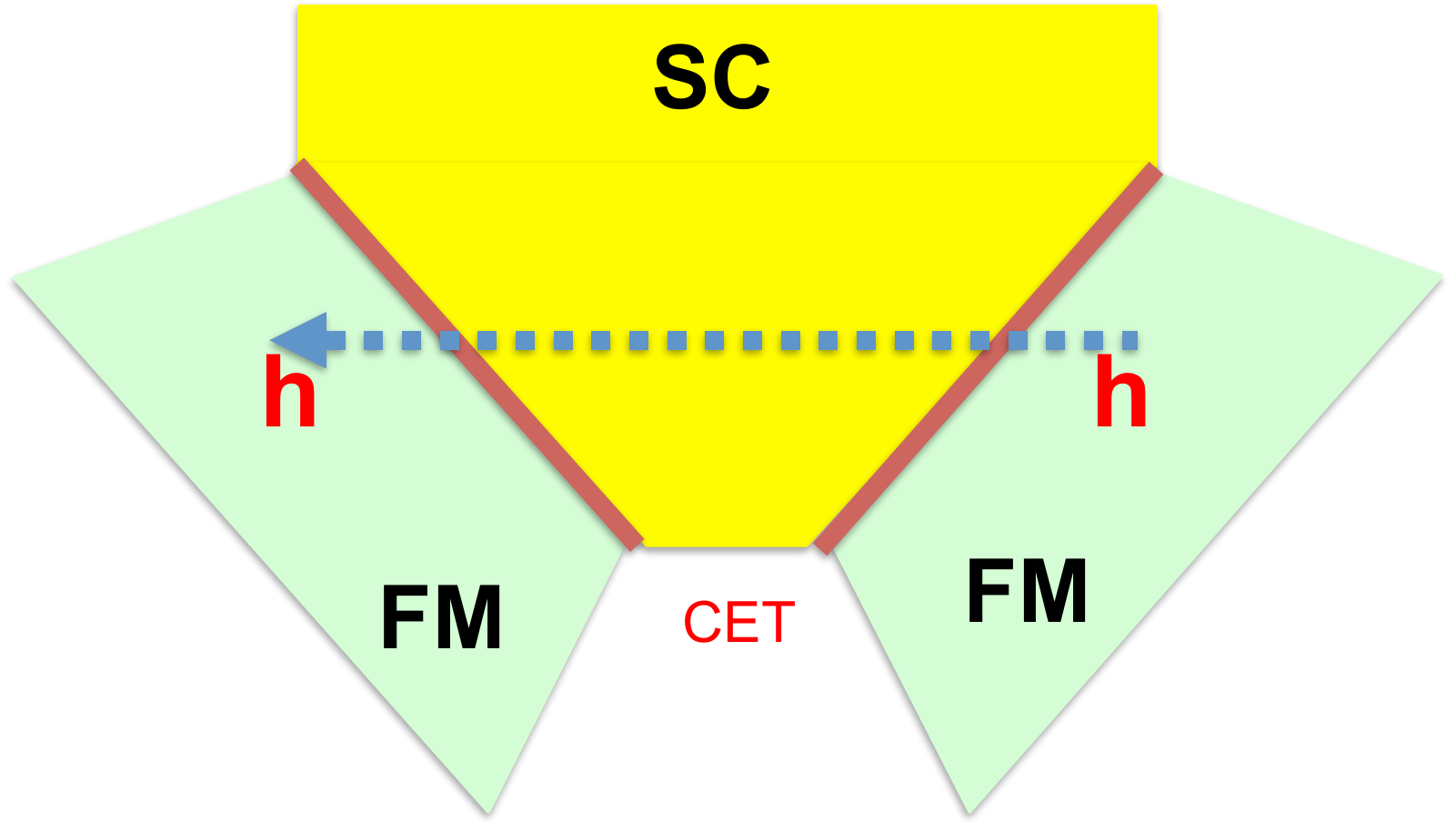}
\includegraphics*[width=0.2\linewidth,clip]{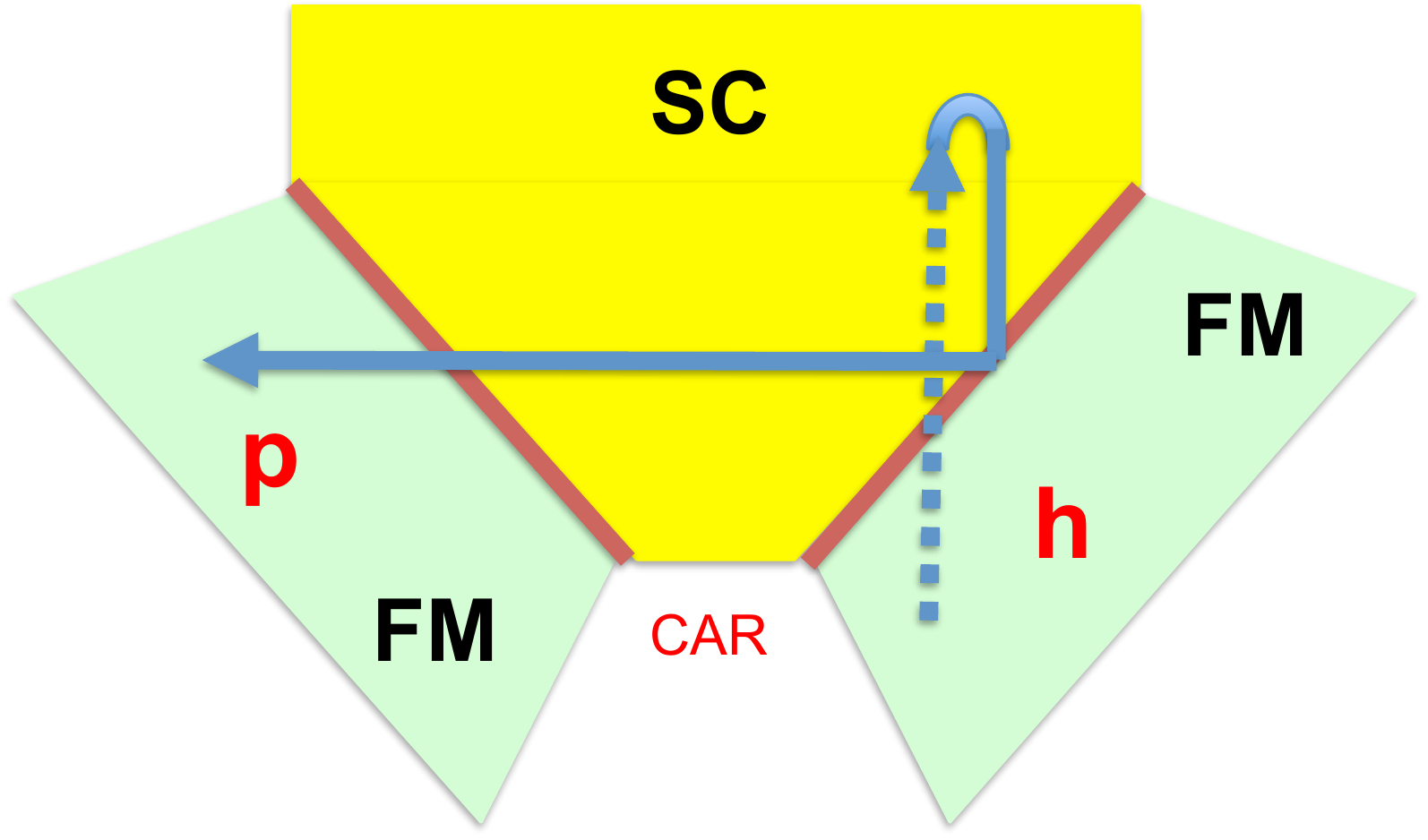}
\includegraphics*[width=0.2\linewidth,clip]{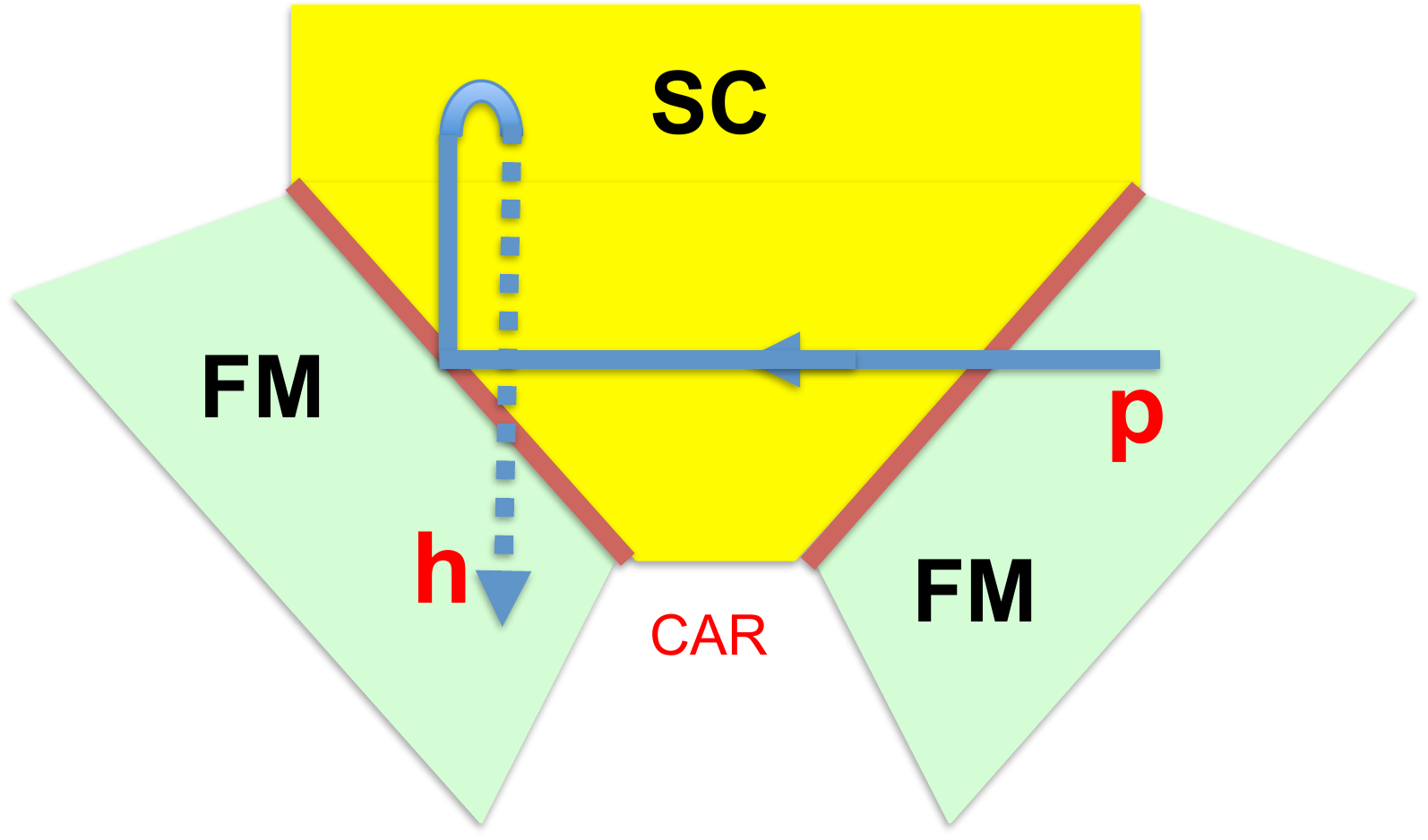}
}
\caption{
Illustration of selected processes contributing to the expressions \eqref{ji}-\eqref{jcar}.
Andreev reflections are denoted as loops, turning particles (p, full lines) into holes (h, dashed lines) or vice versa.
(a)-(b) contributions to the reflection components, equation \eqref{jr};
(c)-(d) contributions to the Andreev reflection components, equation \eqref{jar};
(e)-(f) contributions to the coherent electron transfer components, equation \eqref{jec};
(g)-(h) contributions to the crossed Andreev reflection components, equation \eqref{jcar}.
}
\label{NLAR}
\end{figure}

Using the conservation of $\vec{k}_\parallel $, and that consequently
${\rm d}^2S(\vec{k}_\parallel )={\rm d}^2S(\vec{p}_{\rm F2}) |\hat{\vec{n}} \cdot \vec{v}_{\rm F2}| /|\vec{v}_{\rm F2}| 
={\rm d}^2S(\vec{p}_{\rm F\alpha}) |\hat{\vec{n}}\cdot \vec{v}_{\rm F\alpha}|/|\vec{v}_{\rm F\alpha}| $ (where $\hat{\vec{n}} $ is the contact surface normal),
one can express the currents as
\begin{equation}
\label{def2}
I_{\rm X}=
\frac{{\rm d}^2 S}{{\rm d}\Omega} \Big|_{\vec{p}_{\rm F2}}
\frac{{\cal A}_2 {\cal A}'_2}{(2\pi\hbar)^3L^2}
\int\limits_{-\infty }^{\infty } \frac{{\rm d}\eps}{2}
j_{\rm X}
,
\end{equation}
where 
$\vec{p}_{\rm F2}$ is the particular Fermi momentum in
the superconductor corresponding to a Fermi velocity in direction of the line $2,2'$ (I assume for simplicity that only one such Fermi momentum exists),
and ${\rm d}^2 S/{\rm d}\Omega$ is the differential fraction of the Fermi surface of the superconductor per solid angle $\Omega $ in direction of the Fermi velocity $\vec{v}_{\rm F2}$ that connects the two contacts. Note that ${\rm d}^2 S/{\rm d}\Omega$ is the same at both contacts for superconductors with inversion symmetry, as then this quantity is equal at $\vec{p}_{\rm F2}$ and $-\vec{p}_{\rm F2}$.
Reversed directions are denoted by an overline: $\vec{p}_{\rm F\bar 2}=-\vec{p}_{\rm F2}$ etc.
Let us introduce scattering matrices ${\bf S}(12;34)$ as well as ${\bf S}(\bar 2 \bar 1, \bar 4 \bar3 )$ at the
left interface, the latter being equal to ${\bf S}(12;34) $ for materials with
centrosymmetric symmetry groups, which I consider here.
Analogously, for the right interface let us introduce the scattering matrices
${\bf S}'(2' 1';4'3')={\bf S}'(\bar 1' \bar 2';\bar 3'\bar 4')$.
The scattering matrices for holes are related to the scattering matrices for particles by
$\tilde {\bf S}(21;43)= {\bf S}(\bar 2 \bar 1,\bar 4 \bar 3)^\ast$ etc.
One obtains for this case
\begin{align}
\label{ji}
j_{\rm I} &=  
\sum_\beta x_{\beta } + \sum_{\bar \alpha } \delta x_{\bar \alpha }
\\
\label{jr}
j_{\rm R} &= 
\sum_{\alpha \beta}
|R_{\alpha \beta } - T_{\alpha 2} \; v_2 \; \gamma_2 \; \tilde R_{2 1} \; \tilde \gamma_1 \;
T_{1 \beta } |^2 \delta x_{\beta }
+ \sum_{\bar \beta \bar \alpha }
|R_{\bar \beta \bar \alpha } - T_{\bar \beta \bar 1} \;
v_{\bar 1} \;\gamma_{\bar 1} \; \tilde R_{\bar 1 \bar 2} \; \tilde \gamma_{\bar 2}
\; T_{\bar 2 \bar \alpha } |^2 \delta x_{\bar \alpha }
\\
\label{jar}
j_{\rm AR} &= 
\sum_{\alpha \underline \alpha }
|T_{\alpha 2} \; v_2 \; \gamma_2\;  \tilde T_{2 \underline\alpha } |^2 \delta \tilde x_{\underline \alpha }
+
\sum_{\bar \beta \bar{\underline \beta }}
|T_{\bar \beta \bar 1} \; v_{\bar 1} \; \gamma_{\bar 1} \; \tilde T_{\bar 1 \bar{\underline\beta }} |^2\delta \tilde x_{\bar{\underline\beta }}
\\
\label{jec}
j_{\rm EC} &= 
\sum_{\alpha\alpha' }
|T_{\alpha 2} \; v_2 \; u_{2 2'} \; T'_{2' \alpha'} |^2 \delta x_{\alpha'}
+
\sum_{\bar\beta \bar\beta' }
|T_{\bar \beta \bar 1} \; v_{\bar 1} \; \gamma_{\bar 1} \;
\tilde R_{\bar 1 \bar 2} \; \tilde u_{\bar 2 \bar 2'} \; \tilde R'_{\bar 2' \bar 1'} \; \tilde \gamma_{\bar 1'} \;  T'_{\bar 1' \bar \beta'} |^2 \delta x_{\bar \beta'}
\qquad
\\
j_{\rm CAR} &= 
\sum_{\alpha\beta' }
|T_{\alpha 2} \; v_{2}
\; u_{2 2'} \; R'_{2' 1'} \; \gamma_{1'} \;  \tilde T'_{1' \beta'}|^2\delta \tilde x_{\beta'}
+
\sum_{\bar\beta \bar\alpha' }
|T_{\bar \beta \bar 1} \; v_{\bar 1} \; \gamma_{\bar 1} \;
\tilde R_{\bar 1 \bar 2} \; \tilde u_{\bar 2 \bar 2'} \;  \tilde T'_{\bar 2' \bar \alpha'} |^2\delta \tilde x_{\bar \alpha'}
\label{jcar}
\end{align}
where the vertex corrections due to multiple Andreev processes are
$v_2=(1-\gamma_2 \; \tilde R_{2 1} \; \tilde \gamma_1 \; R_{1 2} )^{-1}$ and $
v_{\bar 1}=(1-\gamma_{\bar 1} \; \tilde R_{\bar 1 \bar 2} \; \tilde \gamma_{\bar 2} \; R_{\bar 2 \bar 1} )^{-1}$.
For unitary order parameters ($\Delta \tilde \Delta \sim \sigma_0$), let us define
$\Omega \, \sigma_0=[-\Delta_{k} \tilde \Delta_{k} -(\varepsilon+\i 0^+)^2 \sigma_0]^{\frac{1}{2}}$
as well as $\gamma=-\Delta/(\eps +\i \Omega )$, $\tilde \gamma= \tilde \Delta/(\eps +\i \Omega )$.
Using the amplitudes
\begin{eqnarray}
\Gamma_{2'}=R'_{2' 1'} \; \gamma_{1'} \; \tilde R'_{1' 2'} , \quad
\tilde \Gamma_{\bar 2'} =
\tilde R'_{\bar 2' \bar 1'} \; \tilde \gamma_{\bar 1'} \; R'_{\bar 1' \bar 2'}, \quad
\end{eqnarray}
and denoting with $L$ the distance between $2$ and $2'$,
\begin{eqnarray}
u_{2 2'}&=& \left[ c_{2'} + \i\frac{ \Gamma_{2'}
\tilde \Delta_{2'} -\varepsilon}{\Omega_{2'}} s_{2'} \right]^{-1} 
,\quad
\gamma_2= u_{2 2'} \left[ \Gamma_{2'} c_{2'} + \i\frac{
\Delta_{2'}+ \Gamma_{2'} \varepsilon }{\Omega_{2'}} s_{2'}
\right] 
\\
\tilde u_{\bar 2 \bar 2'}&=& \left[ c_{2'} - \i\frac{ \tilde \Gamma_{\bar 2'}
\Delta_{\bar 2'} +\varepsilon}{\tilde \Omega_{\bar 2'}} s_{2'} \right]^{-1},  \quad
\tilde \gamma_{\bar 2}= \tilde u_{\bar 2 \bar 2'} \left[ \tilde \Gamma_{\bar 2'}
c_{2'} - \i\frac{ \tilde \Delta_{\bar 2'}- \tilde \Gamma_{\bar 2'}
\varepsilon }{\tilde \Omega_{\bar 2'}} s_{2'} \right]
\end{eqnarray}
with $c_{2'}\equiv \cosh(\Omega_{2'}L/\hbar v_{\rm F,2'})$ and $s_{2'}\equiv \sinh(\Omega_{2'}L/\hbar v_{\rm F,2'})$.
For the distribution functions one obtains for the two leads 
\begin{eqnarray}
\delta x_{\beta }=x_{\bar \alpha } = \delta x(V,T;\eps ),&& \quad
\delta \tilde x_{\alpha }=\delta \tilde x_{\bar \beta } = \delta x(V,T;-\eps ), \\
\delta x_{\alpha'}=\delta x_{\bar \beta' } = x(V',T';\eps ) ,&&\quad
\delta \tilde x_{\beta'}=\delta \tilde x_{\bar \alpha' } = \delta x(V',T';-\eps ).
\end{eqnarray}
Here, $T_{\rm S}$ is the temperature in the superconductor, $T$ and $V$ are temperature and
voltage in the left lead, and $T'$ and $V'$ are temperature and voltage in the right lead.
The voltages are measured with respect to the superconductor.
The expressions appearing in equations \eqref{ji}-\eqref{jcar} have an intuitive interpretation, and selected processes are illustated in figure \ref{NLAR}. 
These terms involve propagation of particles or holes, represented as full lines and dashed lines in the figure. Certain processes involve conversions between particles and holes, accompanied by the creation or destruction of a Cooper pair (loops in the figure), and correspond to the factors $\gamma_{\bar 1} $, $\gamma_{1'}$, $\tilde\gamma_1 $, and $\tilde\gamma_{\bar 1'}$ in \eqref{jr}-\eqref{jcar}. 
Propagation of particles or holes between the left and right interface is represented in these equations by the factors $u_{22'}$ and $\tilde u_{\bar 2 \bar 2'} $. Vertex corrections $v_2$ and $v_{\bar 1}$ correspond to multiple Andreev reflections at either interface. 
The factors $\gamma_2$ and $\tilde \gamma_{\bar 2}$ combine propagation between the two interfaces with Andreev reflections at the other interface.

As an example, for an isotropic singlet superconductor and collinear arrangement of the magnetization directions, one obtains
\begin{align}
\label{ji0}
j_{\rm I} &= 4 \delta x,\quad
j_{\rm R}= \left[2|v_+|^2 \; |r_\uparrow-r_\downarrow e^{\i\vartheta } \gamma_0\gamma_+|^2
+ 2|v_-|^2 \; |r_\downarrow -r_\uparrow e^{-\i \vartheta} \gamma_0 \gamma_-|^2\right] \delta x\\
j_{\rm AR} &= (t_\uparrow t_\downarrow)^2 \left[|v_+|^2 \;
(|\gamma_+|^2+|\gamma_0|^2) 
+ |v_-|^2 \; (|\gamma_-|^2+ |\gamma_0|^2)\right] \delta \tilde x\\
j_{\rm EC} &= \left[ (t_\uparrow t'_\uparrow)^2
|v_+u_+|^2 \left\{ 1+|\gamma_0|^4 (r_\downarrow r'_\downarrow)^2 \right\} 
+ (t_\downarrow t'_\downarrow)^2
|v_-u_-|^2 \left\{ 1+|\gamma_0|^4 (r_\uparrow r'_\uparrow)^2 \right\} \right] \delta x'\\
j_{\rm CAR} &= |\gamma_0|^2 \left[ (t_\uparrow t'_\downarrow)^2
|v_+u_+|^2 \left\{ (r'_\uparrow)^2 +(r_\downarrow)^2 \right\} 
+ (t_\downarrow t'_\uparrow)^2
|v_-u_-|^2 \left\{ (r'_\downarrow)^2 +(r_\uparrow)^2 \right\} \right] \delta \tilde x' 
\label{jcar0}
\end{align}
where I defined
(with $s\equiv s_{2'}$ and $c\equiv c_{2'}$)
\begin{align}
\Gamma'_\pm &= r'_\uparrow r'_\downarrow e^{\pm \i\vartheta' } \gamma_0,\quad
\gamma_\pm= u_\pm \left[ \Gamma'_\pm c +\i s(\Delta + \Gamma'_\pm \varepsilon)/\Omega \right]\\
u_\pm&=\left[c-\i s(\varepsilon+\Gamma'_\pm \Delta)/\Omega \right]^{-1}
,\quad
v_\pm= [ 1-\gamma_\pm \gamma_0 r_\uparrow r_\downarrow e^{\pm \i\vartheta} ]^{-1}.
\end{align}

Early studies of nonlocal transport in F-S-F structures include Ref. \cite{Melin04}.
In Ref. \cite{Metalidis10} the non-local conductance was explained in terms of the processes discussed above for an F-S-F structure with strong spin-polarization. Andreev bound states appear on both ferromagnet-superconductor interfaces, which decay trough the superconductor towards the opposite contact. Parallel and antiparallel alignment of the magnetizations lead to qualitatively different Andreev spectra. The non-local processes have a natural explanation in terms of overlapping spin-polarized Andreev states.
The density of states for the trajectory connecting the two contacts is obtained from
\begin{align}
N_\uparrow(\eps , x)= \frac{N_{\rm F}}{2} \mbox{Re} \frac{1-\gamma_+(\eps,x)\gamma^\ast_-(-\eps^\ast,L-x)}{1+\gamma_+(\eps,x)\gamma^\ast_-(-\eps^\ast,L-x)},
\end{align}
and for $N_\uparrow $ the same expression holds with $+$ and $-$ interchanged.
\begin{figure}[t]
\centering{
\hspace{0.3in}{\tiny (a)} \hspace{2.5in} {\tiny (b)} }\\
\centering{
\includegraphics[width=2.5in]{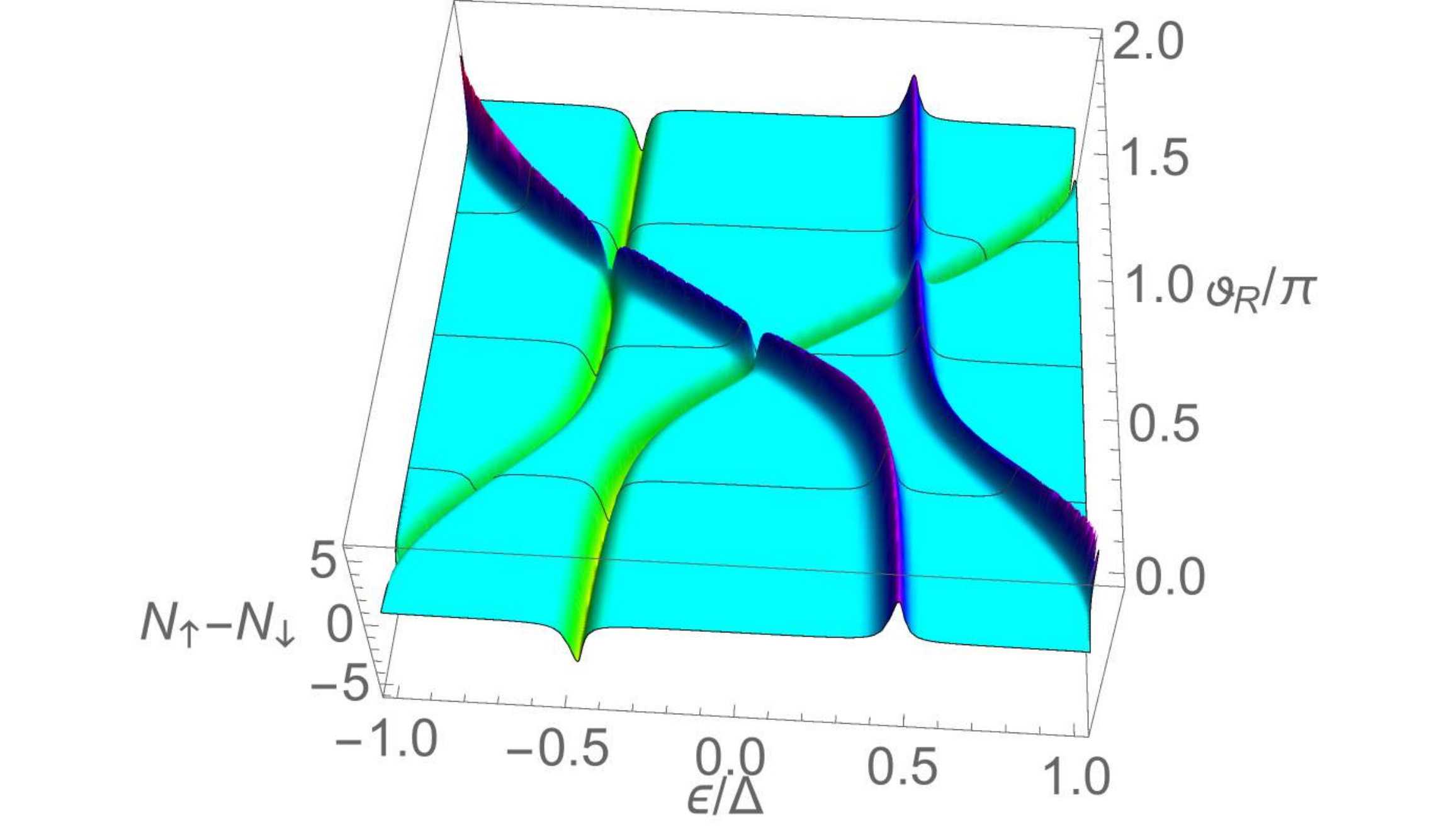}
\includegraphics[width=1.9in]{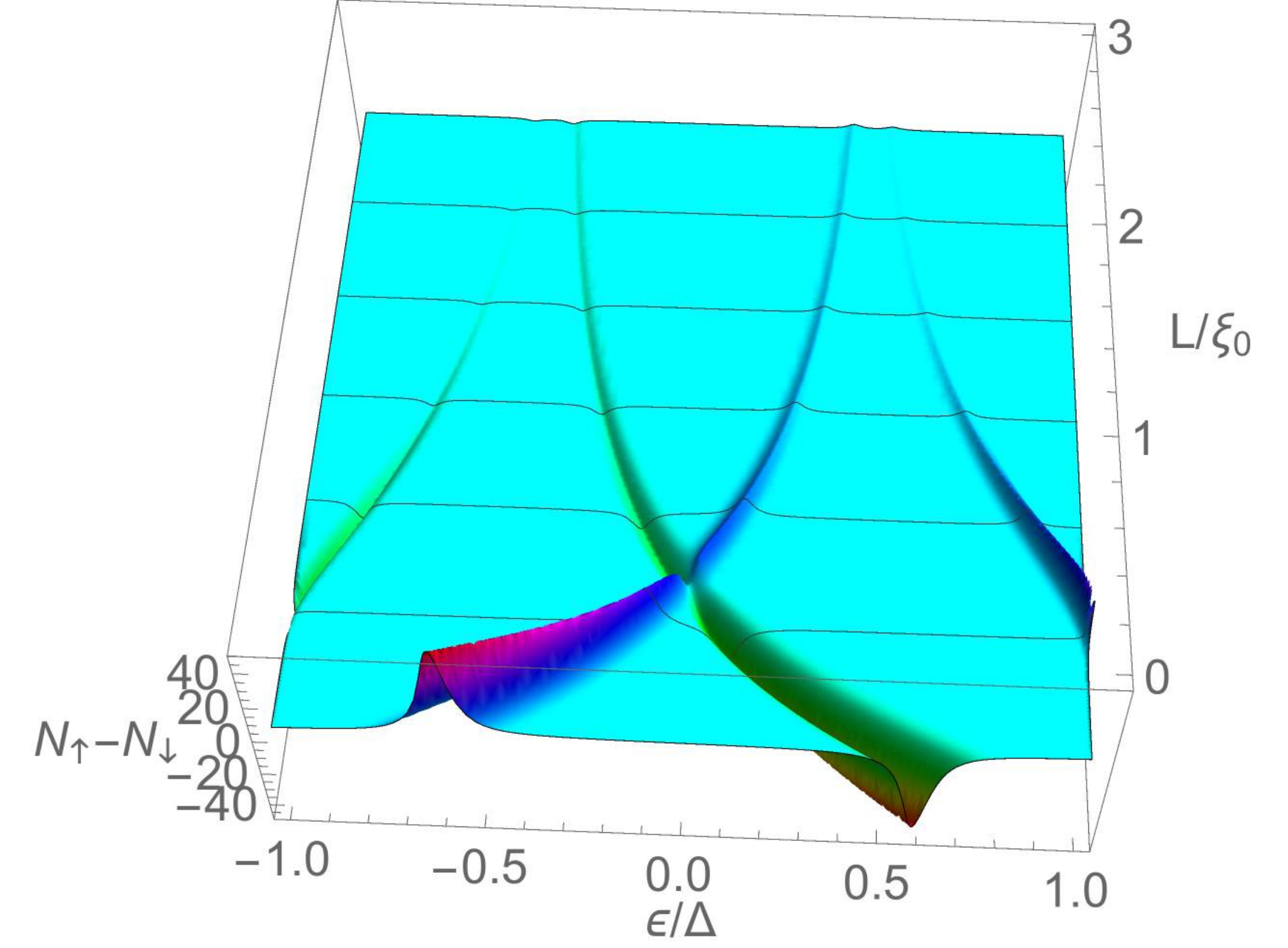}
\hspace{0.4in}$\; $
}\\
\centering{
\hspace{0.3in}{\tiny (c)} \hspace{1.2in} {\tiny (d)} 
\hspace{1.2in}{\tiny (e)} \hspace{1.2in} {\tiny (f)} \hspace{0.5in}$\;$}\\
\centering{
\includegraphics[width=1.25in]{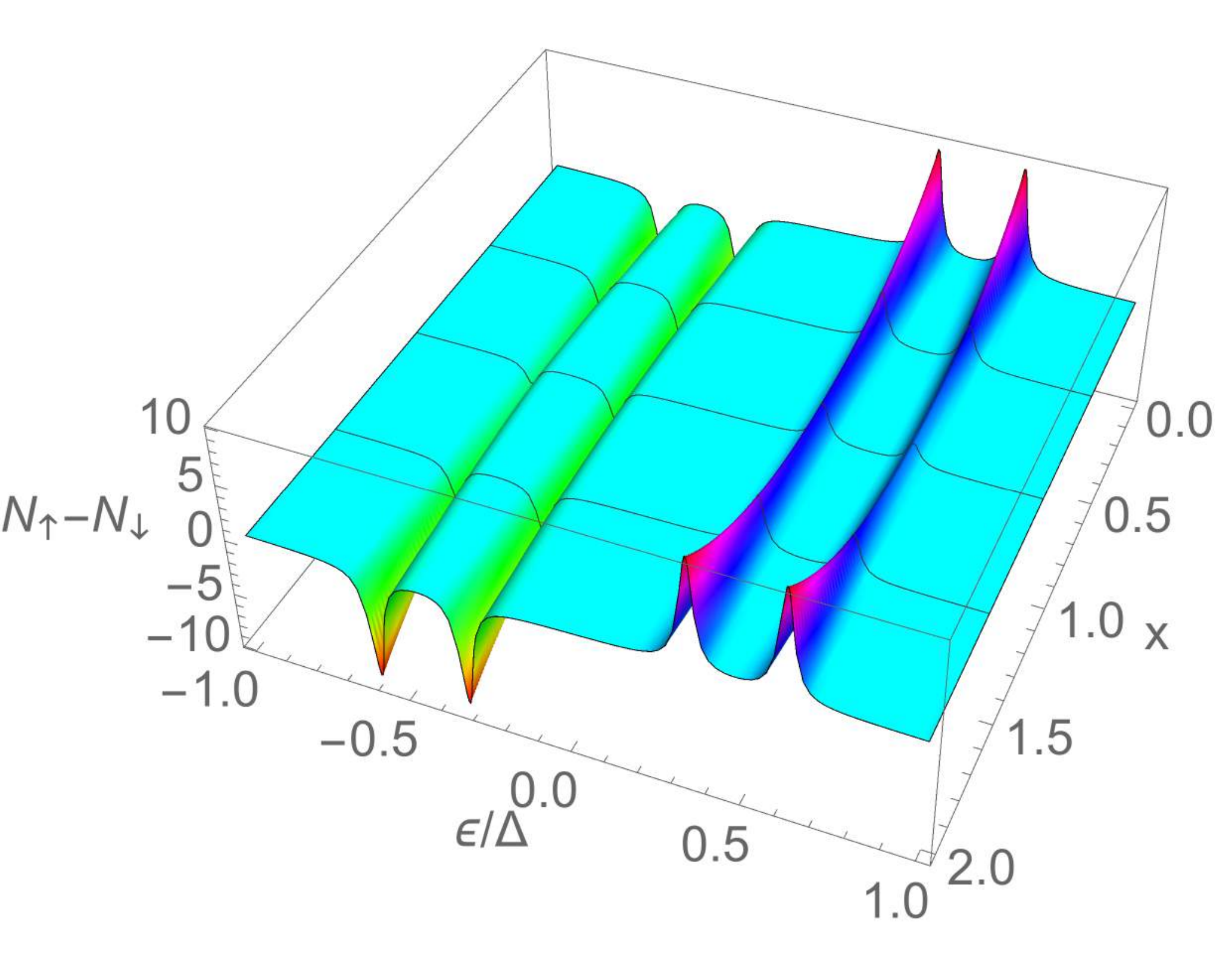}
\includegraphics[width=1.25in]{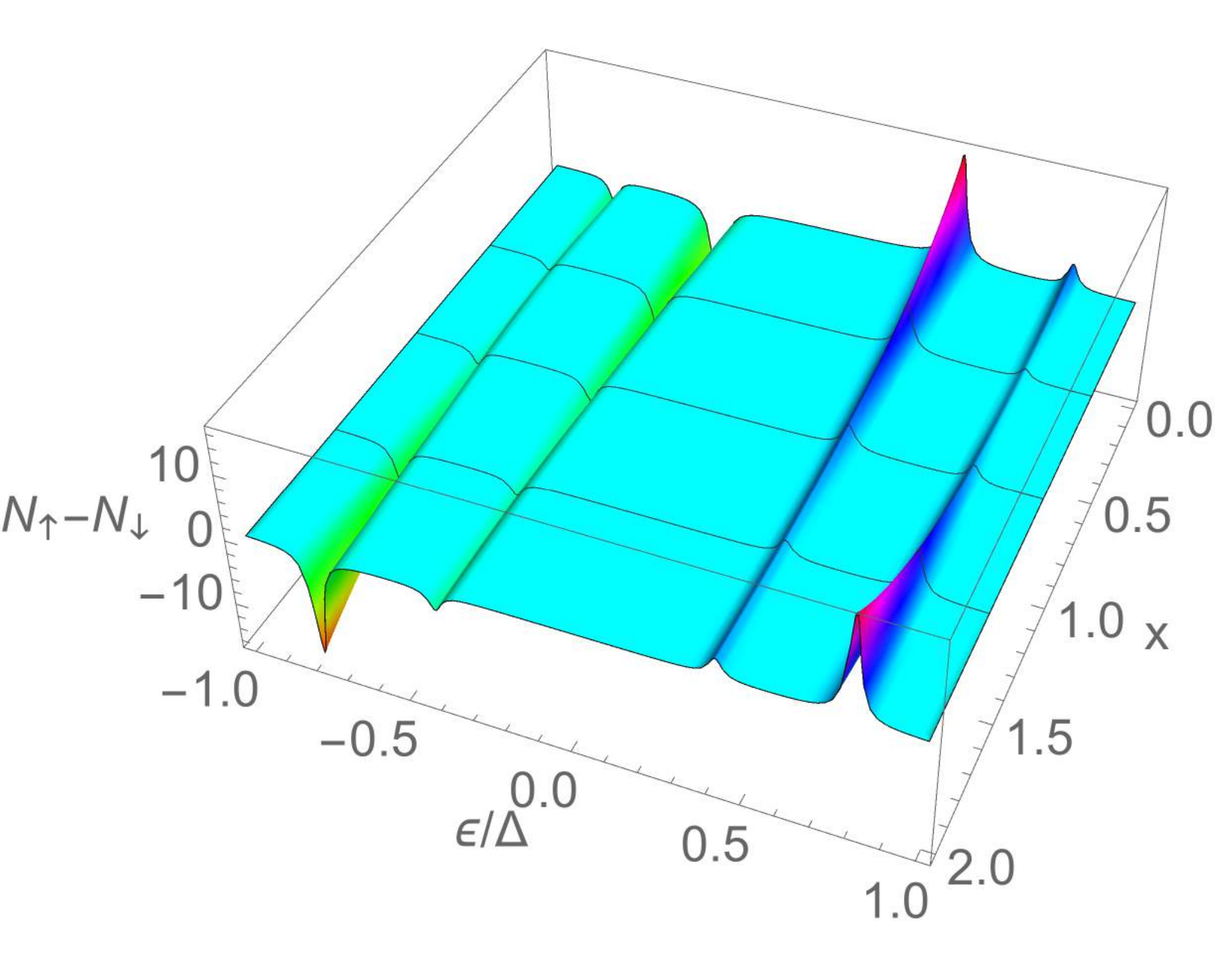}
\includegraphics[width=1.25in]{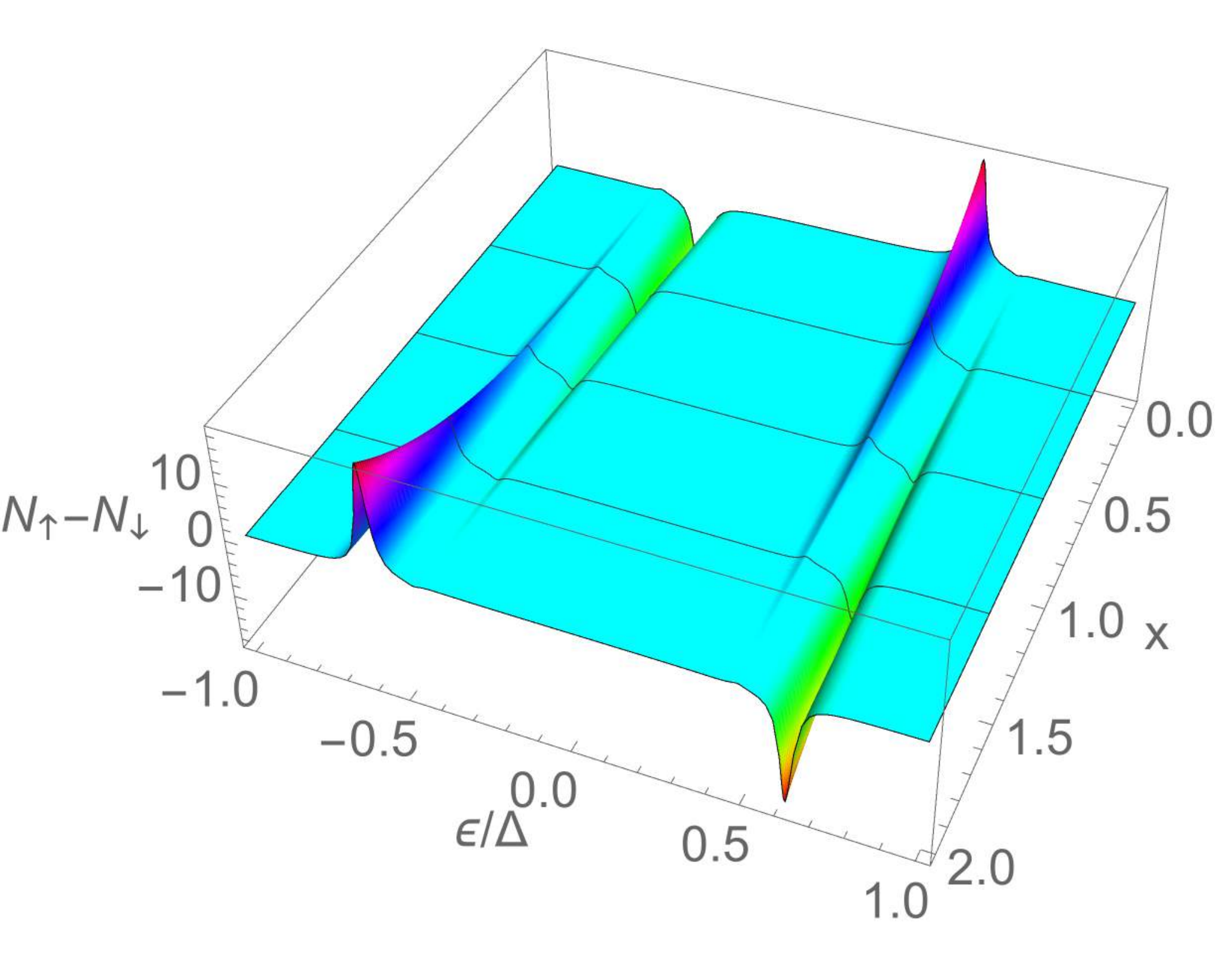}
\includegraphics[width=1.25in]{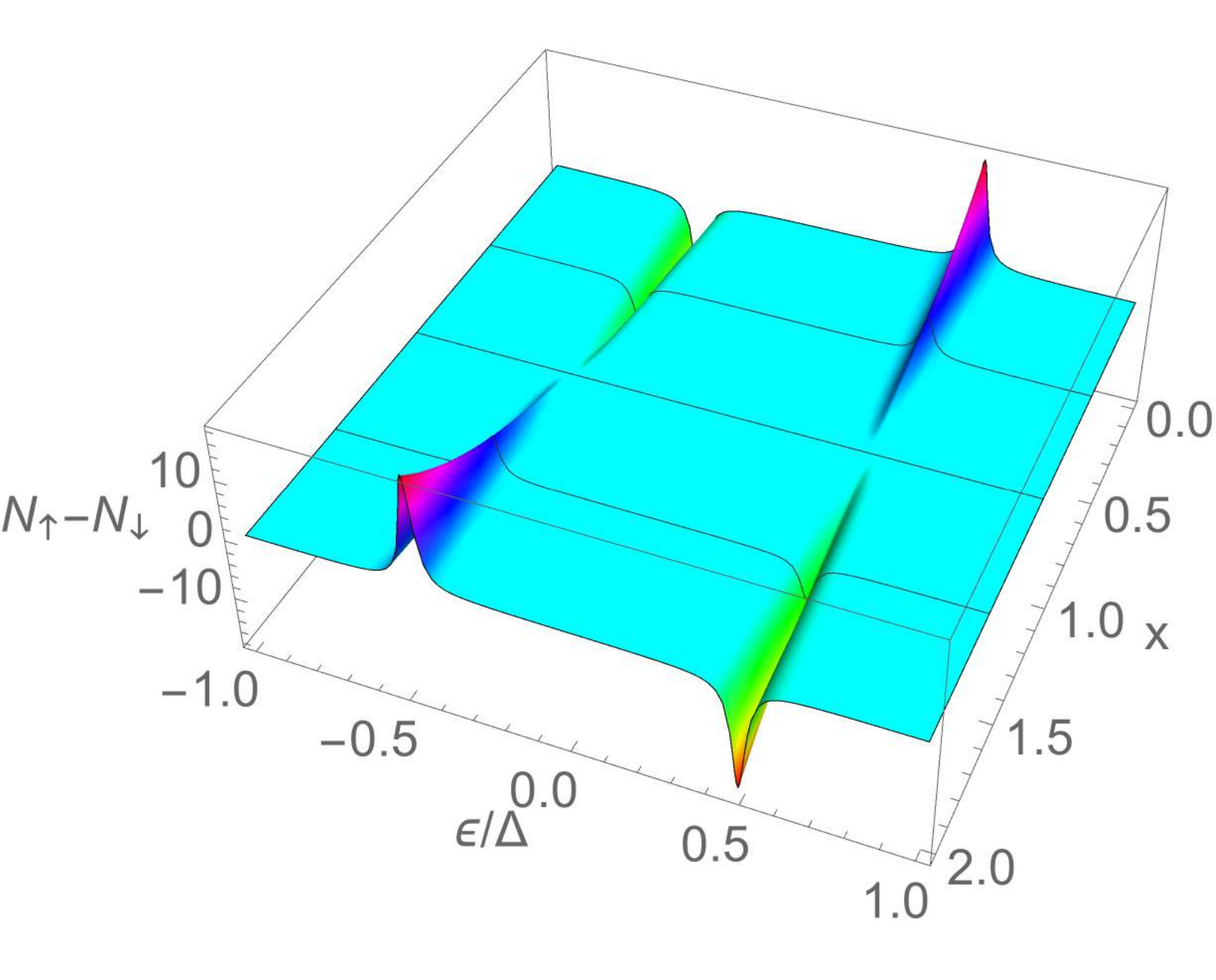}
}
\caption{
Andreev bound states in an F-S-F structure of the type shown in figure \ref{setup}.
(a): 
$P\equiv 
\frac{N_\uparrow - N_\downarrow}{N_\uparrow + N_\downarrow}$ as function of
$\varepsilon$ and $\vartheta_{\rm R}$ at a position in S midway between the contacts, for $L=2\xi_0$ (with the coherence length of the superconductor $\xi_0=\hbar v_{F,S}/|\Delta |$) and $\vartheta_{\rm L}=0.7\pi$.
An avoided crossing appears for equally spin-polarized Andreev states, which is absent for opposite polarization.
(b) dependence on $L/\xi_0$ for fixed $\vartheta_{\rm R}=\vartheta_{\rm L}=0.7\pi$.
(c)-(f): $P$ as function of $\eps $ and $x$ for $L=2\xi_0$ and $\vartheta_{\rm L}=0.7\pi$, and
(c) $\vartheta_{\rm R}=0.7\pi$
(d) $\vartheta_{\rm R}=0.5\pi$
(e) $\vartheta_{\rm R}=-0.6\pi$
(f) $\vartheta_{\rm R}=-0.7\pi$.
At the avoided crossing all bound states have equal weight at both interfaces. In all other cases the bound states for fixed spin projection are localized at one interface only.
In all panels
$(r_\uparrow r_\downarrow)_{\rm L}=(r_\uparrow r_\downarrow)_{\rm R} =0.95$.
}
\label{NL}
\end{figure}
In figure \ref{NL} I show an example of such a setup. As can be seen, avoided crossings of Andreev bound states with equal spin polarization play an important role in such systems. At the avoided crossing the bound states have equal weight at both interfaces.
This contrasts the case when bound states have opposite spin polarization, where no avoided crossings appear, and the case when the two S-F interfaces have markedly different spin-dependent phase shifts, in which case bound states do not overlap and stay localized at one of the two interfaces only. The spin-polarization and weight of the bound states at energies $\eps_{\rm b}$ and $-\eps_{\rm b}$ determine the magitude of the nonlocal currents due to crossed Andreev reflection and elastic co-tunneling. A detailed discussion of how the weights, transmission probabilities, and bound-state geometries influence CAR and EC processes is given in \cite{Metalidis10}.
In diffusive S-F systems a theory for nonlocal transport was developed in \cite{Kalenkov10}.

Equations \eqref{ji0}-\eqref{jcar0} 
have been applied to the study of thermoelectric effects
in non-local setups \cite{Machon13,Machon14}. 
The contributions to the energy current are obtained as
\begin{equation}
I_{\varepsilon }=I_{\varepsilon \rm I}-I_{\varepsilon \rm R} + I_{\varepsilon \rm AR} - I_{\varepsilon \rm EC} + I_{\varepsilon \rm CAR} ,
\end{equation}
where the respective contributions are given by analogous equations
as in equations \eqref{def2}-\eqref{jcar}, however with replacing  the charge e by energy $\varepsilon $.
Spin-dependent interface scattering phases in combination with spin filtering leads to giant thermoelectric effects in F-S-F devices \cite{Machon13,Ozaeta14}.

There have been a number of experimental studies of non-local transport in S-F hybrid structures, e.g. \cite{Beckmann04,Cadden07,Brauer10}. Colci {\it et al.} investigate S-FF-S junctions with two parallel running wires \cite{Colci12}. 
A very recent experiment shows a non-local inverse proximity effect in an (F-N-F')-S-N structure \cite{Flokstra15}. The inverse proximity effect transfers magnetization from the ferromagnet into the superconductor or across a superconductor. For strongly spin-polarized systems this occurs as a result of spin-mixing phases \cite{Eschrig08,Grein13}.
A combination of non-local effects in S-F structures with non-equilibrium Andreev interferometer geometries, in analogy to experiments in S-N structures \cite{Cadden09,Eschrig09NV}, seems to be another exciting avenue for future applications \cite{Noh13}.

\section{Generalized Andreev Equations}
In this section I discuss the physical interpretation of the coherence functions.
To this end, I present a generalized set of Andreev equations which is equivalent to equations \eqref{cricc1}-\eqref{cricc2}.
Let us define for each pair of Fermi momenta $\mvec{p}_F$, $-\mvec{p}_F$ and corresponding Fermi velocities $\mvec{v}_F(\mvec{p}_F)$, $\mvec{v}_F(-\mvec{p}_F)=-\mvec{v}_F(\mvec{p}_F)$
a pair of mutually conjugated trajectories
\begin{align}
\mvec{R}(\rho )=\mvec{R}_0+\hbar \mvec{v}_F(\mvec{p}_F) (\rho-\rho_0),\quad
\tilde{\mvec{R}}(\rho )=\mvec{R}_1-\hbar \mvec{v}_F(\mvec{p}_F) (\rho-\rho_1),\quad \rho_0\le \rho \le \rho_1.
\end{align}
Using $\partial_\rho \equiv \hbar \qpartial $,
let us define the following differential operators
\begin{align}\label{M1}
\hat{{\cal D}} \equiv \left(\begin{array}{cc}
-\i\partial_\rho+\Sigma & \Delta \\ -\tilde\Delta & \i\partial_\rho-\tilde \Sigma
\end{array}\right), \quad
\tilde{\hat{{\cal D}}} \equiv \left(\begin{array}{cc}
-\i\partial_\rho+\tilde \Sigma & \tilde \Delta \\ -\Delta & \i\partial_\rho-\Sigma
\end{array}\right)
\end{align}
which fulfill $\tilde{\hat{{\cal D}}} = -\ta \hat{{\cal D}} \ta $.
Let us also define the adjoint operator $\adj{\hat{{\cal D}}}(\rho,\partial_\rho ) = \hc{\hat{{\cal D}}}(\rho,-\partial_\rho )$.
For a fixed conjugated trajectory pair 
the set of generalized Andreev equations (retarded and advanced) is, 
\begin{align}\label{A1}
\hat{{\cal D}}
\qt 
\left(\begin{array}{cc} u^\ret & \tilde v^\ret\\ v^\ret &\tilde u^\ret \end{array}\right)
= \varepsilon 
\left(\begin{array}{cc} u^\ret & \tilde v^\ret\\ v^\ret &\tilde u^\ret \end{array}\right),
\begin{array}{c} 
\quad v^\ret(\rho_1) = -\tilde \gamma_1 \qt u^\ret (\rho_1)
\\
\quad \tilde v^\ret(\rho_0) = -\gamma_0 \qt \tilde u^\ret(\rho_0)
\end{array}
\\
\label{A2}
\adj{\hat{{\cal D}}}
\qt 
\left(\begin{array}{cc} u^\adv & \tilde v^\adv \\ v^\adv & \tilde u^\adv  \end{array}\right)
= \varepsilon 
\left(\begin{array}{cc} u^\adv & \tilde v^\adv \\ v^\adv & \tilde u^\adv  \end{array}\right),
\begin{array}{c} 
\quad v^\adv(\rho_0)= -\gamma^\dagger_0\qt u^\adv(\rho_0)
\\
\quad \tilde v^\adv(\rho_1) = -\tilde \gamma^\dagger_1 \qt \tilde u^\adv(\rho_1)
\end{array}
\end{align}
where the boundary conditions at $\rho=\rho_0$ and $\rho=\rho_1$
for the solutions fulfill the restrictions shown on the right hand side of the equations.
Then the relation between the Andreev amplitudes $u$, $v$, $\tilde u $, and $\tilde v$ and the coherence amplitudes $\gamma $ and $\tilde \gamma $ is given along the entire trajectories by \cite{Eschrig00}
\begin{align}\label{G1}
\tilde v^\ra = -\gamma^\ra \qt \tilde u^\ra, \quad
v^\ra = -\tilde \gamma^\ra \qt u^\ra 
\end{align}
with $\gar\equiv \ga $, $\gbr\equiv \gb $, $\gaa\equiv \gb^\dagger $, $\gba \equiv \ga^\dagger $.
It is easy to show that the following conservation law along the trajectory holds 
\begin{align}
\partial_\rho \left\{
\hc{\left(\begin{array}{cc} u^\adv & \tilde v^\adv \\ v^\adv & \tilde u^\adv  \end{array}\right)}
\tc \qt
\left(\begin{array}{cc} u^\ret & \tilde v^\ret\\ v^\ret &\tilde u^\ret \end{array}\right)
\right\}=0.
\end{align}
Thus, the matrix inside the curly brackets is given by its value at one point on the trajectory.
The off-diagonal elements are zero due to the conditions in Eq.~\eqref{A1} for $\rho_0$ and $\rho_1$,
leading to $\hc{(u^\adv)} \qt \tilde v^\ret=\hc{(v^\adv )} \qt \tilde u^\ret $ and 
$\hc{(\tilde u^\adv)} \qt v^\ret=\hc{(\tilde v^\adv )} \qt u^\ret $ along the entire trajectory.
The diagonal components $\partial_\rho [\hc{(u^\adv)} \qt u^\ret - \hc{(v^\adv )} \qt v^\ret ] =0$ and $\partial_\rho [\hc{(\tilde u^\adv)} \qt \tilde u^\ret - \hc{(\tilde v^\adv )} \qt \tilde v^\ret ] =0$ translate into
$\partial_\rho [ \hc{(u^\adv)} \qt (1-\gar \qt \gbr ) \qt u^\ret ]=0$ and
$\partial_\rho [ \hc{(\tilde u^\adv)} \qt (1-\gbr \qt \gar ) \qt \tilde u^\ret ]=0$. In particular, for Andreev bound states $1-\gar \qt \gbr =0$ or $1-\gbr \qt \gar=0$, and thus this property is conserved along the entire trajectory.

If one writes Eq.~\eqref{A1} formally as $\hat{\cal D}\qt \hat U=\varepsilon \hat U$, then the conjugated equation $\tilde{\hat{\cal D}} \qt \tilde{\hat U}=-\varepsilon \tilde{\hat U}$ holds with $\tilde {\hat U}=\hat \tau_1 \hat U \hat \tau_1$, which leads, however, to a system identical to Eq.~\eqref{A1}. The adjoint equation $\adj{\hat{\cal D}} \qt \hat{\underline{U}}=\varepsilon \hat{\underline{U}}$ defines adjoined Andreev amplitudes (left eigenvectors) $\underline u$, $\underline{v}$, $\underline{\tilde u}$, and $\underline{\tilde v}$. 
These are, however, equivalent to the advanced eigenvectors in Eq.~\eqref{A2}.

\section{Conclusion}
I have presented theoretical tools for studying Andreev reflection phenomena and Andreev bound states in superconductor-ferromagnet hybrid structures. Concentrating on ballistic heterostructures with strong spin-polarization, I have formulated theories for point contact spectroscopy and for nonlocal transport, as well as for Andreev states in Josephson structures in terms of coherence functions and distribution functions. The connection to coherence amplitudes appearing in the solutions of Andreev equations has been made explicit. The formulas for non-local transport have been given in a general form, allowing for non-collinear geometries, and using the normal-state scattering matrix as input.



\dataccess{All data can be obtained from the author.}

\ack{I acknowledge the hospitality and
the financial support from the Lars Onsager Award committee 
during my stay at the Norwegian University of Science and Technology, as well as stimulating discussions within the Hubbard Theory Consortium.}

\funding{This work is supported by the Engineering and Physical Science
Research Council (EPSRC Grant No. EP/J010618/1).}

\conflict{I have no competing interests.}




\begin{thebibliography}{9}
\bibitem{Cooper56}
Cooper LN. 1956. Bound Electron Pairs in a Degenerate Fermi Gas. 
\textit{Phys. Rev.} {\bf 104}, 1189-1190.

\bibitem{Bardeen57}
Bardeen J, Cooper LN, Schrieffer JR. 1957. Theory of superconductivity.
\textit{Phys. Rev.} {\bf 108}, 1175-1204.

\bibitem{Glover56}
Glover RE, III , Tinkham M. 1956. Transmission of Superconducting Films at Millimeter-Microwave and Far Infrared Frequencies.
\textit{Phys. Rev.} {\bf 104}, 844-845.

\bibitem{Tinkham56}
Tinkham M. 1956. Energy Gap Interpretation of Experiments on Infrared Transmission through Superconducting Films.
\textit{Phys. Rev.} {\bf 104}, 845-846.

\bibitem{Giaever60}
Giaever I. 1960.  Energy Gap in Superconductors Measured by Electron Tunneling.
\textit{Phys. Rev. Lett.} {\bf 5}, 147-148.

\bibitem{Eschrig06}
Eschrig M. 2006. The effect of collective spin-1 excitations on electronic
spectra in high-T$_c$ superconductors.
\textit{Advances in Physics} {\bf 55}, 47-183.

\bibitem{Migdal58}
Migdal AB. 1958. Vzaimodeistvie elektronov i kolebanii kristallicheskoi reshetki v normal'nom metalle.
\textit{Zh. Eksp. Teor.  Fiz.}  {\bf 34}, 1438-1446 (in Russian);
Migdal AB. 1958. Interaction  between electrons and lattice vibrations in a normal metal.
\textit{Sov. Phys. JETP} {\bf 7}, 996-1001 (Engl. transl.)

\bibitem{Eliashberg60}
Eliashberg G. 1960. Vzaimodeistvie elektronov s kolebaniyami reshetki v sverkhprovodnike.
\textit{Zh. Eksp. Teor.  Fiz.}  {\bf 38}, 966-974 (in Russian);
Eliashberg G. 1960. Interactions between electrons and lattice vibrations in a superconductor.
\textit{Sov.Phys.-JETP} {\bf 11}, 696-702 (Engl. transl.)

\bibitem{Scalapino66}
Scalapino DJ, Schrieffer JR, Wilkins JW. 1966.
Strong-Coupling Superconductivity. I.
\textit{Phys. Rev.} {\bf 148}, 263-279.

\bibitem{Giaever62}
Giaever I, Hart HR, Jr., Megerle K. 1962.Tunneling into Superconductors at Temperatures below 1$^{\rm o}$K.
\textit{Phys. Rev. } {\bf 126}, 941-948.

\bibitem{McMillan65}
McMillan WL, Rowell JM.  1965. Lead Phonon Spectrum Calculated from Superconducting Density of States.
\textit{Phys. Rev. Lett.} {\bf 14}, 108-112.

\bibitem{Tomasch65}
Tomasch WJ. 1965. Geometrical resonance in the tunneling characteristics of superconducting Pb.
\textit{Phys. Rev. Lett.} {\bf 15}, 672-675.

\bibitem{McMillan66}
McMillan WL, Anderson PW. 1966. Theory of Geometrical Resonances in the Tunneling Characteristics of Thick Films of Superconductors.
\textit{Phys. Rev. Lett.} {\bf 16}, 85-87.

\bibitem{Wolfram68}
Wolfram T. 1968.  Tomasch Oscillations in the Density of States of Superconducting Films.
\textit{Phys. Rev.} {\bf 170} 481-490.

\bibitem{Rowell66}
Rowell JM, McMillan WL. 1966. Electron Interference in a Normal Metal Induced by Superconducting Contracts.
\textit{Phys. Rev. Lett.} {\bf 16}, 453-456.

\bibitem{Nedellec71}
N\'edellec P, Guyon E. 1971. Effect of the detailed variation of pair potential on the `Tomasch' effect.
\textit{Solid State Comm. } {\bf 9}, 113-116.

\bibitem{deGennes63}
de Gennes PG, Saint-James D. 1963.
Elementary excitations in the vicinity of a normal metal-superconducting metal contact.
\textit{Phys. Lett.} {\bf 4}, 151-152.

\bibitem{James64}
Saint-James D. 1964.
Excitations \'el\'ementaires au voisinage de la surface de s\'eparation d'un m\'etal normal et d'un m\'etal supraconducteur (Elementary excitations in the vicinity of the surface separating a normal metal and a superconducting metal).
\textit{J. Phys. (Paris)} {\bf 25}, 899-905 (in French).
See \href{https://hal.archives-ouvertes.fr/jpa-00205891}{https://hal.archives-ouvertes.fr/jpa-00205891}.

\bibitem{Rowell73}
Rowell JM. 1973.
Tunneling Observation of Bound States in a Normal Metal-Superconductor Sandwich
\textit{Phys. Rev. Lett.} {\bf 30}, 167-170.

\bibitem{Bellanger73}
Bellanger D, Klein J, L\'eger A, Belin M, Defourneau D. 1973. 
Tunneling measurements of electron interference effects in Cu-Pb sandwiches.
\textit{Phys. Lett. A} {\bf 42}, 459-460.

\bibitem{Andreev64}
Andreev AF. 1964. 
Teploprovodnost' promezhutochnogo sostoyaniya sverkhprovodnikov.
\textit{Zh. Eksp. Teor. Fiz.} {\bf 46}, 1823-1828 (in Russian);
Andreev AF. 1964. 
Thermal conductivity of the intermediate state of superconductors.
\textit{Sov.  Phys. JETP} \textbf{19}, 1228-1231 (Engl. transl.)

\bibitem{Klapwijk82}
Klapwijk TM, Blonder GE, Tinkham M. 1982. 
Explanation of subharmonic energy gap structure in superconducting contacts.
\textit{Physica B \& C} {\bf 109}, 1657-1664.

\bibitem{Cuevas06}
Cuevas JC, Hammer J, Kopu J, Viljas JK, Eschrig M. 2006.
Proximity effect and multiple Andreev reflections in diffusive
superconductor-normal-metal-superconductor junctions.
\textit{Phys. Rev. B} {\bf 73}, 184505 (1-6).

\bibitem{Blonder82}
Blonder GE, Tinkham M, Klapwijk TM. 1982.
Transition from metallic to tunneling regimes in superconducting microconstrictions: Excess current, charge imbalance, and supercurrent conversion.
\textit{Phys. Rev. B} {\bf 25}, 4515-4532.

\bibitem{Beenakker92}
Beenakker CWJ. 1992.
Quantum transport in semiconductor-superconductor microjunctions.
\textit{Phys. Rev. B} {\bf 46}, 12841-12844(R).

\bibitem{deGennes63a}
de Gennes  PG, Guyon E. 1963.
Superconductivity in ``normal'' metals.
\textit{Phys. Lett.} {\bf 3}, 168-169.

\bibitem{Werthamer63}
Werthamer NR. 1963.
Theory of the superconducting transition temperature and energy gap function of superposed metal films.
\textit{Phys. Rev.} {\bf 132}, 2440-2445.

\bibitem{Abrikosov58}
Abrikosov AA, Gor'kov LP. 1958.
K teorii sverkhprovodyashchikh splavov. I. Elektrodinamika splavov pri absolyutnom nule.
\textit{Zh. Eksp. Teor. Fiz.} {\bf 35}, 1558-1571 (in Russian);
Abrikosov AA, Gor'kov LP. 1959;
On the theory of superconducting alloys, I. The electrodynamics of alloys at absolute zero;
\textit{Sov. Phys. JETP} {\bf 8}, 1090-1098 (Engl. transl.)

\bibitem{Abrikosov59}
Abrikosov AA, Gor'kov LP. 1959.
Sverkhprovodyashchiye splavy pri temperaturakh vyshe absolyutnogo nulya.
\textit{Zh. Eksp. Teor. Fiz.} {\bf 36}, 319-320 (in Russian);
Abrikosov AA, Gor'kov LP. 1959.
Superconducting alloys at finite temperatures.
\textit{Sov. Phys. JETP} {\bf 9}, 220-221 (Engl. transl.)

\bibitem{Anderson59a}
Anderson PW. 1959.
Theory of dirty superconductors.
\textit{J. Phys. Chem. Solids} {\bf 11}, 26-30.

\bibitem{Finkelstein87}
Finkel'shtein AM. 1987.
O temperature sverkhprovodyashchego perekhoda v amorfnykh plenkakh. 
\textit{Pis'ma Zh. Eksp. Teor. Fiz.} {\bf 45}, 37-40 (in Russian);
Finkel'shtein AM. 1987.
Superconducting transition temperature in amorphous films. 
\textit{JETP Lett.} {\bf 45}, 46-49 (Engl. transl.)

\bibitem{Lifshitz64}
Lifshitz IM. 1964.
O strukture energeticheskogo spektra i kvantovykh sostoyaniyakh neuporyadochennykh kondensirovannykh sistem.
\textit{Uspekhi Fiz. Nauk} {\bf 83}, 617-663 (in Russian);
Lifshitz IM. 1965.
Energy spectrum structure and quantum states of disordered condensed systems.
\textit{Soviet Physics Uspekhi} {\bf 7}, 549-573 (in Russian);
Lifshitz IM. 1964.
The energy spectrum of disordered systems.
\textit{Adv. Phys.} {\bf 13}, 483-536 (Engl. transl.)

\bibitem{Larkin71}
Larkin AI, Ovchinnikov YuN. 1971.
Plotnost' sostoyanii v neodnorodnykh sverkhprovodnikakh.
\textit{Zh. Eksp. Teor. Fiz.} {\bf 61}, 2147-2159 (in Russian).
Larkin AI, Ovchinnikov YuN. 1972.
Density of states in inhomogeneous superconductors.
\textit{Sov. Phys. JETP} {\bf 34}, 1144-1150 (Engl. transl.)

\bibitem{Balatsky97}
Balatsky AV, Trugman SA. 1997.
Lifshitz Tail in the Density of States of a Superconductor with Magnetic Impurities.
\textit{Phys. Rev. Lett.} {\bf 79}, 3767-3770.

\bibitem{Kastalsky91}
Kastalsky A, Kleinsasser AW, Greene LH, Bhat R, Milliken FP, Harbison JP. 1991.
Observation of pair currents in superconductor-semiconductor contacts.
\textit{Phys. Rev. Lett.} {\bf 67}, 3026-3029.

\bibitem{Wees92}
van Wees BJ, de Vries P, Magn\'ee P, Klapwijk TM. 1992.
Excess Conductance of Superconductor-Semiconductor Interfaces Due to Phase Conjugation between Electrons and Holes.
\textit{Phys. Rev. Lett.} {\bf 69}, 510-513.

\bibitem{Zaitsev90}
Za\u{i}tsev FV. 1990. 
Svoystva ``gryaznykh'' S-S$^\ast$-N- i S-S$^\ast$-S-struktur s potentsial'nymi bar'yerami na granitsakh metallov.
\textit{Pis'ma Zh. Eksp. Teor. Fiz.} {\bf 51}, 35-39 (in Russian);
Za\u{i}tsev FV. 1990. 
Porperties of ``dirty'' S-S$^\ast$-N and S-S$^\ast$-S structures with potential barriers at the metal boundaries.
\textit{JETP Lett.} {\bf 51}, 41-46 (Engl. transl.)

\bibitem{Volkov92}
Volkov AF. 1992. 
Ob anomalii provodimosti v kontaktakh sverkhprovodnik-poluprovodnik.
\textit{Pis'ma Zh. Eksp. Teor. Fiz.} {\bf 55}, 713-716 (in Russian);
Volkov AF. 1992. 
Conductance anomaly at superconductor-semiconductor contacts.
\textit{JETP Lett.} {\bf 55}, 746-749 (Engl. transl.)

\bibitem{Abrikosov60}
Abrikosov AA, Gor'kov LP. 1960.
K teorii sverkhprovodyashchikh splavov s paramagnitnymi primesyami.
\textit{Zh. Exp. Teor. Fiz.} {\bf 39}, 1781-1796 (in Russian);
Abrikosov AA, Gor'kov LP. 1961.
Contribution to the theory of superconducting alloys with paramagnetic impurities.
\textit{Sov. Phys. JETP} {\bf 12}, 1243-1253 (Engl. transl.)

\bibitem{Luh65}
Yu Luh. 1965.  Bound state in superconductors with paramagnetic impurities.
\textit{Acta Physica Sinica} {\bf 21}, 75-91.
See \href{http://wulixb.iphy.ac.cn/EN/Y1965/V21/I1/075}{http://wulixb.iphy.ac.cn/EN/Y1965/V21/I1/075}
(in Chinese, Engl. abstract)

\bibitem{Shiba68}
Shiba H. 1968.
Classical Spins in Superconductors.
\textit{Prog. Theor. Phys.}  {\bf 40}, 435-451.

\bibitem{Rusinov69}
Rusinov AI. 1969.
Sverkhprovodimost' vblizi paramagnitnoy primesi.
\textit{Pis'ma Zh. Eksp. Teor. Fiz.} {\bf 9}, 146-149 (in Russian);
Rusinov AI. 1969.
Superconductivity near a paramagnetic impurity.
\textit{JETP Lett.}  {\bf 9}, 85-87 (Engl. transl.)

\bibitem{Caroli64}
Caroli C, de Gennes PG, Matricon J. 1964.
Bound Fermion states on a vortex line in a type II superconductor.
\textit{Phys. Lett.} {\bf 9} 307-309.

\bibitem{Rainer96}
Rainer D, Sauls JA, Waxman D. 1996.
Current carried by bound states of a superconducting vortex.
\textit{Phys. Rev. B} {\bf 54}, 10094-10106.

\bibitem{Eschrig99}
Eschrig M, Sauls JA, Rainer D. 1999.
Electromagnetic Response of a Vortex in Layered Superconductors.
\textit{Phys. Rev. B} {\bf 60}, 10447-10454.

\bibitem{Eschrig02}
Eschrig M, Rainer D, Sauls JA. 2002.
Vortex Core Structure and Dynamics in Layered Superconductors.
in {\it Vortices in Unconventional Superconductors and Superfluids}, ed. R.P. Huebener, N. Schopohl,
and G.E. Volovik, Springer Verlag, ISBN 3-540-42336-2, pp. 175-202.

\bibitem{Sauls09}
Sauls JA, Eschrig M. 2009.
Vortices in chiral, spin-triplet superconductors and superfluids.
\textit{New Journal of Physics} {\bf 11}, 075008 (1-26);
Eschrig M, Sauls JA. 2009.
Charge Dynamics of Vortex Cores in Layered Chiral Triplet Superconductors.
\textit{New Journal of Physics} {\bf 11}, 075009 (1-21);

\bibitem{Thouless72}
Edwards JT, Thouless DJ. 1972.
Numerical studies of localization in disordered systems.
\textit{J. Phys. C: Solid State Phys.} {\bf 5}, 807-820.

\bibitem{McMillan68}
McMillan WL. 1968.
Tunneling Model of the Superconducting Proximity Effect.
\textit{Phys. Rev.} {\bf 175}, 537-542.

\bibitem{Gueron96}
Gu\'eron S, Pothier H, Birge NO, Esteve D, Devoret MH. 1996.
Superconducting Proximity Effect Probed on a Mesoscopic Length Scale.
\textit{Phys. Rev. Lett.}  {\bf 77}, 3025-3028.

\bibitem{Kosztin95}
Kosztin I, Maslov DL, Goldbart PM. 1995.
Chaos in Andreev Billiards.
\textit{Phys. Rev.  Lett.} {\bf 75}, 1735-1738.

\bibitem{Lodder98}
Lodder A, Nazarov YuV. 1998. 
Density of states and the energy gap in Andreev billiards.
\textit{Phys. Rev. B} {\bf 58}, 5783-5788.

\bibitem{Kulik69}
Kulik IO. 1969.
Prostranstvennoye kvantovaniye i effekt blizosti v S-N-S-kontaktakh.
\textit{Zh. Eksp. Teor. Fiz.} {\bf 57}, 1745-1759 (in Russian);
Kulik IO. 1970.
Macroscopic quantization and the proximity effect in S-N-S junctions.
\textit{Soviet Physics JETP} {\bf 30}, 944-950 (Engl. transl.)

\bibitem{Zhou98}
Zhou F, Charlat P, Spivak B, Pannetier B. 1998.
Density of States in Superconductor-Normal Metal-Superconductor Junctions.
\textit{Journ. Low Temp. Phys.} {\bf 110}, 841-850.

\bibitem{leSeur08}
le Sueur H, Joyez P, Pothier H, Urbina C, Esteve D. 2008.
Phase Controlled Superconducting Proximity Effect Probed by Tunneling Spectroscopy.
\textit{Phys. Rev. Lett.} {\bf 100}, 197002 (1-4).

\bibitem{Byers95}
Byers C, Flatte T. 1995.
Probing Spatial Correlations with Nanoscale Two-Contact Tunneling.
\textit{Phys. Rev. Lett. } {\bf 74}, 306-309.

\bibitem{Deutscher00}
Deutscher G, Feinberg D. 2002.
Coupling superconducting-ferromagnetic point contacts by Andreev reflections.
\textit{Appl. Phys. Lett.} {\bf 76}, 487-489.

\bibitem{Beckmann04}
Beckmann D, Weber  HB, von L\"ohneysen H. 2004.
Evidence for Crossed Andreev Reflection in Superconductor-Ferromagnet Hybrid Structures.
\textit{Phys. Rev. Lett. } {\bf 93}, 197003 (1-4).

\bibitem{Hu94}
Hu C-R. 1994.  
Midgap surface states as a novel signature for $d_{x_a^2-x^2_b}$-wave superconductivity.
\textit{Phys. Rev. Lett.}  {\bf 72} 1526-1529.

\bibitem{Tanaka95}
Tanaka Y and Kashiwaya S. 1995.
Theory of Tunneling Spectroscopy of $d$-Wave Superconductors.
\textit{Phys. Rev. Lett.} {\bf 74}, 3451-3454.

\bibitem{Volovik99}
Volovik GE. 1999.
Fermion zero modes on vortices in chiral superconductors.
\textit{Pis'ma Zh. Eksp. Teor. Fiz.} {\bf 70} 601-606; 
\textit{JETP Lett.} {\bf 70} 609-614 (Reprinted).

\bibitem{Kitaev01}
Kitaev AYu. 2001.
Unpaired Majorana fermions in quantum wires.
\textit{Uspekhi Fiz. Nauk} {\bf 171} (\textit{Phys. Usp.} {\bf 44}) (suppl.) 131-136.

\bibitem{Volovik03}
Volovik GE. 2003.
\textit{The Universe in a Helium Droplet}.
Clarendon, UK: Clarendon Press.

\bibitem{Tokuyasu88}
Tokuyasu T, Sauls JA, Rainer D. 1988. 
Proximity effect of a ferromagnetic insulator in contact with a superconductor.
\textit{Phys. Rev. B} {\bf 38}, 8823-8833.

\bibitem{Cottet05}
Cottet A, Belzig W. 2005.
Superconducting proximity effect in a diffusive ferromagnet with spin-active interfaces.
\textit{Phys. Rev. B} {\bf 72}, 180503(R) (1-4).

\bibitem{Brataas00}
Brataas A, Nazarov YuV, Bauer GEW. 2000.
Finite-Element Theory of Transport in Ferromagnet-Normal Metal Systems.
\textit{Phys. Rev. Lett.} {\bf 11}, 2481-2484.

\bibitem{Fogelstrom00}
Fogelstr\"om M. 2000. 
Josephson currents through spin-active interfaces.
\textit{Phys. Rev. B} {\bf 62}, 11812-11819.

\bibitem{Eschrig03}
Eschrig M, Kopu J, Cuevas JC, Sch\"on G. 2003.
Theory of Half-Metal/Superconductor Heterostructures.
\textit{Phys. Rev. Lett.} {\bf 90}, 137003 (1-4).

\bibitem{Krawiec04}
Krawiec M, Gy\"orffy BL, Annett JF. 2004.
Current-carrying Andreev bound states in a superconductor-ferromagnet proximity system.
\textit{Phys. Rev. B} {\bf 70}, 134519 (1-5).

\bibitem{Zhao04}
Zhao E, L\"ofwander T, Sauls JA. 2004.
Nonequilibrium superconductivity near spin-active interfaces.
\textit{Phys. Rev. B} {\bf 70}, 134510 (1-12).

\bibitem{Annett06}
Annett JF, Krawiec M, Gy\"orffy BL. 2006.
Origin of spontaneous currents in a superconductor-ferromagnetic proximity system.
\textit{Physica C: Superconductivity} {\bf 437-438}, 7-10.

\bibitem{Lofwander10}
L\"ofwander T, Grein R, Eschrig M. 2010.
Is CrO$_2$ Fully Spin Polarized? Analysis of Andreev Spectra and Excess Current.
\textit{Phys. Rev. Lett.} {\bf 105}, 207001 (1-4).

\bibitem{Cottet08a}
Cottet A, Belzig W. 2008.
Conductance and current noise of a superconductor/ferromagnet quantum point contact.
\textit{Phys. Rev. B} {\bf 77}, 064517 (1-8).

\bibitem{Cottet08}
Cottet A, Dou\c{c}ot B, Belzig W. 2008.
Finite Frequency Noise of a Superconductor-Ferromagnet Quantum Point Contact.
\textit{Phys. Rev. Lett.} {\bf 101}, 257001 (1-4).

\bibitem{Kalenkov07}
Kalenkov M, Zaikin A D. 2007.
Nonlocal Andreev reflection at high transmissions.
\textit{Phys. Rev. B} {\bf 75}, 172503 (1-4).

\bibitem{Melin04}
M\'elin R, Feinberg D. 2004.
Sign of the crossed conductances at a ferromagnet/ superconductor/ferromagnet 
double interface. 
\textit{Phys. Rev. B} {\bf 70}, 174509 (1-9).

\bibitem{Metalidis10}
Metalidis G, Eschrig M, Grein R, Sch\"on G. 2010.
Nonlocal conductance via overlapping Andreev bound states in ferromagnet-superconductor heterostructures.
\textit{Phys. Rev. B} {\bf 82}, 180503(R) (1-4).

\bibitem{Huebler12}
H\"ubler F, Wolf MJ, Scherer T, Wang D, Beckmann D, von L\"ohneysen H. 2012.
Observation of Andreev Bound States at Spin-active Interfaces.
\textit{Phys. Rev. Lett.} {\bf 109}, 087004 (1-5).

\bibitem{Linder09}
Linder J, Yokoyama T, Sudb{\o} A, Eschrig M. 2009. 
Pairing Symmetry Conversion by Spin-Active Interfaces in Magnetic Normal-Metal-Superconductor Junctions.
\textit{Phys. Rev. Lett.} {\bf 102}, 107008 (1-4)

\bibitem{Linder10}
Linder J, Yokoyama T, Sudb{\o} A, Eschrig M. 2010. 
Signature of odd-frequency pairing correlations induced by a magnetic interface.
\textit{Phys. Rev. B} {\bf 81}, 214504 (1-13).

\bibitem{Cottet11}
Cottet A. 2011.
Inducing Odd-Frequency Triplet Superconducting Correlations in a Normal Metal.
\textit{Phys. Rev. Lett.} {\bf 107}, 177001 (1-5).

\bibitem{Eschrig09}
Eschrig M. 2009.
Scattering problem in nonequilibrium quasiclassical theory of metals and superconductors:
General boundary conditions and applications.
\textit{Phys. Rev. B} {\bf 80}, 134511 (1-22).

\bibitem{Yokoyama07}
Yokoyama T, Tanaka Y, Golubov AA. 2007.
Manifestation of the odd-frequency spin-triplet pairing state in diffusive ferromagnet/superconductor junctions.
\textit{Phys. Rev. B} {\bf 75}, 134510 (1-8).

\bibitem{Asano07}
Asano Y, Tanaka Y, Golubov AA. 2007.
Josephson Effect due to Odd-Frequency Pairs in Diffusive Half Metals.
\textit{Phys. Rev. Lett.} {\bf 98}, 107002 (1-4).

\bibitem{Braude07}
Braude V, Nazarov YuV. 2007.
Fully Developed Triplet Proximity Effect.
\textit{Phys. Rev. Lett.} {\bf 98}, 077003 (1-4).

\bibitem{Hernando02}
Huertas-Hernando D,  Nazarov YuV, Belzig W. 2002.
Absolute Spin-Valve Effect with Superconducting Proximity Structures.
\textit{Phys. Rev. Lett.} {\bf 88}, 047003 (1-4).

\bibitem{Eschrig07}
Eschrig M, L\"ofwander T, Champel T, Cuevas JC, Kopu J, Sch\"on G. 2007.
Symmetries of Pairing Correlations in Superconductor-Ferromagnet Nanostructures.
\textit{J. Low Temp. Phys.} {\bf 147} 457-476.

\bibitem{Kopu04}
Kopu J, Eschrig M, Cuevas JC, Fogelstr\"om M. 2004.
Transfer-matrix description of heterostructures involving superconductors and ferromagnets.
\textit{Phys. Rev. B} {\bf 69}, 094501 (1-9).

\bibitem{Eschrig08}
Eschrig M, L\"ofwander T. 2008.
Triplet supercurrents in clean and disordered half-metallic ferromagnets.
\textit{Nature Physics } {\bf 4}, 138-143;
Eschrig M, L\"ofwander T. 2006.
arXiv:cond-mat/0612533.

\bibitem{Bergeret01}
Bergeret FS, Volkov AF, Evetov KB. 2001.
Long-Range Proximity Effects in Superconductor-Ferromagnet Structures.
\textit{Phys. Rev. Lett.} {\bf 86}, 4096-4099.

\bibitem{Kadigrobov01}
Kadigrobov A, Shekhter RI, Jonson M. 2001.
Quantum spin fluctuations as a source of long-range proximity effects in diffusive ferromagnet-super conductor structures.
\textit{Europhys. Lett.} {\bf 54}, 394-400.

\bibitem{Volkov03}
Volkov AF, Bergeret FS, Efetov KB. 2003.
Odd Triplet Superconductivity in Superconductor-Ferromagnet Multilayered Structures.
\textit{Phys. Rev. Lett.} {\bf 90}, 117006 (1-4).

\bibitem{Houzet07}
Houzet M, Buzdin AI. 2007.
Long range triplet Josephson effect through a ferromagnetic trilayer.
\textit{Phys. Rev. B} {\bf 76}, 060504(R) (1-4).

\bibitem{Izyumov02}
Izyumov YuA, Proshin YuN, Khusainov MG. 2002.
Konkurentsiya sverkhprovodimosti i magnetizma v geterostrukturakh ferromagnetik/sverkhprovodnik
(in Russian).
\textit{Usp. Fiz. Nauk} {\bf 172}, 113-154 (in Russian).
Izyumov YuA, Proshin YuN, Khusainov MG. 2002.
Competition between superconductivity and magnetism in ferromagnet/superconductor heterostructures.
\textit{Phys. Usp.} {\bf 45} 109-148 (Engl. transl.)

\bibitem{Fominov03}
Fominov YaV, Kupriyanov MYu, Feigel'man MV. 2003.
Kommentariy k obzoru Yu.A. Izyumova, Yu.N. Proshina, M.G. Khusainova ``Konkurentsiya sverkhprovodimosti i magnetizma v geterostrukturakh ferromagnetik/sverkhprovodnik''.
\textit{Usp. Fiz. Nauk} {\bf 173}, 113-115 (in Russian).
Fominov YaV, Kupriyanov MYu, Feigel'man MV. 2003.
A comment on the paper ``Competition between superconductivity and magnetism in ferromagnet/superconductor heterostructures'' by Yu A Izyumov, Yu N Proshin, and M G Khusainov.
\textit{Phys. Usp.} {\bf 46} 105-107 (Engl. transl.)

\bibitem{Eschrig04}
Eschrig M, Kopu J, Konstandin A, Cuevas JC, Fogelstr\"om M, Sch\"on G. 2004.
Singlet-Triplet Mixing in Superconductor-Ferromagnet Hybrid Devices.
\textit{Adv. in Solid State Phys.} {\bf 44}, 533-545.

\bibitem{Golubov04}
Golubov AA, Kupriyanov MYu, Il'ichev E. 2004.
The current-phase relation in Josephson junctions.
\textit{Rev. Mod. Phys.} {\bf 76}, 411-469.

\bibitem{Buzdin05}
Buzdin AI. 2005.
Proximity effects in superconductor-ferromagnet heterostructures.
\textit{Rev. Mod. Phys.} {\bf 77}, 935-976.

\bibitem{Bergeret05}
Bergeret FS, Volkov AF, Evetov KB. 2005.
Odd triplet superconductivity and related phenomena in superconductor-ferromagnet structures.
\textit{Rev. Mod. Phys.} {\bf 77}, 1321-1373.

\bibitem{Lyuksyutov07}
Lyuksyutov IF, Pokrovsky VL. 2007.
Ferromagnet-superconductor hybrids.
{\it Adv. Phys.} {\bf 54}, 67-136.

\bibitem{Eschrig11}
Eschrig M. 2011.
Spin-polarized supercurrents for spintronics.
\textit{Physics Today} {\bf 64}, 43-49.

\bibitem{Blamire14}
Blamire MG, Robinson JWA. 2014.
The interface between superconductivity and magnetism: understanding and device prospects.
{\it J. Phys. Condens. Matter} {\bf 26} 453201 (1-13).

\bibitem{Linder15}
Linder J, Robinson JWA. 2015.
Superconducting spintronics.
\textit{Nature Physics} {\bf 11}, 307-315.

\bibitem{Eschrig15}
Eschrig M. 2015.
Spin-polarized supercurrents for spintronics: a review of current progress.
\textit{Reports on Progress in Physics} {\bf 78}, 104501 (1-50); arXiv: 1509.02242, 1-95.

\bibitem{Eschrig00}
Eschrig M. 2000.
Distribution functions in nonequilibrium theory of superconductivity and Andreev spectroscopy in unconventional superconductors.
\textit{Phys. Rev. B} {\bf 61}, 9061-9076.

\bibitem{Grein09}
Grein R, Eschrig M, Metalidis G, Sch\"on G. 2009.
Spin-Dependent Cooper Pair Phase and Pure Spin Supercurrents in Strongly Polarized Ferromagnets.
\textit{Phys. Rev. Lett.} {\bf 102} 226005 (1-4).

\bibitem{Grein13}
Grein R, L\"ofwander T, Eschrig M. 2013.
Inverse proximity effect and influence of disorder on triplet supercurrents in strongly spin-polarized ferromagnets.
\textit{Phys. Rev. B} {\bf 88}, 054502 (1-11).

\bibitem{Nagato93}
Nagato Y, Nagai K, Hara J. 1993.
\textit{J. Low Temp. Phys.} {\bf 93}, 33-56.

\bibitem{Schopohl95}
Schopohl N, Maki K. 1995.
Quasiparticle spectrum around a vortex line in a $d$-wave superconductor.
\textit{Phys. Rev. B} {\bf 52}, 490-493.

\bibitem{Halterman09}
Halterman K, Valls OT. 2009.
Emergence of triplet correlations in superconductor/half-metallic nanojunctions with spin-active interfaces.
\textit{Phys. Rev. B} {\bf 80} 104502 (1-13).

\bibitem{Galaktionov08}
Galaktionov AV, Kalenkov MS, Zaikin AD. 2008. 
Josephson current and Andreev states in superconductor-half metal-superconductor heterostructures.
\textit{Phys. Rev. B} {\bf 77}, 094520 (1-11).

\bibitem{Deutscher05}
Deutscher G. 2005. 
Andreev-Saint-James reflections: A probe of cuprate superconductors. 
\textit{Rev. Mod. Phys.} {\bf 77}, 109-135.

\bibitem{Heeger88}
Heeger AJ, Kivelson S, Schrieffer JR, Su W-P. 1988.
Solitons in conducting polymers.
\textit{Rev. Mod. Phys.} {\bf 60}, 781-850.

\bibitem{Jackiw76}
Jackiw R, Rebbi C. 1976. 
Solitons with fermion number 1/2.
\textit{Phys. Rev. D} {\bf 13} 3398-3409.

\bibitem{Atiyah75}
Atiyah MF, Patodi VK, Singer IM. 1975.
Spectral asymmetry and Riemannian geometry. I.
\textit{Math. Proc. Cambridge Philos. Soc.} {\bf 77}, 43-69. 

\bibitem{Eschrig15a}
Eschrig M, Cottet A, Belzig W, Linder J. 2015.
General Boundary Conditions for Quasiclassical Theory of Superconductivity in the Diffusive Limit: Application to Strongly Spin-polarized Systems.
\textit{New Journal of Physics} {\bf 17}, 083037 (1-21).

\bibitem{Slonc96}
Slonczewski JC. 1996.
Current-driven excitation of magnetic multilayers.
\textit{Journal of Magnetism and Magnetic Materials} {\bf 159}, L1-L7.

\bibitem{Ralph08}
Ralph DC, Stiles MD. 2008.
Spin transfer torques.
\textit{Journal of Magnetism and Magnetic Materials} {\bf 320}, 1190-1216.

\bibitem{Brataas12}
Brataas A, Kent A D, Ohno H. 2012.
Current-induced torques in magnetic materials.
\textit{Nature Materials} {\bf 11}, 372-381.

\bibitem{Locatelli14}
Locatelli N, Cros V, Grollier J. 2014.
Spin-torque building blocks.
\textit{Nature Materials} {\bf 13}, 11-20.

\bibitem{Zhao07}
Zhao E, Sauls JA. 2007.
Dynamics of Spin Transport in Voltage-Biased Josephson Junctions.
\textit{Phys. Rev. Lett.} {\bf 98}, 206601 (1-4).

\bibitem{Zhao08}
Zhao E, Sauls JA. 2008.
Theory of nonequilibrium spin transport and spin-transfer torque
in superconducting-ferromagnetic nanostructures.
\textit{Phys. Rev. B} {\bf 78}, 174511 (1-15).

\bibitem{Waintal02}
Waintal X, Brouwer PW. 2002.
Magnetic exchange interaction induced by a Josephson current.
\textit{Phys. Rev. B} {\bf 65}, 054407 (1-11).

\bibitem{Kummel85}
K\"ummel R, Senftinger W. 1985.
Andreev-Reflected Wave Packets in Voltage-Biased Superconducting Quantum Wells.
\textit{Z. Phys. B - Condensed Matter} {\bf 59}, 275-281.

\bibitem{Arnold87}
Arnold GB. 1987.
Superconducting tunneling without the tunneling Hamiltonian. II. Subgap harmonic structure.
\textit{J. Low Temp. Phys.} {\bf 68}, 1-27.

\bibitem{Averin95}
Averin DV, Bardas A. 1995. 
ac Josephson Effect in a Single Quantum Channel.
\textit{Phys. Rev. Lett.} {\bf 75}, 1831-1834.

\bibitem{Wang10}
Wang S, Tang L, Xia K. 2010.
Spin transfer torque in the presence of Andreev reflections.
\textit{Phys. Rev. B} {\bf 81}, 094404 (1-7).

\bibitem{Shomali11}
Shomali Z, Zareyan M, Belzig W. 2011.
Spin supercurrent in Josephson contacts with noncollinear ferromagnets.
\textit{New Journal of Physics} {\bf 13}, 083033 (1-14).

\bibitem{Linder11}
Linder J, Yokoyama T. 2011.
Supercurrent-induced magnetization dynamics in a Josephson junction
with two misaligned ferromagnetic layers.
\textit{Phys. Rev. B} {\bf 83}, 012501 (1-4).

\bibitem{Mai11}
Mai S, Kandelaki E, Volkov AF, Efetov KB. 2011.
Interaction of Josephson and magnetic oscillations in Josephson tunnel junctions with a
ferromagnetic layer.
\textit{Phys. Rev. B} {\bf 84}, 144519 (1-12).

\bibitem{Linder14}
Kulagina I, Linder J. 2014.
Spin supercurrent, magnetization dynamics, and $\phi$-state in spin-textured Josephson junctions.
\textit{Phys. Rev. B} {\bf 90}, 054504 (1-14).

\bibitem{Holmqvist11}
Holmqvist C, Teber S, Fogelstr\"om M. 2011.
Nonequilibrium effects in a Josephson junction coupled to a precessing spin.
\textit{Phys. Rev. B} {\bf 83}, 104521 (1-18).

\bibitem{Holmqvist14}
Holmqvist C, Fogelstr\"om M, Belzig W. 2014.
Spin-polarized Shapiro steps and spin-precession-assisted multiple Andreev reflection.
\textit{Phys. Rev. B} {\bf 90}, 014516 (1-9).

\bibitem{Jong95}
de Jong MJM, Beenakker CWJ. 1995.
Andreev Reflection in Ferromagnet-Superconductor Junctions.
\textit{Phys. Rev. Lett.} {\bf 74}, 1657-1660.

\bibitem{Kashiwaya99}
Kashiwaya S, Tanaka Y, Yoshida N, Beasley MR. 1999.
Spin current in ferromagnet-insulator-superconductor junctions.
\textit{Phys. Rev. B} {\bf 60}, 3572-3580.

\bibitem{Zutic00}
\v{Z}uti\'c I, Valls OT. 2000.
Tunneling spectroscopy for ferromagnet/superconductor junctions.
\textit{Phys. Rev. B} {\bf 61}, 1555-1566; Erratum: \textit{Phys. Rev. B} {\bf 61}, 14845.

\bibitem{Mazin01}
Mazin II, Golubov AA, Nadgorny B. 2001.
Probing spin polarization with Andreev reflection: A theoretical basis.
\textit{J. Appl. Phys.} {\bf 89}, 7576-7578.

\bibitem{Perez04}
P\'erez-Willard F, Cuevas JC, S\"urgers C, Pfundstein P, Kopu J, Eschrig M, von L\"ohneysen H. 2004.
Determining the current polarization in Al/Co nanostructured point contacts.
\textit{Phys. Rev. B} {\bf 69}, 140502 (1-4).

\bibitem{Eschrig13}
Eschrig M, Golubov AA, Mazin II, Nadgorny B, Tanaka Y, Valls OT, and \v{Z}uti\'c I. 2013.
Comment on ``Unified Formalism of Andreev Reflection at a Ferromagnet/Superconductor Interface''.
\textit{Phys. Rev. Lett.} {\bf 111}, 139703 (1-2).

\bibitem{Grein10}
Grein R, L\"ofwander T, Metalidis G, Eschrig M. 2010.
Theory of superconductor-ferromagnet point-contact spectra: The case of strong spin polarization.
\textit{Phys. Rev. B} {\bf 81}, 094508 (1-17).

\bibitem{Piano11}
Piano S, Grein R, Mellor CJ, V\'yborn\'y K, Campion R, Wang M, Eschrig M, Gallagher BL. 2011.
Spin polarization of (Ga,Mn)As measured by Andreev spectroscopy: The role of spin-active scattering.
\textit{Phys. Rev. B} {\bf 83}, 081305(R) (1-4).

\bibitem{Kupferschmidt11}
Kupferschmidt JN, Brouwer PW. 2011.
Andreev reflection at half-metal/superconductor interfaces with nonuniform magnetization.
\textit{Phys. Rev. B} {\bf 83}, 014512 (1-14).

\bibitem{Wilken12}
Wilken FB, Brouwer PW. 2012.
Impurity-assisted Andreev reflection at a spin-active half metal-superconductor interface.
\textit{Phys. Rev. B} {\bf 85}, 134531 (1-10).

\bibitem{Yates13}
Yates KA, Anwar MS, Aarts J, Conde O, Eschrig M, L\"ofwander T, Cohen LF. 2013.
Andreev spectroscopy of CrO$_2$ thin films on TiO$_2$ and Al$_2$O$_3$.
\textit{Europhys. Lett.} {\bf 103}, 67005 (1-5).

\bibitem{Sun15}
Kuei Sun, Shah N. 2015.
General framework for transport in spin-orbit-coupled superconducting heterostructures: Nonuniform spin-orbit coupling and spin-orbit-active interfaces.
\textit{Phys. Rev. B} {\bf 91}, 144508 (1-8).

\bibitem{Visani12}
Visani C, Sefrioui Z, Tornos J, Leon C, Briatico J, Bibes M, Barth\'el\'emy A, Santamar\'ia J, Villegas JE. 2012.
Equal-spin Andreev reflection and long-range coherent transport in high-temperature superconductor/half-metallic ferromagnet junctions.
\textit{Nature Physics} {\bf 8}, 539-543.

\bibitem{Kalenkov10}
Kalenkov MS, Zaikin AD. 2010.
Nonlocal spin-sensitive electron transport in diffusive proximity heterostructures.
\textit{Phys. Rev. B} {\bf 82}, 024522 (1-8).

\bibitem{Machon13}
Machon P, Eschrig M, Belzig W. 2013.
Nonlocal Thermoelectric Effects and Nonlocal Onsager relations in a Three-Terminal Proximity-Coupled Superconductor-Ferromagnet Device.
\textit{Phys. Rev. Lett.} {\bf 110}, 047002 (1-5).

\bibitem{Machon14}
Machon P, Eschrig M, Belzig W. 2014.
Giant thermoelectric effects in a proximity-coupled superconductor-ferromagnet device.
\textit{New Journal of Physics} {\bf 16}, 073002 (1-19).

\bibitem{Ozaeta14}
Ozaeta A, Virtanen P, Bergeret FS, Heikkil\"a TT. 2014.
Predicted Very Large Thermoelectric Effect in Ferromagnet-Superconductor Junctions in the Presence of a Spin-Splitting Magnetic Field.
\textit{Phys. Rev. Lett.} {\bf 112}, 057001 (1-5).

\bibitem{Cadden07}
Cadden-Zimansky P, Jiang Z, Chandrasekhar V. 2007.
Charge imbalance, crossed Andreev reflection and elastic co-tunnelling in
ferromagnet/superconductor/normal-metal structures. 
{\it New Journal of Physics} {\bf 9}, 116 (1-25).

\bibitem{Brauer10}
Brauer J, H\"ubler F, Smetanin M, Beckmann D, von L\"ohneysen H. 2010.
Nonlocal transport in normal-metal/superconductor hybrid structures: Role of interference and interaction.
\textit{Phys. Rev. B} {\bf 81}, 024515 (1-7).

\bibitem{Colci12}
Colci M, Kuei Sun, Shah N, Vishveshwara S, Van Harlingen DJ. 2012.
Anomalous polarization-dependent transport in nanoscale double-barrier superconductor/ferromagnet/ superconductor junctions.
\textit{Phys. Rev. B} {\bf 85}, 180512 (1-5).

\bibitem{Flokstra15}
Flokstra MG, Satchell N, Kim J, Burnell G, Curran PJ, Bending SJ, Cooper JFK, Kinane CJ, Langridge S, Isidori A, Pugach N, Eschrig M, Luetkens H, Suter A, Prokscha T, Lee SL. 2015.
Remotely induced magnetism in a normal metal using a superconducting spin-valve.
\textit{Nature Physics} accepted for publication; arXiv:1505.03565.

\bibitem{Cadden09}
Cadden-Zimansky P, Wei J, Chandrasekhar V. 2009.
Cooper-pair-mediated coherence between two normal metals.
\textit{Nature Physics} {\bf 5}, 393-397.

\bibitem{Eschrig09NV}
Eschrig M. 2009.
Superconductor-metal heterostructures: Coherent conductors at a distance.
\textit{Nature Physics} {\bf 5}, 384-385.

\bibitem{Noh13}
Noh T, Houzet M, Meyer JS, Chandrasekhar V. 2013.
Nonlocal spin correlations mediated by a superconductor.
\textit{Phys. Rev. B} {\bf 87}, 220502(R) (1-4).


\end{thebibliography}
\end{document}